\documentclass[preprint,3p]{elsarticle}

\usepackage{amssymb}
\usepackage{amsmath}
\usepackage{graphicx}
\usepackage{dcolumn}
\usepackage{bm}
\usepackage{psfrag}
\usepackage{tikz}
\usepackage{pgfplots}
\usepackage{enumitem}

\usepackage{hyperref}
\allowdisplaybreaks

\begin{document}

\begin{frontmatter}



\title{A Kinetic Scheme based on Positivity Preservation with Exact Shock Capture}


\author[1]{Shashi Shekhar Roy\corref{cor1}%
\fnref{fn1}}
\author[2]{S. V. Raghurama Rao\fnref{fn2}}
\cortext[cor1]{Corresponding author}
\fntext[fn1]{E-mail addresses: shashi@iisc.ac.in, shashisroy@gmail.com}
\fntext[fn2]{E-mail address: raghu@iisc.ac.in}
\affiliation[1]{organization={Research Scholar, Department of Aerospace Engineering, Indian Institute of Science},
            city={Bangalore},
            country={India}}
						
\affiliation[2]{organization={Department of Aerospace Engineering, Indian Institute of Science},
            city={Bangalore},
            country={India}}

\begin{abstract}
In this paper, we present a kinetic model with flexible velocities that satisfy positivity preservation conditions for the Euler equations. Our 1D kinetic model consists of two velocities and employs both the asymmetrical and symmetrical models. Switching between the two models is governed by our formulation of kinetic relative entropy along with an additional criterion to ensure an accurate, entropic, and robust scheme. In 2D, we introduce a novel three-velocity kinetic model, defined to ensure a locally one-dimensional formulation for the resulting macroscopic normal flux. For first order accuracy, we also obtain a limit on the time step which ensures positivity preservation. The resulting numerical scheme captures grid-aligned steady shocks exactly. Several benchmark compressible flow test cases are solved in 1D and 2D to demonstrate the efficacy of the proposed solver.
\end{abstract}

%


\end{frontmatter}



\section{Introduction}
Among the different types of numerical schemes for solving the hyperbolic Euler equations, kinetic schemes stand out, as they do not directly discretize the Euler equations. Instead, kinetic schemes start by discretizing the Boltzmann equation, which is the governing equation at the kinetic level. Suitable moments of the Boltzmann equation give us the macroscopic conservation laws and this strategy leads to efficient kinetic or Boltzmann schemes. Some of the well-known early generation kinetic schemes are described in \cite{chu1965kinetic,sanders1974possible,pullin1980direct,reitz1981one,deshpande1986second,deshpande1986kinetic,mandal1994kinetic,kaniel1988kinetic,perthame1990boltzmann,prendergast1993numerical,raghurama1995peculiar}. While there are many more in this category, the schemes of particular interest is the vector kinetic framework with discrete velocities, developed by Natalini \cite{natalini1998discrete} and Aregba-Driollet and Natalini \cite{aregba2000discrete}. Shrinath {\em et al.} \cite{shrinath2023kinetic} have utilized the conservation form of vector kinetic equations and flexible velocities satisfying Rankine-Hugoniot (R-H) jump conditions at the interface to develop a low diffusive kinetic scheme. Recently, Shashi Shekhar Roy and S. V. Raghurama Rao \cite{ROY2023106016} have described a kinetic model for which the equilibrium distribution has flexible velocities and ranges of velocities, and its moments are utilized in a vector kinetic framework. The velocities are used to satisfy the R-H conditions, while the ranges of velocities are used to provide additional numerical diffusion in smoothly varying flow regions. For the purpose of identifying smooth flow regions, a novel formulation for relative entropy is also presented.  The last two of the above works focus on the exact capture of steady discontinuities, which could be enforced in the discretization only in macroscopic schemes before them.    

Positivity preservation is another important feature of numerical schemes and refers to the preservation of positivity of density and pressure (or internal energy) at all points and at all later times, given the initial solution with positive density and pressure. It is a desirable feature in numerical schemes, since most schemes fail once negative pressure is encountered. This generally occurs at high Mach numbers, in near-vacuum conditions, and in the presence of strong gradients, in general. In the category of kinetic schemes, Perthame \cite{perthame1990boltzmann} first demonstrated the positivity of density and pressure in his kinetic scheme. Later, Estivalezes and Villedieu \cite{estivalezes1996high} showed that the Kinetic Flux Vector splitting scheme\cite{deshpande1986kinetic} preserves the positivity of density and pressure under a CFL-like condition. They also extended the positivity property to second order accuracy for a positivity preserving flux vector splitting scheme. Gressier {\em et al.} \cite{GRESSIER1999199} showed that flux vector splitting schemes like those of van Leer and Steger-Warming are positivity preserving under a CFL-like condition. Thus, in summary, many flux vector splitting schemes are positivity preserving. But they have an inherent drawback of being very diffusive. In the category of approximate Riemann solvers, positivity preservation was attempted by Einfeldt {\em et al.} \cite{EINFELDT1991273} by imposing bounds on the wave speeds in HLL (Harten-Lax-Leer) scheme. While there are others in this category, more recent work includes that by Parent\cite{PARENT2013194}, who has obtained a positivity preserving scheme by modifying the Roe's scheme such that the coefficient matrices of the discretized equations are positive. However, many of these solvers are strongly dependent on the eigenstructure of the underlying hyperbolic systems. Further, most of the Riemann solvers suffer from drawbacks like admitting entropy-violating solutions, as well as numerical instability issues like odd-even decoupling, carbuncle phenomenon, kinked Mach stems, etc. \cite{quirk1997contribution}. Since kinetic schemes are not dependent on the eigenstructure, they become attractive for enforcing additional features such as positivity and for application to flows with more realistic and complicated equations of state.

In the present work, we describe a kinetic model for the Euler equations with two flexible velocities in 1D, which are set to satisfy the conditions for positivity preservation. Our objective is to obtain a low-diffusive kinetic scheme which is also positivity preserving. Our numerical scheme is not based on the solution of any Riemann problem, and it neither requires the computation of Roe averages nor is dependent strongly on the eigenstructure. Our 2D formulation describes a novel three-velocity kinetic model which is defined to ensure that the resulting macroscopic normal flux at the interface takes a locally one-dimensional form. We use a reconstruction-based flux-limited approach and a higher order Runge-Kutta method to extend our basic scheme to second order accuracy. An extensive set of benchmark test cases is solved to demonstrate that our numerical scheme is robust, accurate, entropic, and free from numerical instability issues. As a final exercise, the basic scheme is extended to viscous flows, and some benchmark viscous problems are solved to demonstrate the ability of our numerical scheme to resolve the associated flow features.  

\section{Gas-kinetic theory}
According to kinetic theory, the dynamics of gases is governed by the Boltzmann equation, given by
\begin{equation}
\frac{\partial f}{\partial t}+ \textbf{v} \cdot\frac{\partial f}{\partial \textsl{\textbf{x}}}= Q(f)
\label{eq:gkt_1}
\end{equation}
Here, $f(t,\textbf{x}, \textbf{v})$ is the velocity distribution function, $\textbf{v}$ is molecular velocity, and $Q(f)$ is the rate of change of $f$ due to collisions. The linear advection term in Equation \eqref{eq:gkt_1} describes the temporal and spatial evolution of the molecular velocity distribution function, which moves the state away from equilibrium. The collision term, on the other hand, is non-linear and it drives the distribution function towards equilibrium, vanishing in a limit (better described with the BGK model mentioned later). The equilibrium distribution function given by gas-kinetic theory is the Maxwell-Boltzmann equilibrium distribution, defined by 
\begin{equation}
f^{eq}_{Maxwell}= \frac{\rho}{I_{0}} \left(\frac{\beta}{\pi}\right)^{N/2} exp\left(-\beta|\textbf{v}-\textbf{u}|^{2}\right) exp\left(-I/I_{0}\right) 
\label{eq:gkt_2}
\end{equation}
where $\beta$= $\frac{1}{2RT}$, $\textbf{u}(t,\textbf{x})$ is the macroscopic velocity, $N$ is the translational degrees of freedom, $I$ is the internal energy variable corresponding to non-translational degrees of freedom, $I_{0}~=~\frac{2- N(\gamma -1)}{2(\gamma -1)}RT$ and $\gamma$= $\frac{c_{p}}{c_{v}}$. The combined mass, momentum, and total energy of the particles is conserved during collisions. Thus, 1, $\textbf{v}$ and $I+ \frac{|\textbf{v}|^{2}}{2}$ are the collisional invariants. By multiplying the Boltzmann equation with the moment vector $\bm{\Psi}= \left[1, v_{1}, ..,v_{N}, I+ \frac{|\textbf{v}|^{2}}{2}\right]^{T}$ and integrating w.r.t. $\textbf{v}$ and $I$, {\em i.e.}, by taking moments, we obtain the macroscopic conservation laws of mass, momentum, and energy. Further, under the assumption that $f$ relaxes instantaneously to $f^{eq}$ (justified through operator splitting and the BGK model), the moments of the Boltzmann equation give us the inviscid Euler equations, as follows.  
\begin{equation}
\int_{\mathbb{R}^{N}}d\textbf{v} \int_{\mathbb{R}^{+}}dI \ \bm{\Psi}\left(\frac{\partial f}{\partial t}+ \frac{\partial (v_{i}f)}{\partial x_{i}}= 0,f= f^{eq}\right) \Rightarrow \frac{\partial \textbf{U}}{\partial t}+ \frac{\partial \textbf{G}_{i}}{\partial x_{i}}= 0
\label{eq:gkt_3}
\end{equation}
Here, the conserved variable vector $\textbf{U}$ and the inviscid flux vector along $i$ direction, $\textbf{G}_{i}$ for the Euler equations are given by
\begin{equation}
\textbf{U}= \begin{bmatrix} \rho \\ \rho u_{j} \\ \rho E \end{bmatrix}, \textbf{G}_{i}= \begin{bmatrix} \rho u_{i} \\ \rho u_{i}u_{j}+ p \delta_{ij} \\ (\rho E+ p)u_{i} \end{bmatrix}, \ E= e+ \frac{|\textbf{u}|^{2}}{2}, \ e= \frac{p}{(\gamma -1)\rho}
\label{eq:gkt_4}
\end{equation}
The moment relations can be written as
\begin{subequations}
	\label{eq:gkt_5}
 \begin{equation}
    \int_{\mathbb{R}^{N}}d\textbf{v} \int_{\mathbb{R}^{+}}dI \ \bm{\Psi} f= \int_{\mathbb{R}^{N}}d\textbf{v} \int_{\mathbb{R}^{+}}dI \ \bm{\Psi} f^{eq}=  \textbf{U}
 \end{equation}
 \begin{equation}
     \int_{\mathbb{R}^{N}}v_{i}d\textbf{v} \int_{\mathbb{R}^{+}}dI \ \bm{\Psi} f^{eq}= \textbf{G}_{i}
 \end{equation}
 \end{subequations}
	Multiplying the Boltzmann equation by $\ln f$ and taking its moment, we get the kinetic entropy inequality, {\em i.e.}, the H-theorem
	\begin{equation}
	\frac{\partial H}{\partial t}+ \frac{\partial H_{v,i}}{\partial x_{i}} \leq 0
	\label{eq:gkt_6}
	\end{equation}
	with
	\begin{subequations}
	\begin{equation}
	\int_{\mathbb{R}^{N}}d\textbf{v} \int_{\mathbb{R}^{+}}dI \ f \ln f= H,
	\label{eq:gkt_7a}
	\end{equation}
	\begin{equation}
	\int_{\mathbb{R}^{N}}v_{i}d\textbf{v} \int_{\mathbb{R}^{+}}dI \ f \ln f= H_{v,i},
	\label{eq:gkt_7b}
	\end{equation}
	\begin{equation}
	\int_{\mathbb{R}^{N}}d\textbf{v} \int_{\mathbb{R}^{+}}dI \ Q(f) \ln f \leq 0
	\label{eq:gkt_7c}
	\end{equation}
	\end{subequations}
	Next, we introduce a popular simplification to the collision term, called the BGK model \cite{bhatnagar1954model}. For this model, the collision term is approximated by the following expression.
	\begin{equation}
	Q(f)= -\frac{1}{\epsilon}\left[f- f^{eq}\right]
	\label{eq:gkt_8}
	\end{equation}
	Here, $\epsilon$ is the relaxation time. The description of Boltzmann equations thus far has assumed that molecular velocity is continuous, encompassing all possible real values. In contrast, in a discrete velocity Boltzmann system, the distribution function is typically a vector, with each component satisfying a Boltzmann equation while being advected by a single velocity. The Discrete Velocity Boltzmann-BGK Equations corresponding to the $i^{th}$ macroscopic equation take the following form.
	\begin{equation}
	\frac{\partial f_{ji}}{\partial t}+ \lambda_{jk}\frac{\partial f_{ji}}{\partial x_{k}}= -\frac{1}{\epsilon}\left[f_{ji}- f^{eq}_{ji}\right], \ j=1,..,N_{d}, \ k=1,...,N
	\label{eq:gkt_9}
	\end{equation}
	The number of discrete velocities, $N_{d}$ satisfies the condition, $N_{d}\geq N+1$. The moment relations become
	\begin{equation}
	\sum^{N_{d}}_{j=1}f_{ji}= \sum^{N_{d}}_{j=1}f^{eq}_{ji}= U_{i}, \ \sum^{N_{d}}_{j=1} \lambda_{jk}f^{eq}_{ji}= (G_{k})_{i}
	\label{eq:gkt_10}
	\end{equation}
	Thus, in a discrete kinetic framework, the moment relations get simplified, as complex integrals are replaced by simple summations. Consequently, the equilibrium distributions often reduce to simple linear combinations of the conserved variable vector and inviscid flux vectors. The work in this paper is based on further developments based on the {\em flexible velocity framework} introduced in \cite{shrinath2023kinetic,ROY2023106016}.    
	
\section{Kinetic model for 1D Euler Equations}
Our kinetic model in 1D comprises of two velocities, $\lambda_{p} \geq 0$ and $\lambda_{m} \leq 0$. For this model, we numerically solve Flexible Velocity Boltzmann Equations for $f_{1i}$ and $f_{2i}$, which are being advected by velocities $\lambda_{p}$ and $\lambda_{m}$ respectively. These Boltzmann equations, which correspond to the $i^{th}$ macroscopic equation, are given by
\begin{equation}
\frac{\partial \textbf{f}_{i}}{\partial t}+ \frac{\partial (\Lambda\textbf{f}_{i})}{\partial x}= 
- \frac{1}{\epsilon} \left[\textbf{f}_{i}- \textbf{f}_{i}^{eq}\right] 
\label{eq:1d_euler_4}
\end{equation}
Here,
\begin{equation}
\textbf{f}_{i}^{eq}= \begin{bmatrix} f^{eq}_{1i} \\ f^{eq}_{2i} \end{bmatrix}, \Lambda= \begin{bmatrix} \lambda_{p} & 0\\ 0 & \lambda_{m} \end{bmatrix},
\label{eq:1d_euler_5}
\end{equation}
Now, given the row vector $\textbf{P}_{i}= \begin{bmatrix}1 & 1\end{bmatrix}$, the moment relations become
\begin{subequations}
\label{eq:1d_euler_6}
\begin{equation}
\textbf{P}_{i}\textbf{f}_{i}^{eq}= f^{eq}_{1i}+ f^{eq}_{2i}= U_{i}
\end{equation}
\begin{equation}
\textbf{P}_{i}\Lambda\textbf{f}_{i}^{eq}= \lambda_{p}f^{eq}_{1i}+ \lambda_{m}f^{eq}_{2i}= G_{i}
\end{equation}
\end{subequations}
The moment relations in \eqref{eq:1d_euler_6} can be solved for $f^{eq}_{1i}$ and $f^{eq}_{2i}$ to give
\begin{equation}
f^{eq}_{1i}= \frac{-\lambda_{m}}{\lambda_{p}- \lambda_{m}}U_{i} + \frac{1}{\lambda_{p}- \lambda_{m}}G_{i},  \ \ f^{eq}_{2i}= \frac{\lambda_{p}}{\lambda_{p}- \lambda_{m}}U_{i} - \frac{1}{\lambda_{p}- \lambda_{m}}G_{i}
\label{eq:1d_euler_7}
\end{equation}
For the special case when the velocities are $\lambda$ and -$\lambda$, the expressions in \eqref{eq:1d_euler_7} simplify and become
\begin{equation}
f^{eq}_{1i}= \frac{U_{i}}{2}+ \frac{G_{i}}{2\lambda}, \ \ f^{eq}_{2i}= \frac{U_{i}}{2}- \frac{G_{i}}{2\lambda}
\label{eq:1d_euler_8}
\end{equation}
We work in a finite volume framework and numerically solve the Boltzmann Equations \eqref{eq:1d_euler_4}, written in conservation form, for the $j^{th}$ cell. A uniform cell size ($\Delta x$) is assumed in 1D.
Operator-splitting strategy is used to solve the Boltzmann equations. At the end of $n^{th}$ time step, the distribution function is relaxed instantaneously to the equilibrium distribution function. In the next step, the advective part of Boltzmann equations is discretized and solved numerically to obtain the distribution function for the next time step, as follows.
\begin{subequations}
\label{eq:1d_euler_9}
\[\text{Relaxation step: Instantaneous, i.e. }\epsilon\rightarrow 0\text{. Thus,} \]
\begin{equation}
(\textbf{f}_{i})^{n}_{j}=  (\textbf{f}_{i}^{eq})^{n}_{j}
\end{equation}
\[\text{Advection step: The advective part of Boltzmann equation is,}\]
\begin{equation}
\frac{\partial (\textbf{f}_{i})_{j}}{\partial t}+ \frac{\partial (\textbf{h}_{i})_{j}}{\partial x}= 0; \textbf{h}_{i}= \Lambda\textbf{f}_{i}^{eq}\text{. In integral form,}\nonumber
\end{equation}
\begin{equation}
\frac{d (\textbf{f}_{i})_{j}}{dt} = -\frac{1}{\Delta x}\left[ (\textbf{h}_{i})_{j+1/2}^{n}- (\textbf{h}_{i})_{j-1/2}^{n}\right]
\end{equation}
\end{subequations}
In the present work, we are using flux difference splitting to define the interface kinetic flux $(\textbf{h}_{i})_{j+\frac{1}{2}}$ as follows.
\begin{equation}
(\textbf{h}_{i})_{j+\frac{1}{2}} = \frac{1}{2}\left\{(\textbf{h}_{i})_{j}+ (\textbf{h}_{i})_{j+1}\right\}-\frac{1}{2}\left\{(\Delta \textbf{h}_{i}^{+})_{j+\frac{1}{2}}- (\Delta \textbf{h}_{i}^{-})_{j+\frac{1}{2}}\right\}
\label{eq:1d_euler_11}
\end{equation}
with
\begin{subequations}
\label{eq:1d_euler_12}
\begin{equation}
\Lambda^{\pm}= \frac{\Lambda \pm |\Lambda|}{2}
\end{equation}
\begin{equation}
(\Delta \textbf{h}_{i}^{+})_{j+ \frac{1}{2}}= \left(\Lambda^{+}\Delta \textbf{f}_{i}^{eq}\right)_{j+ \frac{1}{2}}= \begin{bmatrix}\left(\lambda_{p}\Delta f^{eq}_{1i}\right)_{j+ \frac{1}{2}} \\ 0 \end{bmatrix}= \begin{bmatrix}(\lambda_{p})_{j+\frac{1}{2}}\left\{(f^{eq}_{1i})_{j+1} -(f^{eq}_{1i})_{j}\right\} \\ 0 \end{bmatrix}
\end{equation}
\begin{equation}
(\Delta \textbf{h}_{i}^{-})_{j+ \frac{1}{2}}= \left(\Lambda^{-}\Delta \textbf{f}_{i}^{eq}\right)_{j+ \frac{1}{2}}= \begin{bmatrix} 0 \\ \left(\lambda_{m}\Delta f^{eq}_{2i}\right)_{j+ \frac{1}{2}} \end{bmatrix}= \begin{bmatrix} 0 \\ (\lambda_{m})_{j+\frac{1}{2}}\left\{(f^{eq}_{2i})_{j+1} -(f^{eq}_{2i})_{j}\right\}  \end{bmatrix}
\end{equation}
\end{subequations}
The temporal derivative is approximated using the forward Euler method. The discretized equations thus become
\begin{equation}
(\textbf{f}_{i})_{j}^{n+1} = (\textbf{f}_{i}^{eq})_{j}^{n} -\frac{\Delta t}{\Delta x}\left[ (\textbf{h}_{i})_{j+\frac{1}{2}}^{n}- (\textbf{h}_{i})_{j-\frac{1}{2}}^{n}\right]
\label{eq:1d_euler_13}
\end{equation}
The macroscopic update formula, obtained by taking moments of Equation \eqref{eq:1d_euler_13}, is given by
\begin{equation}
(U_{i})_{j}^{n+1} = (U_{i})_{j}^{n} -\frac{\Delta t}{\Delta x}\left[ (G_{i})_{j+\frac{1}{2}}^{n}- (G_{i})_{j-\frac{1}{2}}^{n}\right]
\label{eq:1d_euler_14}
\end{equation}
where
\begin{equation}
(G_{i})_{j+\frac{1}{2}}= \textbf{P}_{i} (\textbf{h}_{i})_{j+\frac{1}{2}}= \frac{1}{2}\left\{(G_{i})_{j}+ (G_{i})_{j+1}\right\}-\frac{1}{2}\left\{(\Delta G_{i}^{+})_{j+\frac{1}{2}}- (\Delta G_{i}^{-})_{j+\frac{1}{2}}\right\}
\label{eq:1d_euler_15}
\end{equation}
with
\begin{subequations}
\label{eq:1d_euler_16}
\begin{eqnarray}
(\Delta G_{i}^{+})_{j+\frac{1}{2}} &&= \textbf{P}_{i}(\Delta \textbf{h}_{i}^{+})_{j+ \frac{1}{2}}= \left(\lambda_{p}\Delta f^{eq}_{1i}\right)_{j+ \frac{1}{2}} \nonumber\\
&&= \left( \frac{\lambda_{p}}{\lambda_{p}- \lambda_{m}}\right)_{j+\frac{1}{2}} \left\{ (G_{i})_{j+1}- (G_{i})_{j}\right\}- \left(\frac{\lambda_{p}\lambda_{m}}{\lambda_{p}- \lambda_{m}} \right)_{j+\frac{1}{2}} \left\{ (U_{i})_{j+1}- (U_{i})_{j} \right\}
\end{eqnarray}
\begin{eqnarray}
(\Delta G_{i}^{-})_{j+\frac{1}{2}} &&= \textbf{P}_{i}(\Delta \textbf{h}_{i}^{-})_{j+ \frac{1}{2}}= \left(\lambda_{m}\Delta f^{eq}_{2i}\right)_{j+ \frac{1}{2}} \nonumber\\
&&= \left(\frac{-\lambda_{m}}{\lambda_{p}- \lambda_{m}}\right)_{j+\frac{1}{2}} \left\{ (G_{i})_{j+1}- (G_{i})_{j}\right\}+ \left(\frac{\lambda_{p}\lambda_{m}}{\lambda_{p}- \lambda_{m}}\right)_{j+\frac{1}{2}} \left\{ (U_{i})_{j+1}- (U_{i})_{j}\right\}
\end{eqnarray}
\end{subequations}
The macroscopic flux vector at the interface can be written in the familiar HLL type flux form as,
\begin{equation}
\textbf{G}_{j+\frac{1}{2}}=\left(\frac{\lambda_{p}}{\lambda_{p}- \lambda_{m}}\right)_{j+\frac{1}{2}}\textbf{G}_{j} - \left(\frac{\lambda_{m}}{\lambda_{p}- \lambda_{m}}\right)_{j+\frac{1}{2}}\textbf{G}_{j+1}+\left( \frac{\lambda_{p}\lambda_{m}}{\lambda_{p}- \lambda_{m}}\right)_{j+\frac{1}{2}}\left(\textbf{U}_{j+1}-\textbf{U}_{j}\right)
\label{eq:1d_euler_17}
\end{equation}
For the simplified model where the velocities are $\lambda$ and -$\lambda$ (with $\lambda >$0), the macroscopic flux simplifies to a scalar numerical diffusion model with the coefficient of numerical diffusion being $\lambda$, as follows.
\begin{equation}
\textbf{G}_{j+\frac{1}{2}}= \frac{1}{2} \left( \textbf{G}_{j}+ \textbf{G}_{j+1} \right) - \frac{\lambda_{j+\frac{1}{2}}}{2}\left(\textbf{U}_{j+1}-\textbf{U}_{j}\right)
\label{eq:1d_euler_18}
\end{equation}

\subsection{Positivity analysis}
Let us assume that the solution at the initial time is physically admissible, {\em i.e.}, the initial solution has positive density and pressure throughout the domain. A numerical scheme is then positively conservative/ positivity preserving if the numerical solution at all later times also has positive pressure and density. That is,
\begin{equation}
\rho\left(x,t_{0}\right)> 0, p\left(x,t_{0}\right)> 0\Rightarrow \rho\left(x,t\right)> 0, p\left(x,t\right)> 0, \forall \ t> t_{0}
\label{eq:1d_pos_1}
\end{equation}
Let \textbf{W} be the set of all physically admissible conserved variable vectors \textbf{U}. Then, the condition for positivity preservation can be reformulated as
\begin{equation}
\text{For positivity: }\textbf{U}\left(x,t_{0}\right) \in \textbf{W} \Rightarrow \textbf{U}\left(x,t\right) \in \textbf{W}, \forall \ t> t_{0}
\label{eq:1d_pos_2}
\end{equation}
We start the positivity analysis for our first order accurate numerical scheme by writing the macroscopic update formula in vector form as follows.
\begin{eqnarray}
&&\textbf{U}^{n+1}_{j}= \textbf{U}^{n}_{j}- \frac{\Delta t}{\Delta x}\left(\textbf{G}^{n}_{j+\frac{1}{2}}- \textbf{G}^{n}_{j-\frac{1}{2}}\right) \nonumber\\
&&= \textbf{U}^{n}_{j}- \frac{\Delta t}{\Delta x} \left[ \left\{ -\left( \frac{\lambda_{p}\lambda_{m}}{\lambda_{p}- \lambda_{m}}\right)^{n}_{j+\frac{1}{2}}\textbf{U}^{n}_{j} + \left(\frac{\lambda_{p}}{\lambda_{p}- \lambda_{m}}\right)^{n}_{j+\frac{1}{2}}\textbf{G}^{n}_{j} \right\} \right. \nonumber\\
 && + \left. \left\{ \left( \frac{\lambda_{p}\lambda_{m}}{\lambda_{p}- \lambda_{m}}\right)^{n}_{j+\frac{1}{2}}\textbf{U}^{n}_{j+1} - \left(\frac{\lambda_{m}}{\lambda_{p}- \lambda_{m}}\right)^{n}_{j+\frac{1}{2}}\textbf{G}^{n}_{j+1} \right\} \right] \nonumber\\
&& + \frac{\Delta t}{\Delta x} \left[ 
\left\{ -\left( \frac{\lambda_{p}\lambda_{m}}{\lambda_{p}- \lambda_{m}}\right)^{n}_{j-\frac{1}{2}}\hspace{-0.3cm}\textbf{U}^{n}_{j-1} + \left(\frac{\lambda_{p}}{\lambda_{p}- \lambda_{m}}\right)^{n}_{j-\frac{1}{2}}\hspace{-0.3cm}\textbf{G}^{n}_{j-1} \right\} +\right.
 \left. \left\{ \left( \frac{\lambda_{p}\lambda_{m}}{\lambda_{p}- \lambda_{m}}\right)^{n}_{j-\frac{1}{2}}\hspace{-0.3cm}\textbf{U}^{n}_{j} - \left(\frac{\lambda_{m}}{\lambda_{p}- \lambda_{m}}\right)^{n}_{j-\frac{1}{2}}\hspace{-0.3cm}\textbf{G}^{n}_{j} \right\} \right] \nonumber\\
&&= - \frac{\Delta t}{\Delta x}\left(\frac{\lambda_{m}}{\lambda_{p}- \lambda_{m}}\right)^{n}_{j+\frac{1}{2}} \underbrace{\left\{ \left(\lambda_{p}\right)^{n}_{j+\frac{1}{2}}\textbf{U}^{n}_{j+1} - \textbf{G}^{n}_{j+1} \right\}}_{\text{Term 1}} +
 \frac{\Delta t}{\Delta x}\left(\frac{\lambda_{p}}{\lambda_{p}- \lambda_{m}}\right)^{n}_{j-\frac{1}{2}} \underbrace{\left\{ -\left(\lambda_{m}\right)^{n}_{j-\frac{1}{2}}\textbf{U}^{n}_{j-1} + \textbf{G}^{n}_{j-1} \right\}}_{\text{Term 2}} \nonumber\\
&& + \underbrace{\textbf{U}^{n}_{j}- \frac{\Delta t}{\Delta x} \left[ \left\{\left(\frac{\lambda_{p}}{\lambda_{p}- \lambda_{m}}\right)^{n}_{j+\frac{1}{2}}\hspace{-0.25cm}+ \left(\frac{\lambda_{m}}{\lambda_{p}- \lambda_{m}}\right)^{n}_{j-\frac{1}{2}} \right\} \textbf{G}^{n}_{j} -\right.
 \left.\left\{ \left( \frac{\lambda_{p}\lambda_{m}}{\lambda_{p}- \lambda_{m}}\right)^{n}_{j+\frac{1}{2}}\hspace{-0.25cm}+ \left( \frac{\lambda_{p}\lambda_{m}}{\lambda_{p}- \lambda_{m}}\right)^{n}_{j-\frac{1}{2}}\right\} \textbf{U}^{n}_{j}  \right]}_{\text{Term 3}}\\
\label{eq:1d_pos_3}
\end{eqnarray}   
Let $\textbf{U}^{n}_{j} \in \textbf{W}$, $\forall$ j. Then, the numerical scheme is positivity preserving, {\em i.e.}, $\textbf{U}^{n+1}_{j} \in \textbf{W}$, if Terms 1, 2 and 3 in Equation \eqref{eq:1d_pos_3} are all positive. Thus, our numerical method is positivity preserving if the following conditions are all satisfied.
\begin{enumerate}
	\item $\left\{\left(\lambda_{p}\right)_{j+\frac{1}{2}}\textbf{U}_{j+1} - \textbf{G}_{j+1}\right\} \in \textbf{W}$ . Let $\left(\lambda_{p}\right)_{j+\frac{1}{2}}\textbf{U}_{j+1} - \textbf{G}_{j+1}$= $\begin{bmatrix} G_{1} & G_{2} & G_{3} \end{bmatrix}^{T}$. The positivity of density and pressure requires that $G_{1}\geq 0$ and $2G_{1}G_{3}-G^{2}_{2}\geq 0$. This gives us (as derived in \ref{appendix:a0}),\label{condition1}
\begin{equation}
\left(\lambda_{p}\right)_{j+\frac{1}{2}} \geq \left(u_{j+1}+ \sqrt{\frac{\gamma -1}{2\gamma}} a_{j+1}\right). 
\label{eq:1d_pos_4}
\end{equation}

\item $\left\{-\left(\lambda_{m}\right)_{j-\frac{1}{2}}\textbf{U}_{j-1} + \textbf{G}_{j-1}\right\} \in \textbf{W}$. Similarly, $\left\{-\left(\lambda_{m}\right)_{j+\frac{1}{2}}\textbf{U}_{j} + \textbf{G}_{j}\right\} \in \textbf{W}$. From this condition, we get,\label{condition2}
\begin{equation}
\left(\lambda_{m}\right)_{j+\frac{1}{2}} \leq \left(u_{j}- \sqrt{\frac{\gamma -1}{2\gamma}} a_{j}\right)
\label{eq:1d_pos_5}
\end{equation}
We note that for the special case of scalar numerical diffusion model, the positivity conditions in \eqref{eq:1d_pos_4} and \eqref{eq:1d_pos_5} can be combined to give,
\begin{equation}
\left(\lambda\right)_{j+\frac{1}{2}} \geq max\left( -u_{j}+ \sqrt{\frac{\gamma -1}{2\gamma}} a_{j}, u_{j+1}+ \sqrt{\frac{\gamma -1}{2\gamma}} a_{j+1}\right)
\label{eq:1d_pos_6}
\end{equation}

\item Term 3 in Equation \eqref{eq:1d_pos_3} is positive if it can be written as a positive coefficient matrix (see \cite{PARENT2011238}) multiplied by $\textbf{U}^{n}_{j}$. This requirement leads to a limit on the time step and is explored in Section \ref{section:1d_dt}.\label{condition3}
\end{enumerate}

\subsection{\texorpdfstring{Fixing $\lambda$'s}{Fixing λ's}}
We first define a non-negative numerical wave speed at a cell-interface $x_{j+\frac{1}{2}}$, which satisfies the Rankine-Hugoniot jump conditions (used in \cite{ROY2023106016}), as follows.
\begin{equation}
\left(\lambda_{RH}\right)_{j+\frac{1}{2}}= min_{i}\left(\frac{\left|\Delta G_{i}\right|}{\left|\Delta U_{i}\right|+ \epsilon_{0}}\right), \ \Delta= ()_{j+1}- ()_{j}
\label{eq:1d_lambda_1}
\end{equation}
Here, $\epsilon_{0}$ is a small positive term that prevents the denominator from going to zero. We have taken $\epsilon_{0}= 10^{-10}$ in the present work.\\
\begin{figure}
\centering
\includegraphics[width=15cm]{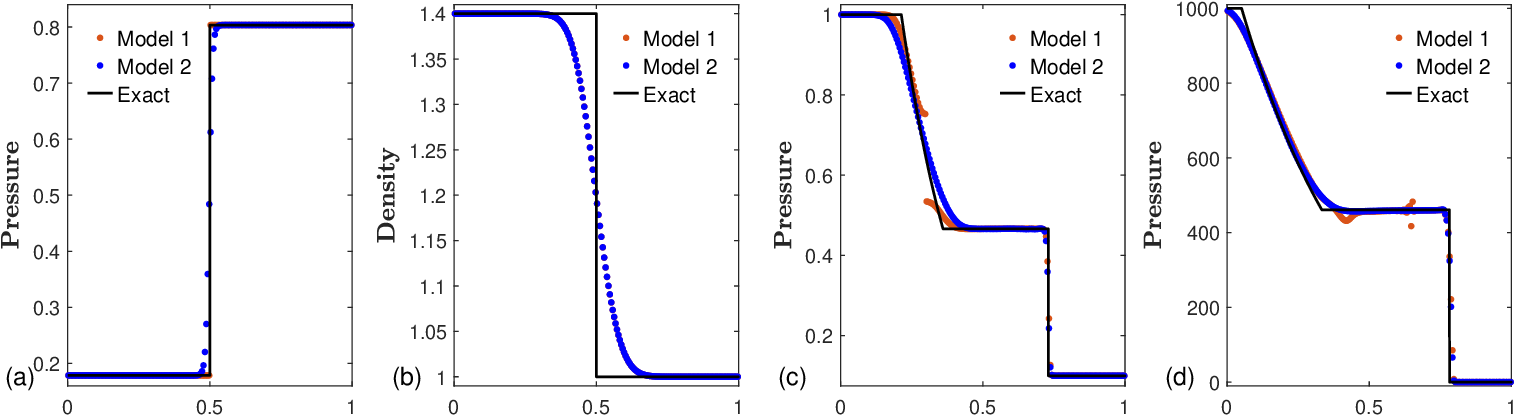}
\caption{\label{fig:1d_model_comparison} First order accurate results for Model 1 and Model 2 for a) Steady shock problem, b) Steady contact-discontinuity problem, c) Sod's shock tube problem, d) Left half portion of Woodward and Colella's blast wave problem.}
\end{figure} 
\textbf{Model 1}: For this unsymmetrical kinetic model, the velocities $\lambda_{p}$($\geq$0) and $\lambda_{m}$($\leq$0) are allowed to take different magnitudes. The resulting macroscopic flux is given by Equation \eqref{eq:1d_euler_17}. We define $\lambda_{p}$ and $\lambda_{m}$ as follows.
\begin{subequations}
\label{eq:1d_lambda_2}
\begin{equation}
\left(\lambda_{p}\right)_{j+\frac{1}{2}}= max\left(\left(\lambda_{RH}\right)_{j+\frac{1}{2}},u_{j+1}+ \sqrt{\frac{\gamma -1}{2\gamma}} a_{j+1}\right)
\end{equation}
\begin{equation}
\left(\lambda_{m}\right)_{j+\frac{1}{2}}= min\left(-\left(\lambda_{RH}\right)_{j+\frac{1}{2}},u_{j}- \sqrt{\frac{\gamma -1}{2\gamma}} a_{j}\right)
\end{equation}
\end{subequations}
Definitions in \eqref{eq:1d_lambda_2} ensure that $\lambda_{p}\geq$ 0 and $\lambda_{m}\leq$ 0, and that the positivity conditions \eqref{eq:1d_pos_4} and \eqref{eq:1d_pos_5} are also satisfied. Following observations (see Figure \ref{fig:1d_model_comparison} for reference) are made regarding the resulting numerical scheme for the 1D Euler equations.
\begin{enumerate}
	\item[a.]Numerical diffusion is optimal for the exact capture of a steady shock. We can show this analytically by considering a steady shock ($\lambda_{RH}$= 0) at cell-interface at $x_{j+\frac{1}{2}}$, with flow from left to right (backward facing shock). Across the shock, we have
	\begin{equation}
(u-a)_{j}> 0 \left( > (u-a)_{j+1}\right) \Rightarrow \left(u- \sqrt{\frac{\gamma -1}{2\gamma}}a\right)_{j} >(u-a)_{j}> 0
\label{eq:1d_lambda_3}
\end{equation}
Thus, at the interface,
\begin{equation}
\left(\lambda_{m}\right)_{j+\frac{1}{2}}= min \left(0,u_{j}- \sqrt{\frac{\gamma -1}{2\gamma}} a_{j}\right)= 0
\label{eq:1d_lambda_4}
\end{equation}
whereas $\left(\lambda_{p}\right)_{j+\frac{1}{2}} >$0. The interface numerical flux in Equation \eqref{eq:1d_euler_17} then simplifies to,
\begin{equation}
\textbf{G}_{j+\frac{1}{2}}= \textbf{G}_{j}
\label{eq:1d_lambda_4a}
\end{equation}
Similarly, for a forward facing steady shock, we have
\begin{equation}
\left(\lambda_{p}\right)_{j+\frac{1}{2}}= 0, \ \left(\lambda_{m}\right)_{j+\frac{1}{2}} <0\text{. Thus, }\textbf{G}_{j+\frac{1}{2}}= \textbf{G}_{j+1}
\label{eq:1d_lambda_5}
\end{equation}  
 Thus, similar to other HLL-type schemes that satisfy this property (see \cite{EINFELDT1988},\cite{EINFELDT1991273}), the numerical diffusion for our scheme is optimal at a steady shock wave, leading to its exact capture.  
\item[b.]A steady contact-discontinuity is captured sharply, but not exactly.   
\item[c.]Entropy-violating expansion shock is formed in the numerical results for Sod's shock tube problem. Further, there is a significant difference between the numerical and exact solutions for Woodward and Colella's blast wave problem.
\end{enumerate}

\textbf{Model 2}: For this model, the velocities taken are $\lambda$ and -$\lambda$, with $\lambda >$0. This leads to an interface flux with scalar numerical diffusion (Equation \eqref{eq:1d_euler_18}). We define $\lambda$ so that it satisfies the positivity conditions \eqref{eq:1d_pos_4} and \eqref{eq:1d_pos_5}, as follows.
\begin{eqnarray}
\left(\lambda\right)_{j+\frac{1}{2}}&=& max\left( \left(\lambda_{RH}\right)_{j+\frac{1}{2}}, -u_{j}+ \sqrt{\frac{\gamma -1}{2\gamma}} a_{j}, u_{j+1}+ \sqrt{\frac{\gamma -1}{2\gamma}} a_{j+1}\right)\\
&=& max\left( \lambda_{p}, -\lambda_{m}\right)_{j+\frac{1}{2}}\nonumber
\label{eq:1d_lambda_6}
\end{eqnarray}
Following observations (see Figure \ref{fig:1d_model_comparison}) are made for this numerical scheme.  
\begin{enumerate}
	\item[a.]Numerical solutions for steady shock and steady contact-discontinuity problems are diffused.
	\item[b.]No entropy-violating shocks are formed for Sod's shock tube problem, {\em i.e.}, the solution is entropic. Further, the numerical results for Woodward and Colella's problem are agreeable.
\end{enumerate}

\textbf{Final kinetic model}: Based on the observations made, we take advantage of both the kinetic models by utilizing the asymmetrical model in large gradient regions and the symmetrical model in smoothly varying regions of flow. To identify expansions and other smooth flow regions, the following kinetic relative entropy based criterion (previously used in \cite{ROY2023106016,SHASHI_HYP}) is used.
\begin{equation}
\text{For smoothly varying flow regions: }d^{2} > 0 \ \textrm{and} \ |\Delta s| \leq k(s_{max}-s_{min}); \Delta=()_{R}- ()_{L}
\label{eq:1d_lambda_7}
\end{equation}
where $s_{max}$ and $s_{min}$ are the maximum and minimum entropy in the domain at a given time level. $k$ is a fraction, taken as small as possible. We have taken $k= 0.1$ for all our test cases. The kinetic relative entropy $d^{2}$ is given by
\begin{eqnarray}
d^{2} &&= \left\langle \Delta \left(\frac{\partial H(f^{eq})}{\partial f^{eq}}\right)\Delta f^{eq}\right\rangle \\
&&= \Delta \left\{R\left(\frac{\gamma -s/c_{v}}{\gamma -1} - \frac{\rho u^{2}}{2p}\right)\right\} \Delta \left(\rho\right)+ \Delta \left(R\frac{\rho u}{p}\right) \Delta \left(\rho u\right)+ \Delta \left(-R\frac{\rho}{p}\right) \Delta \left(\rho E\right)
\label{eq:1d_lambda_8}
\end{eqnarray}
The derivation of the final expression \eqref{eq:1d_lambda_8} for $d^{2}$ for the present kinetic model is shown in \ref{appendix:a3}. Thus, the lambdas for our final numerical scheme are defined as follows.
\begin{subequations}
\label{eq:1d_lambda_9}
\begin{equation}
\text{In smoothly varying flow regions: }\left(\lambda\right)_{j+\frac{1}{2}}= max\left( \left(\lambda_{RH}\right)_{j+\frac{1}{2}}, -u_{j}+ \sqrt{\frac{\gamma -1}{2\gamma}} a_{j}, u_{j+1}+ \sqrt{\frac{\gamma -1}{2\gamma}} a_{j+1}\right)
\end{equation}
\begin{eqnarray}
\text{Everywhere else: }\left(\lambda_{p}\right)_{j+\frac{1}{2}}&=& max\left(\left(\lambda_{RH}\right)_{j+\frac{1}{2}},u_{j+1}+ \sqrt{\frac{\gamma -1}{2\gamma}} a_{j+1}\right),\nonumber\\
\left(\lambda_{m}\right)_{j+\frac{1}{2}}&=& min\left(-\left(\lambda_{RH}\right)_{j+\frac{1}{2}},u_{j}- \sqrt{\frac{\gamma -1}{2\gamma}} a_{j}\right)
\end{eqnarray}
\end{subequations}
Following are the features of the resulting numerical scheme.
\begin{enumerate}
	\item[a.]In our scheme, we switch from Model 1 to Model 2 in smoothly varying regions of flow. Since the positivity conditions \eqref{eq:1d_pos_4} and \eqref{eq:1d_pos_5} are satisfied by both formulations, our numerical flux thus satisfies the two positivity conditions.
	\item[b.]Our numerical scheme preserves the exact shock-capturing property of Model 1.
	\item[c.]Our numerical scheme switches over to the scalar numerical diffusion model in smoothly varying flow regions, which include expansive sonic points. No entropy-violating expansion shocks are formed, making our scheme entropic. Further, the numerical results for the Woodward and Colella problem are reasonably accurate.
	\item[d.]We use our discrete kinetic formulation of relative entropy and an additional condition to identify smoothly varying flow regions.
\end{enumerate}

\subsection{Time step restrictions}
\label{section:1d_dt}
Both positivity and numerical stability considerations impose limits on the maximum allowable time step. The final time step $\Delta t$ is calculated by taking the minimum of the two limiting time steps.

\subsubsection{Maximum time step based on positivity}
The positivity condition \ref{condition3} gives us a limit on the time step. To begin, we rewrite Term 3 in Equation \eqref{eq:1d_pos_3} as
\begin{equation}
\text{Term 3= }\left[\textbf{I}- \frac{\Delta t}{\Delta x} \left[ \left\{ \left(\frac{\lambda_{p}}{\lambda_{p}- \lambda_{m}}\right)^{n}_{j+\frac{1}{2}} \right. \right. \hspace{-0.47cm}+ \left. \left(\frac{\lambda_{m}}{\lambda_{p}- \lambda_{m}}\right)^{n}_{j-\frac{1}{2}} \right\} \textbf{A}_{j}^{n}  -
\left\{ \left( \frac{\lambda_{p}\lambda_{m}}{\lambda_{p}- \lambda_{m}}\right)^{n}_{j+\frac{1}{2}} \right. \hspace{-0.47cm}+ \left. \left. \left( \frac{\lambda_{p}\lambda_{m}}{\lambda_{p}- \lambda_{m}}\right)^{n}_{j-\frac{1}{2}} \right\}\textbf{I} \right]\right]\textbf{U}_{j}^{n}
\label{eq:1d_dt_1}
\end{equation}  
Here, we have used the fact that Euler fluxes are homogeneous functions of degree 1, {\em i.e.}, we can write $\textbf{G}= \textbf{A}\textbf{U}$, with $\textbf{A}$ being the advective flux Jacobian matrix. Now, Term 3 is positive if the coefficient matrix multiplied to $\textbf{U}_{j}^{n}$ is positive, {\em i.e.}, if it has non-negative eigenvalues, and its eigenvectors correspond to the eigenvectors of the advective flux Jacobian $\textbf{A}^{n}_{j}= \left(\partial \textbf{G}/\partial \textbf{U}\right)_{j}^{n}$, for $\gamma\in\left[1,3\right]$ (see \cite{PARENT2011238}). This condition is met when the following criterion is satisfied.
\begin{equation}
1- \frac{\Delta t}{\Delta x} max_{r} \left[ \left\{ \left(\frac{\lambda_{p}}{\lambda_{p}- \lambda_{m}}\right)_{j+\frac{1}{2}} \right. \right. + \left. \left(\frac{\lambda_{m}}{\lambda_{p}- \lambda_{m}}\right)_{j-\frac{1}{2}} \right\} eig_{r}\left(\textbf{A}_{j}\right)  -
\left\{ \left( \frac{\lambda_{p}\lambda_{m}}{\lambda_{p}- \lambda_{m}}\right)_{j+\frac{1}{2}} \right. + \left. \left. \left( \frac{\lambda_{p}\lambda_{m}}{\lambda_{p}- \lambda_{m}}\right)_{j-\frac{1}{2}} \right\} \right]  \geq 0
\label{eq:1d_dt_2}
\end{equation}
Here, $eig_{r}\left(\textbf{A}_{j}\right)$ is the $r^{th}$ eigenvalue of $\textbf{A}_{j}$. Thus, we obtain the following limit on the global time step based on positivity consideration.
\begin{equation}
\Delta t \leq \Delta t_{p}= min_{j}\left[ \frac{\Delta x}{max_{r}\left[ \left\{ \left(\frac{\lambda_{p}}{\lambda_{p}- \lambda_{m}}\right)_{j+\frac{1}{2}} + \left(\frac{\lambda_{m}}{\lambda_{p}- \lambda_{m}}\right)_{j-\frac{1}{2}} \right\} eig_{r}\left(\textbf{A}_{j}\right) -
\left\{ \left( \frac{\lambda_{p}\lambda_{m}}{\lambda_{p}- \lambda_{m}}\right)_{j+\frac{1}{2}} + \left( \frac{\lambda_{p}\lambda_{m}}{\lambda_{p}- \lambda_{m}}\right)_{j-\frac{1}{2}} \right\} \right]} \right]
\label{eq:1d_dt_3}
\end{equation} 
We note that for the scalar numerical diffusion model, the expression in \eqref{eq:1d_dt_3} gets simplified to give
\begin{equation}
\Delta t \leq  min_{j}\left( \frac{2 \Delta x}{\lambda_{j+\frac{1}{2}}+ \lambda_{j-\frac{1}{2}}} \right)
\label{eq:1d_dt_4}
\end{equation}

\subsubsection{Maximum time step based on Stability}
A linear stability analysis of the advective part of 1D Boltzmann equations (\ref{appendix:a1}) gives us the following criteria for numerical stability.
\begin{equation}
	\frac{\lambda_{p} \Delta t}{\Delta x} \leq 1\text{ and } -\frac{\lambda_{m} \Delta t}{\Delta x} \leq 1 \Rightarrow \frac{\lambda \Delta t}{\Delta x} \leq 1, \ \lambda= max (\lambda_{p}, -\lambda_{m})
	\label{eq:1d_dt_5}
	\end{equation}
	Thus, Model 1 with velocities $\lambda_{p}$ and $\lambda_{m}$ is linearly stable if Model 2 with velocities $\lambda$ and -$\lambda$, where $\lambda$~=~max($\lambda_{p},-\lambda_{m}$), is linearly stable. We use this result to simplify our non-linear stability analysis by having our equilibrium distribution functions correspond solely to the scalar numerical diffusion model. For the scalar numerical diffusion model, the equilibrium distribution functions can be written as
		\begin{equation}
\textbf{f}^{eq}_{1}= \frac{\textbf{U}}{2}+ \frac{\textbf{G}}{2\lambda}, \ \textbf{f}^{eq}_{2}= \frac{\textbf{U}}{2}- \frac{\textbf{G}}{2\lambda}
\label{eq:1d_dt_6}
\end{equation}
	To perform stability analysis for the non-linear Euler equations, we use Bouchut's stability criterion \cite{bouchut1999construction}. According to Bouchut, for stability
	\begin{equation}
	eig\left(\frac{\partial \textbf{f}^{eq}_{1,2}}{\partial \textbf{U}}\right) \subset\left[0,\infty\right)
	\label{eq:1d_dt_7}
	\end{equation}
	Here $eig$ refers to the eigenspectrum. Substituting \eqref{eq:1d_dt_6} into \eqref{eq:1d_dt_7}, we get,
	\begin{equation}
	\lambda \geq max \left( \left|u-a\right|, \left|u\right|, \left|u+a\right|\right)
	\label{eq:1d_dt_8}
	\end{equation}
	We use the stability criterion in \eqref{eq:1d_dt_8} to impose the following limit on the global time step.
	\begin{equation}
	\Delta t \leq \Delta t _{s}= min_{j}\left(\frac{\Delta x}{\lambda_{max,j}}\right), \ \lambda_{max,j}= max\left(\left|u-a\right|, \left|u\right|, \left|u+a\right|\right)_{j}
	\label{eq:1d_dt_9}
	\end{equation}
\textbf{Global time step }: Global time step is given by
\begin{equation}
	\Delta t= \sigma \ min (\Delta t_{p}, \Delta t_{s}),  \ 0< \sigma \leq 1
	\label{eq:1d_dt_10}
	\end{equation}
	where $\Delta t_{p}$ and $\Delta t_{s}$ are defined in \eqref{eq:1d_dt_3} and \eqref{eq:1d_dt_9} respectively. Throughout this work, we will refer to $\sigma$ as the CFL no., while noting that our calculation of the time step takes into account not just stability but also positivity.
	
	\subsection{Extension to second order accuracy}
 For second order accuracy, we use a flux-limited approach to add anti-diffusion terms to the first order kinetic flux, in the following manner.
\begin{equation}
(\textbf{h}_{i})_{j+\frac{1}{2}, 2O}= (\textbf{h}_{i})_{j+\frac{1}{2}}+ \frac{1}{2}\Phi\left( \frac{(\Delta \textbf{h}_{i}^{+})_{j+ \frac{1}{2}}}{(\Delta \textbf{h}_{i}^{+})_{j- \frac{1}{2}}}\right)(\Delta \textbf{h}_{i}^{+})_{j- \frac{1}{2}}- \frac{1}{2}\Phi\left(\frac{(\Delta \textbf{h}_{i}^{-})_{j+ \frac{1}{2}}}{(\Delta \textbf{h}_{i}^{-})_{j+ \frac{3}{2}}}\right)(\Delta \textbf{h}_{i}^{-})_{j+ \frac{3}{2}}
\label{eq:1d_2O_1}
\end{equation}
The definition in \eqref{eq:1d_2O_1} is applicable term-wise. $(\textbf{h}_{i})_{j+\frac{1}{2}}$ is the first order flux defined in \eqref{eq:1d_euler_11}, the flux differences $\Delta \textbf{h}_{i}^{\pm}$ are given in \eqref{eq:1d_euler_12}, and $\Phi$ is a diagonal matrix of a limiter function. By setting the limiters to 1, we get a (semi-discrete) Beam Warming type flux.
\begin{equation}
(\textbf{h}_{i})_{j+\frac{1}{2}, 2O}= (\textbf{h}_{i})_{j+\frac{1}{2}}+ \frac{1}{2}(\Delta \textbf{h}_{i}^{+})_{j- \frac{1}{2}}- \frac{1}{2}(\Delta \textbf{h}_{i}^{-})_{j+ \frac{3}{2}}
\label{eq:1d_2O_2}
\end{equation}
Equation \eqref{eq:1d_2O_1} can be rewritten as
\begin{equation}
\begin{bmatrix} h_{1i} \\ h_{2i}\end{bmatrix}_{j+ \frac{1}{2}, 2O}= \begin{bmatrix} h_{1i} \\ h_{2i}\end{bmatrix}_{j+ \frac{1}{2}}+ \frac{1}{2} \begin{bmatrix}\phi\left\{\frac{\left(\lambda_{p}\Delta f^{eq}_{1i}\right)_{j+ \frac{1}{2}}}{\left(\lambda_{p}\Delta f^{eq}_{1i}\right)_{j- \frac{1}{2}}}\right\}\left(\lambda_{p}\Delta f^{eq}_{1i}\right)_{j- \frac{1}{2}} \\0 \end{bmatrix}- \frac{1}{2} \begin{bmatrix}0 \\ \phi\left\{\frac{\left(\lambda_{m}\Delta f^{eq}_{2i}\right)_{j+ \frac{1}{2}}}{\left(\lambda_{m}\Delta f^{eq}_{2i}\right)_{j+ \frac{3}{2}}}\right\}\left(\lambda_{m}\Delta f^{eq}_{2i}\right)_{j+ \frac{3}{2}} \end{bmatrix}
\label{eq:1d_2O_3}
\end{equation}
Here $\phi(r)$ is the limiter function and $r$ is a ratio. To prevent the need for division, we express the limiter function as a function of two variables and write it as $\phi(1,r)$. We then select a limiter function that satisfies the following multiplication property.
\begin{equation}
\alpha \phi(x,y)= \phi(\alpha x, \alpha y)
\label{eq:eq:1d_2O_3a}
\end{equation}
Well known limiters that satisfy this property are the minmod, van Leer and superbee limiters. We have used minmod limiter in the present work. Equation \eqref{eq:1d_2O_3} can then be written as, 
\begin{subequations}
\label{eq:1d_2O_4}
\begin{equation}
\begin{bmatrix} h_{1i} \\ h_{2i}\end{bmatrix}_{j+ \frac{1}{2}, 2O}= \begin{bmatrix} h_{1i} \\ h_{2i}\end{bmatrix}_{j+ \frac{1}{2}}+ \frac{1}{2} \begin{bmatrix}\phi\left\{\left(\lambda_{p}\Delta f^{eq}_{1i}\right)_{j+ \frac{1}{2}},\left(\lambda_{p}\Delta f^{eq}_{1i}\right)_{j- \frac{1}{2}}\right\} \\0 \end{bmatrix}- \frac{1}{2} \begin{bmatrix}0 \\ \phi\left\{\left(\lambda_{m}\Delta f^{eq}_{2i}\right)_{j+ \frac{1}{2}},\left(\lambda_{m}\Delta f^{eq}_{2i}\right)_{j+ \frac{3}{2}}\right\} \end{bmatrix}
\end{equation}
\begin{equation}
\phi(x,y)= minmod(x,y)= \left\{\begin{array}{cc}
x,\text{ if }|x|<|y|\text{ and }xy>0\\
y,\text{ if }|x|>|y|\text{ and }xy>0\\
0,\text{ if }xy<0\end{array} \right\}
\end{equation}
\end{subequations}
The macroscopic flux at the interface then becomes
\begin{eqnarray}
(G_{i})_{j+\frac{1}{2}, 2O}&=& \textbf{P}_{i}(\textbf{h}_{i})_{j+\frac{1}{2}, 2O}\nonumber\\
&=& (G_{i})_{j+\frac{1}{2}}+ \frac{1}{2}\phi\left\{\left(\lambda_{p}\Delta f^{eq}_{1i}\right)_{j+ \frac{1}{2}},\left(\lambda_{p}\Delta f^{eq}_{1i}\right)_{j- \frac{1}{2}}\right\}
- \frac{1}{2}\phi\left\{\left(\lambda_{m}\Delta f^{eq}_{2i}\right)_{j+ \frac{1}{2}},\left(\lambda_{m}\Delta f^{eq}_{2i}\right)_{j+ \frac{3}{2}}\right\}\nonumber\\
&=& (G_{i})_{j+\frac{1}{2}}+ \frac{1}{2}\phi\left\{(\Delta G^{+}_{i})_{j+\frac{1}{2}},(\Delta G^{+}_{i})_{j-\frac{1}{2}}\right\}- \frac{1}{2}\phi\left\{(\Delta G^{-}_{i})_{j+\frac{1}{2}},(\Delta G^{-}_{i})_{j+\frac{3}{2}}\right\}
\label{eq:1d_2O_5}
\end{eqnarray}
The temporal derivative is approximated using Strong Stability Preserving Runge Kutta \cite{gottlieb2001strong} (SSPRK) Method. The update formula is,
\begin{subequations}
\label{eq:1d_2O_6}
\begin{equation}
(U_{i})^{1}_{j}= (U_{i})^{n}_{j}- \Delta t \ \text{R}((U_{i})^{n}_{j})
\end{equation}
\begin{equation}
(U_{i})^{2}_{j}= \frac{1}{4}(U_{i})^{1}_{j}+ \frac{3}{4}(U_{i})^{n}_{j} -\frac{1}{4}\Delta t \ \text{R}((U_{i})^{1}_{j})
\end{equation}
\begin{equation}
(U_{i})^{n+1}_{j}= \frac{2}{3}(U_{i})^{2}_{j}+ \frac{1}{3}(U_{i})^{n}_{j} -\frac{2}{3}\Delta t \ \text{R}((U_{i})^{2}_{j})
\end{equation}
\end{subequations} 
Here, R is the residual, i.e. $\text{R}((U_{i})^{n}_{j})= \frac{1}{\Delta x}\left[ (G_{i})_{j+\frac{1}{2}}^{n}- (G_{i})_{j-\frac{1}{2}}^{n}\right]$. At this point, we emphasize that positivity analysis has not been done for second order accuracy in this work. Thus, our second order scheme is not necessarily positivity preserving. However, the results obtained with the second order version are encouraging. The time step $\Delta t$ for our second order method is approximated by 
\begin{equation}
\left(\Delta t_{2O}\right)= \sigma \ min(\frac{\Delta t_{p}}{2}, \Delta t_{s})
\label{eq:1d_2O_7}
\end{equation}
Thus, the time step based on positivity for second order accuracy is approximated as half of that corresponding to first order accuracy.

\section{Kinetic model for 2D Euler Equations}
In 2D, we consider three velocities for our kinetic model. We utilize this construction in such a way that positivity analysis done in 1-D is easily extended to 2-D and, further, the model recognizes the steady shocks aligned with the grid lines. Thus, $N_{d}$= 3, which meets the minimum number of velocity requirement (given by $N_{d}\geq$N+1). Corresponding Boltzmann Equations for the $i^{th}$ macroscopic equation are given by
\begin{subequations}
\begin{equation}
\frac{\partial \textbf{f}_{i}}{\partial t}+ \frac{\partial (\Lambda_{1}\textbf{f}_{i})}{\partial x_{1}}+ \frac{\partial (\Lambda_{2}\textbf{f}_{i})}{\partial x_{2}}= 
- \frac{1}{\epsilon} \left[\textbf{f}_{i}- \textbf{f}_{i}^{eq}\right],
\label{eq:2d_euler_1a}
\end{equation}
\begin{equation}
\textbf{f}_{i}^{eq}= \begin{bmatrix} f^{eq}_{1i} \\ f^{eq}_{2i} \\ f^{eq}_{3i}\end{bmatrix}, \Lambda_{1}= \begin{bmatrix} \lambda_{1,1} & 0 & 0 \\  0 & \lambda_{2,1} & 0 \\ 0 & 0 & \lambda_{3,1} \end{bmatrix}, \Lambda_{2}= \begin{bmatrix} \lambda_{1,2} & 0 & 0\\  0 & \lambda_{2,2} & 0\\ 0 & 0 & \lambda_{3,2}\end{bmatrix}
\label{eq:2d_euler_1b}
\end{equation}
\end{subequations}
Defining the row vector  $\textbf{P}_{i}= \begin{bmatrix}1 & 1 & 1\end{bmatrix}$, the moment relations can be written as
\begin{subequations}
\label{eq:2d_euler_2}
\begin{equation}
\textbf{P}_{i}\textbf{f}_{i}^{eq}= f^{eq}_{1i}+ f^{eq}_{2i}+ f^{eq}_{3i}= U_{i}
\end{equation}
\begin{equation}
\textbf{P}_{i} \Lambda_{1}\textbf{f}_{i}^{eq}= \lambda_{1,1}f^{eq}_{1i}+ \lambda_{2,1}f^{eq}_{2i}+ \lambda_{3,1}f^{eq}_{3i}= G_{1,i}
\end{equation}
\begin{equation}
\textbf{P}_{i} \Lambda_{2}\textbf{f}_{i}^{eq}= \lambda_{1,2}f^{eq}_{1i}+ \lambda_{2,2}f^{eq}_{2i}+ \lambda_{3,2}f^{eq}_{3i}= G_{2,i}
\end{equation}
\end{subequations}
Thus, we have three equations and three unknowns, which we can solve to get expressions for $f^{eq}_{1i}$, $f^{eq}_{2i}$ and $f^{eq}_{3i}$ in terms of $U_{i}$, $G_{1,i}$, $G_{2,i}$ and lambdas. Here, $U_{i}$, $G_{1,i}$ and $G_{2,i}$ represent the $i^{th}$ component of the conserved variable vector \textbf{U} and flux vectors $\textbf{G}_{1}$ and $\textbf{G}_{2}$ respectively for the 2D Euler equations given by 
\begin{eqnarray}
    \frac{\partial \textbf{U}}{\partial t} + \frac{\partial \textbf{G}_{1}}{\partial x_{1}} + \frac{\partial \textbf{G}_{2}}{\partial x_{2}} = 0, \ \textrm{with} \ \textbf{U} = \left[ \begin{array}{c} \rho \\ \rho u_{1} \\ \rho u_{2} \\ \rho E \end{array} \right], \ 
    \textbf{G}_{1} = \left[ \begin{array}{c} \rho u_{1} \\ \rho u_{1}^{2}+ p \\ \rho u_{1} u_{2} \\ (\rho  E +p)u_{1}  \end{array} \right] \ \textrm{and} \ 
    \textbf{G}_{2} = \left[ \begin{array}{c} \rho u_{2} \\  \rho u_{2} u_{1} \\ \rho u_{2}^{2}+ p \\(\rho  E +p)u_{2} \end{array} \right]
    \label{eq:2d_euler_2a}
\end{eqnarray}

Next, we solve the Boltzmann equations \eqref{eq:2d_euler_1a} numerically by using the operator-splitting strategy, leading to instantaneous relaxation and advection steps for $(j,k)^{th}$ cell, as follows.  
\begin{subequations}
\[\text{Relaxation step: Instantaneous, {\em i.e.}, }\epsilon\rightarrow 0\text{. Thus,} \]
\begin{equation}
(\textbf{f}_{i})^{n}_{j,k}=  (\textbf{f}_{i}^{eq})^{n}_{j,k}
\label{eq:2d_euler_3a}
\end{equation}
\[\text{Advection step: The advective part of Boltzmann equation is,}\]
\begin{equation}
\frac{\partial \textbf{f}_{i}}{\partial t}+ \frac{\partial \textbf{h}_{1i}}{\partial x_{1}}+ \frac{\partial \textbf{h}_{2i}}{\partial x_{2}}= 0; \ \textbf{h}_{1i}= \Lambda_{1}\textbf{f}_{i}^{eq}, \ \textbf{h}_{2i}= \Lambda_{2}\textbf{f}_{i}^{eq}
\label{eq:2d_euler_3b}
\end{equation}
\end{subequations}
Rewriting \eqref{eq:2d_euler_3b} in integral form for $(j,k)^{th}$ cell, we get
\begin{subequations}
\begin{equation}
A_{j,k}\frac{d \left(\textbf{f}_{i}\right)_{j,k}}{dt}+  \oint \textbf{h}_{\perp i}dl =0; \ \textbf{h}_{\perp i}= \Lambda_{\perp}\textbf{f}_{i}^{eq}, \ \Lambda_{\perp}= \Lambda_{1}n_{1}+ \Lambda_{2}n_{2}
\label{eq:2d_euler_4a}
\end{equation}
\begin{equation}
\Rightarrow A_{j,k}\frac{d \left(\textbf{f}_{i}\right)_{j,k}}{dt}+  \sum_{s=1}^{4} (\textbf{h}_{\perp i})_{s}l_{s} =0\text{ (mid-point quadrature)}
\label{eq:2d_euler_4b}
\end{equation}
\end{subequations} 
Here, $A_{j,k}$ is the area of $(j,k)^{th}$ cell. The normal and tangential unit vectors at a cell interface $s$ are given by $\widehat{e}_{\perp}$= ($n_{1}$, $n_{2}$) and $\widehat{e}_{\parallel}$= (-$n_{2}$, $n_{1}$) respectively. For first order accuracy, we discretize Equations \eqref{eq:2d_euler_4b} as follows.
\begin{subequations}
\begin{equation}
\left(\textbf{f}_{i}\right)^{n+1}_{j,k}= \left(\textbf{f}^{eq}_{i}\right)^{n}_{j,k} -\frac{\Delta t}{A_{j,k}}\sum_{s=1}^{4} (\textbf{h}_{\perp i})_{s}l_{s}
\label{eq:2d_euler_5a}
\end{equation}
\begin{equation}
(\textbf{h}_{\perp i})_{s}= \frac{1}{2}\left\{(\textbf{h}_{\perp i})_{L}+ (\textbf{h}_{\perp i})_{R}\right\}-\frac{1}{2}\left\{(\Delta \textbf{h}_{\perp i}^{+})_{s}- (\Delta \textbf{h}_{\perp i}^{-})_{s}\right\}
\label{eq:2d_euler_5b}
\end{equation}
\begin{equation}
(\Delta \textbf{h}_{\perp i}^{\pm})_{s} = \left(\Lambda_{\perp}^{\pm}\Delta\textbf{f}_{i}^{eq}\right)_{s}
\label{eq:2d_euler_5c}
\end{equation}
\end{subequations}
The 2D kinetic normal flux $\textbf{h}_{\perp i}$ has three components (since we are considering three velocities in 2D), and their sum ({\em i.e.}, moment) gives us the macroscopic flux, which in general has a formulation different than that for the 1D flux. We would, however, like to preserve the 1D flux structure, for the ease of extending the positivity analysis and further retain the advantage of exact capture of grid-aligned steady shocks. Thus, we define our three flexible velocities ($\lambda_{1,1}$,$\lambda_{1,2}$), ($\lambda_{2,1}$,$\lambda_{2,2}$) and ($\lambda_{3,1}$,$\lambda_{3,2}$) such that the resulting macroscopic flux has a locally 1D formulation. We define the three velocities as shown in Figure \ref{fig:2d_eq} and given in Equation \eqref{eq:2d_euler_6} below.  

\begin{figure}[h!] 
\centering
\begin{tikzpicture}
\small\begin{axis}
 [every axis plot post/.append style={
  mark=none,domain=0:5,samples=50,smooth},
  xmin=0,xmax=5,ymin=0,ymax= 4,
	axis x line*=bottom, 
  axis y line*=left,
	axis line style={draw=none},
	xtick={},
	unit vector ratio*=1 1 1,
	xticklabels={\empty},
	yticklabel={\empty},
	tick style={draw=none}
	]

\draw [ultra thick](axis cs:1.8,3.2) --(axis cs:2.8, 1) node [anchor= north west]{$s$} ;
\draw [ultra thick,-stealth](axis cs:2.3, 2.1) --(axis cs:4.5,3.1)node [anchor=south east]{$(\textbf{$\lambda$}_{1,1},\textbf{$\lambda$}_{1,2})$};
\draw [ultra thick,-stealth](axis cs:2.3, 2.1) --(axis cs:0.5,2.3)node [anchor=south]{$(\textbf{$\lambda$}_{2,1},\textbf{$\lambda$}_{2,2})$};
\draw [ultra thick,-stealth](axis cs:2.3, 2.1) --(axis cs:1.26712,0.61232)node [anchor=north west]{$(\textbf{$\lambda$}_{3,1},\textbf{$\lambda$}_{3,2})$};
\draw [ultra thick](axis cs:2.24, 2.232) --(axis cs:2.46,2.332);
\draw [ultra thick](axis cs:2.46,2.332) --(axis cs:2.52,2.2);

\draw[black,thick,dashed,<->] (axis cs:2.4, 1.88) --(axis cs:4.6,2.88)node[midway,sloped,fill=white]{$\textbf{$\lambda$}_{p,\perp}$};
\draw[black,thick,dashed,<->] (axis cs:0.88356, 1.45616) --(axis cs:2.3,2.1)node[midway,sloped,fill=white]{$\textbf{$\lambda$}_{m,\perp}$};
\draw[black,thick,dashed,<->] (axis cs:2, 1.963636) --(axis cs:1.6,2.843636)node[midway,fill=white]{$\textbf{$\lambda$}_{\parallel}$};
\draw[black,thick,dashed,<->] (axis cs:2, 1.963636) --(axis cs:2.4,1.083636)node[midway,fill=white]{$\textbf{$\lambda$}_{\parallel}$};
\draw [ultra thick,-stealth](axis cs:2.6,1.44) --(axis cs:2.8,1.5309)node [anchor=west] {$\widehat{e}_{\perp}$};
\end{axis}
\end{tikzpicture}
\caption{Velocities for 2D equilibrium distribution}
\label{fig:2d_eq}%
\end{figure}
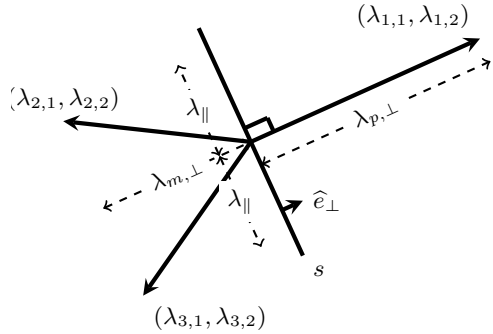

\begin{subequations}
\label{eq:2d_euler_6}
\begin{equation}
\lambda_{1,1}= \lambda_{p,\perp}n_{1},\hspace{0.5cm} \lambda_{2,1}= (\lambda_{m,\perp}n_{1}-\lambda_{\parallel}n_{2}),\hspace{0.5cm} \lambda_{3,1}= (\lambda_{m,\perp}n_{1}+\lambda_{\parallel}n_{2})
\end{equation}
\begin{equation}
\lambda_{1,2}= \lambda_{p,\perp}n_{2},\hspace{0.5cm} \lambda_{2,2}= (\lambda_{m,\perp}n_{2}+\lambda_{\parallel}n_{1}),\hspace{0.5cm} \lambda_{3,2}= (\lambda_{m,\perp}n_{2}-\lambda_{\parallel}n_{1})
\end{equation}
\end{subequations}
where $\lambda_{p,\perp}\geq$0, $\lambda_{m,\perp}\leq$0. Now, substituting the lambda's as defined in \eqref{eq:2d_euler_6} into the moment relations \eqref{eq:2d_euler_2} and solving for the equilibrium distributions, we get
\begin{equation}
\begin{bmatrix}f^{eq}_{1i}\\ f^{eq}_{2i}\\f^{eq}_{3i}\end{bmatrix}= \frac{1}{(\lambda_{p,\perp}- \lambda_{m,\perp})}\begin{bmatrix} - \lambda_{m,\perp} & n_{1} & n_{2}\\ \frac{\lambda_{p,\perp}}{2} & \frac{-\lambda_{p,\perp}n_{2}+ \lambda_{m,\perp}n_{2}- \lambda_{\parallel}n_{1}}{2\lambda_{\parallel}} & \frac{\lambda_{p,\perp}n_{1}- \lambda_{m,\perp}n_{1}- \lambda_{\parallel}n_{2}}{2\lambda_{\parallel}} \\ \frac{\lambda_{p,\perp}}{2} & \frac{\lambda_{p,\perp}n_{2}- \lambda_{m,\perp}n_{2}- \lambda_{\parallel}n_{1}}{2\lambda_{\parallel}} & \frac{-\lambda_{p,\perp}n_{1}+ \lambda_{m,\perp}n_{1}- \lambda_{\parallel}n_{2}}{2\lambda_{\parallel}}\end{bmatrix} \begin{bmatrix}U_{i} \\ G_{1,i} \\ G_{2,i} \end{bmatrix}
\label{eq:2d_euler_7}
\end{equation}
Further,
\begin{subequations}
\label{eq:2d_euler_8}
\begin{equation}
\Lambda_{\perp}= \Lambda_{1}n_{1}+ \Lambda_{2}n_{2}= \begin{bmatrix} \lambda_{p,\perp} &0 &0\\ 0& \lambda_{m,\perp} &0\\ 0& 0& \lambda_{m,\perp}\end{bmatrix}
\end{equation}
\begin{equation}
\left(\Delta \textbf{h}_{\perp i}^{+}\right)_{s} = \left(\Lambda_{\perp}^{+}\Delta\textbf{f}_{i}^{eq}\right)_{s}= \begin{bmatrix} (\lambda_{p,\perp} \Delta f^{eq}_{1i})_{s} \\0 \\0 \end{bmatrix}= \begin{bmatrix} (\lambda_{p,\perp})_{s}\left\{ (f^{eq}_{1i})_{R}- (f^{eq}_{1i})_{L}\right\} \\0 \\0 \end{bmatrix}
\end{equation}
\begin{equation}
\left(\Delta \textbf{h}_{\perp i}^{-}\right)_{s} = \left(\Lambda_{\perp}^{-}\Delta\textbf{f}_{i}^{eq}\right)_{s}= \begin{bmatrix} 0 \\ (\lambda_{m,\perp} \Delta f^{eq}_{2i})_{s} \\ (\lambda_{m,\perp} \Delta f^{eq}_{3i})_{s} \end{bmatrix}= \begin{bmatrix} 0 \\ (\lambda_{m,\perp})_{s}\left\{ (f^{eq}_{2i})_{R}- (f^{eq}_{2i})_{L}\right\} \\ (\lambda_{m,\perp})_{s}\left\{ (f^{eq}_{3i})_{R}- (f^{eq}_{3i})_{L}\right\} \end{bmatrix}
\end{equation}
\end{subequations}
The macroscopic update formula, obtained by taking moments of Equation \eqref{eq:2d_euler_5a}, is given by
\begin{subequations}
\begin{equation}
\left(U_{i}\right)^{n+1}_{j,k}= \left(U_{i}\right)^{n}_{j,k} -\frac{\Delta t}{A_{j,k}}\sum_{s=1}^{4} (G_{\perp i})_{s}l_{s}
\label{eq:2d_euler_9a}
\end{equation}
\begin{equation}
(G_{\perp i})_{s}= \textbf{P}_{i}(\textbf{h}_{\perp i})_{s} =\frac{1}{2}\left\{(G_{\perp i})_{L}+ (G_{\perp i})_{R}\right\}-\frac{1}{2}\left\{(\Delta G_{\perp i}^{+})_{s}- (\Delta G_{\perp i}^{-})_{s}\right\}; \ G_{\perp i}= G_{1i}n_{1}+ G_{2i}n_{2}
\label{eq:2d_euler_9b}
\end{equation}
\begin{equation}
(\Delta G_{\perp i}^{+})_{s}= \textbf{P}_{i}(\Delta \textbf{h}_{\perp i}^{+})_{s}= (\lambda_{p,\perp} \Delta f^{eq}_{1i})_{s}
\label{eq:2d_euler_9c}
\end{equation}
\begin{equation}
(\Delta G_{\perp i}^{-})_{s}= \textbf{P}_{i}(\Delta \textbf{h}_{\perp i}^{-})_{s}= \left(\lambda_{m,\perp} \Delta \left(f^{eq}_{2i}+ f^{eq}_{3i}\right)\right)_{s}
\label{eq:2d_euler_9d}
\end{equation}
\end{subequations}
The macroscopic normal flux at interface $s$ in \eqref{eq:2d_euler_9b} can now be rewritten in vector form as
\begin{equation}
(\textbf{G}_{\perp})_{s}= \left(\frac{\lambda_{p,\perp}}{\lambda_{p,\perp}- \lambda_{m,\perp}}\right)_{s}(\textbf{G}_{\perp})_{L} - \left(\frac{\lambda_{m,\perp}}{\lambda_{p,\perp}- \lambda_{m,\perp}}\right)_{s}(\textbf{G}_{\perp})_{R}+\left( \frac{\lambda_{p,\perp}\lambda_{m,\perp}}{\lambda_{p,\perp}- \lambda_{m,\perp}}\right)_{s}\left(\textbf{U}_{R}-\textbf{U}_{L}\right)
\label{eq:2d_euler_10}
\end{equation}

\subsection{Positivity analysis}
In 2D, we use a structured grid, and we assume that normals at interfaces point towards +$\xi$ and +$\eta$ directions respectively. For first order accuracy, the macroscopic update formula in $(j,k)^{th}$ cell is given by
\begin{eqnarray}
&&\textbf{U}^{n+1}_{j,k}= \textbf{U}^{n}_{j,k}- \frac{\Delta t}{A_{j,k}}\left[(\textbf{G}_{\perp})^{n}_{j+\frac{1}{2},k}l_{j+\frac{1}{2},k}- (\textbf{G}_{\perp})^{n}_{j-\frac{1}{2},k}l_{j-\frac{1}{2},k}+ (\textbf{G}_{\perp})^{n}_{j,k+\frac{1}{2}}l_{j,k+\frac{1}{2}}- (\textbf{G}_{\perp})^{n}_{j,k-\frac{1}{2}}l_{j,k-\frac{1}{2}}\right]\nonumber\\
&&= - \frac{\Delta t}{A_{j,k}}\left[\frac{\lambda_{m,\perp}l}{\lambda_{p,\perp}-\lambda_{m,\perp}}\left\{-(\textbf{G}_{\perp})_{R}+ \lambda_{p,\perp} (\textbf{U})_{R}\right\}\right]^{n}_{j+\frac{1}{2},k}
+\frac{\Delta t}{A_{j,k}}\left[\frac{\lambda_{p,\perp}l}{\lambda_{p,\perp}-\lambda_{m,\perp}}\left\{(\textbf{G}_{\perp})_{L}- \lambda_{m,\perp} (\textbf{U})_{L}\right\}\right]^{n}_{j-\frac{1}{2},k}\nonumber\\
&& - \frac{\Delta t}{A_{j,k}}\left[\frac{\lambda_{m,\perp}l}{\lambda_{p,\perp}-\lambda_{m,\perp}}\left\{-(\textbf{G}_{\perp})_{R}+ \lambda_{p,\perp} (\textbf{U})_{R}\right\}\right]^{n}_{j,k+\frac{1}{2}}
+\frac{\Delta t}{A_{j,k}}\left[\frac{\lambda_{p,\perp}l}{\lambda_{p,\perp}-\lambda_{m,\perp}}\left\{(\textbf{G}_{\perp})_{L}- \lambda_{m,\perp} (\textbf{U})_{L}\right\}\right]^{n}_{j,k-\frac{1}{2}}\nonumber\\
&&+\underbrace{\textbf{U}^{n}_{j,k}-\frac{\Delta t}{A_{j,k}} \left[\left[\frac{\lambda_{p,\perp}l}{\lambda_{p,\perp}-\lambda_{m,\perp}}\left\{(\textbf{G}_{\perp})_{L}- \lambda_{m,\perp} (\textbf{U})_{L}\right\}\right]^{n}_{j+\frac{1}{2},k}- \left[\frac{\lambda_{m,\perp}l}{\lambda_{p,\perp}-\lambda_{m,\perp}}\left\{-(\textbf{G}_{\perp})_{R}+ \lambda_{p,\perp} (\textbf{U})_{R}\right\}\right]^{n}_{j-\frac{1}{2},k}\right. }\nonumber\\
&&\underbrace{\left.+ \left[\frac{\lambda_{p,\perp}l}{\lambda_{p,\perp}-\lambda_{m,\perp}}\left\{(\textbf{G}_{\perp})_{L}- \lambda_{m,\perp} (\textbf{U})_{L}\right\}\right]^{n}_{j,k+\frac{1}{2}}- \left[\frac{\lambda_{m,\perp}l}{\lambda_{p,\perp}-\lambda_{m,\perp}}\left\{-(\textbf{G}_{\perp})_{R}+ \lambda_{p,\perp} (\textbf{U})_{R}\right\}\right]^{n}_{j,k-\frac{1}{2}}\right]}_{\text{Term 5}}
\label{eq:2d_pos_1}
\end{eqnarray}

Now, for the numerical scheme to be positively conservative, {\em i.e.}, for $\textbf{U}^{n+1}_{j,k} \in \textbf{W}$, the following conditions have to be satisfied.  
\begin{enumerate}
	\item $\left\{\lambda_{p,\perp}\textbf{U}_{R} - (\textbf{G}_{\perp})_{R}\right\}_{s} \in \textbf{W}$. Let $\lambda_{p,\perp}\textbf{U}_{R} - (\textbf{G}_{\perp})_{R}$= $\begin{bmatrix} G_{\perp,1} & G_{\perp,2} & G_{\perp,3} & G_{\perp,4} \end{bmatrix}^{T}$. The positivity of density and pressure requires that $G_{\perp,1}\geq 0$ and $2G_{\perp,1}G_{\perp,4}-(G^{2}_{\perp,2}+ G^{2}_{\perp,3})\geq 0$. This gives us \label{2d_condition1}
\begin{equation}
\left(\lambda_{p,\perp}\right)_{s} \geq \left((u_{\perp})_{R}+ \sqrt{\frac{\gamma -1}{2\gamma}} a_{R}\right)_{s}, \ u_{\perp}= u_{1}n_{1}+ u_{2}n_{2}
\label{eq:2d_pos_2}
\end{equation}

	\item $\left\{-\lambda_{m,\perp}\textbf{U}_{L} + (\textbf{G}_{\perp})_{L}\right\}_{s} \in \textbf{W}$. From this condition, we get \label{2d_condition2}
	\begin{equation}
\left(\lambda_{m,\perp}\right)_{s} \leq \left((u_{\perp})_{L}- \sqrt{\frac{\gamma -1}{2\gamma}} a_{L}\right)_{s}
\label{eq:2d_pos_3}
\end{equation}
For the special case of scalar numerical diffusion model, the positivity conditions in \eqref{eq:2d_pos_2} and \eqref{eq:2d_pos_3} can be combined to give,
\begin{equation}
\left(\lambda_{\perp}\right)_{s} \geq max\left( -u_{\perp,L}+ \sqrt{\frac{\gamma -1}{2\gamma}} a_{L}, u_{\perp,R}+ \sqrt{\frac{\gamma -1}{2\gamma}} a_{R}\right)_{s}
\label{eq:2d_pos_4}
\end{equation}

	\item Term 5 can be written as a positive term multiplied with $\textbf{U}^{n}_{j,k}$. This condition is used to obtain a limit on time step (see section \ref{section:2d_dt}).\label{2d_condition3}
\end{enumerate}

\subsection{\texorpdfstring{Fixing $\lambda$'s}{Fixing λ's}}
We define our lambdas to satisfy the positivity conditions using a similar approach that we used in 1D, as follows.
\begin{subequations}
\label{eq:2d_lambda_1}
\begin{equation}
\text{In smoothly varying flow regions: }\left(\lambda_{\perp}\right)_{s}= max\left( \lambda_{RH}, -(u_{\perp})_{L}+ \sqrt{\frac{\gamma -1}{2\gamma}} a_{L}, (u_{\perp})_{R}+ \sqrt{\frac{\gamma -1}{2\gamma}} a_{R}\right)_{s}
\end{equation}
\begin{eqnarray}
\text{Everywhere else: }\left(\lambda_{p,\perp}\right)_{s}&=& max\left(\lambda_{RH},(u_{\perp})_{R}+ \sqrt{\frac{\gamma -1}{2\gamma}} a_{R}\right)_{s},\nonumber\\
\left(\lambda_{m,\perp}\right)_{s}&=& min\left(-\lambda_{RH},(u_{\perp})_{L}- \sqrt{\frac{\gamma -1}{2\gamma}} a_{L}\right)_{s}
\end{eqnarray}
\end{subequations}
where,
\begin{equation}
\left(\lambda_{RH}\right)_{s}= min_{i}\left(\frac{\left|\Delta G_{\perp i}\right|}{\left|\Delta U_{i}\right|+ \epsilon_{0}}\right), \ \Delta = ()_{R}- ()_{L} 
\label{eq:2d_lambda_2}
\end{equation}

\subsection{Time step restrictions}
\label{section:2d_dt}
\subsubsection{Time step restriction based on positivity}
To start, we rewrite Term 5 in the macroscopic update formula \eqref{eq:2d_pos_1} as follows
\begin{equation}
\begin{split}
&\left[\textbf{I}\right.-\frac{\Delta t}{A_{j,k}}\left\{\left(\frac{n_{1}\lambda_{p,\perp}l}{\lambda_{p,\perp}-\lambda_{m,\perp}}\right)^{n}_{j+\frac{1}{2},k}\hspace{-0.45cm}+ \left(\frac{n_{1}\lambda_{p,\perp}l}{\lambda_{p,\perp}-\lambda_{m,\perp}}\right)^{n}_{j,k+\frac{1}{2}}\hspace{-0.45cm}+ \left(\frac{n_{1}\lambda_{m,\perp}l}{\lambda_{p,\perp}-\lambda_{m,\perp}}\right)^{n}_{j-\frac{1}{2},k}\hspace{-0.45cm}+ \left(\frac{n_{1}\lambda_{m,\perp}l}{\lambda_{p,\perp}-\lambda_{m,\perp}}\right)^{n}_{j,k-\frac{1}{2}}\right\}(\textbf{A}_{1})^{n}_{j,k}\\
&-\frac{\Delta t}{A_{j,k}}\left\{\left(\frac{n_{2}\lambda_{p,\perp}l}{\lambda_{p,\perp}-\lambda_{m,\perp}}\right)^{n}_{j+\frac{1}{2},k}\hspace{-0.35cm}+ \left(\frac{n_{2}\lambda_{p,\perp}l}{\lambda_{p,\perp}-\lambda_{m,\perp}}\right)^{n}_{j,k+\frac{1}{2}}\hspace{-0.35cm}+ \left(\frac{n_{2}\lambda_{m,\perp}l}{\lambda_{p,\perp}-\lambda_{m,\perp}}\right)^{n}_{j-\frac{1}{2},k}\hspace{-0.35cm}+ \left(\frac{n_{2}\lambda_{m,\perp}l}{\lambda_{p,\perp}-\lambda_{m,\perp}}\right)^{n}_{j,k-\frac{1}{2}}\right\}(\textbf{A}_{2})^{n}_{j,k}\\
&+\left.\frac{\Delta t}{A_{j,k}}\left\{\left(\frac{\lambda_{p,\perp}\lambda_{m,\perp}l}{\lambda_{p,\perp}-\lambda_{m,\perp}}\right)^{n}_{j+\frac{1}{2},k}\hspace{-0.35cm}+ \left(\frac{\lambda_{p,\perp}\lambda_{m,\perp}l}{\lambda_{p,\perp}-\lambda_{m,\perp}}\right)^{n}_{j,k+\frac{1}{2}}\hspace{-0.35cm}+ \left(\frac{\lambda_{p,\perp}\lambda_{m,\perp}l}{\lambda_{p,\perp}-\lambda_{m,\perp}}\right)^{n}_{j-\frac{1}{2},k}\hspace{-0.35cm}+ \left(\frac{\lambda_{p,\perp}\lambda_{m,\perp}l}{\lambda_{p,\perp}-\lambda_{m,\perp}}\right)^{n}_{j,k-\frac{1}{2}}\right\}\textbf{I}\right]\textbf{U}^{n}_{j,k}
\end{split}
\label{eq:2d_dt_1}
\end{equation} 
Now, defining
\begin{subequations}
\label{eq:2d_dt_2}
\begin{equation}
l_{1}= \left(\frac{n_{1}\lambda_{p,\perp}l}{\lambda_{p,\perp}-\lambda_{m,\perp}}\right)_{j+\frac{1}{2},k}+ \left(\frac{n_{1}\lambda_{p,\perp}l}{\lambda_{p,\perp}-\lambda_{m,\perp}}\right)_{j,k+\frac{1}{2}}+ \left(\frac{n_{1}\lambda_{m,\perp}l}{\lambda_{p,\perp}-\lambda_{m,\perp}}\right)_{j-\frac{1}{2},k}+ \left(\frac{n_{1}\lambda_{m,\perp}l}{\lambda_{p,\perp}-\lambda_{m,\perp}}\right)_{j,k-\frac{1}{2}}
\end{equation}
\begin{equation}
l_{2}= \left(\frac{n_{2}\lambda_{p,\perp}l}{\lambda_{p,\perp}-\lambda_{m,\perp}}\right)_{j+\frac{1}{2},k}+ \left(\frac{n_{2}\lambda_{p,\perp}l}{\lambda_{p,\perp}-\lambda_{m,\perp}}\right)_{j,k+\frac{1}{2}}+ \left(\frac{n_{2}\lambda_{m,\perp}l}{\lambda_{p,\perp}-\lambda_{m,\perp}}\right)_{j-\frac{1}{2},k}+ \left(\frac{n_{2}\lambda_{m,\perp}l}{\lambda_{p,\perp}-\lambda_{m,\perp}}\right)_{j,k-\frac{1}{2}}
\end{equation}
\begin{equation}
l_{0}=\sqrt{l_{1}^{2}+ l_{2}^{2}}; \hspace{0.2cm}n^{0}_{1}= \frac{l_{1}}{l_{0}}; \hspace{0.2cm}n^{0}_{2}= \frac{l_{2}}{l_{0}}; \hspace{0.2cm} \textbf{A}_{0}= \textbf{A}_{1}n^{0}_{1}+ \textbf{A}_{2}n^{0}_{2}
\end{equation}
\begin{equation}
b= \left(\frac{\lambda_{p,\perp}\lambda_{m,\perp}l}{\lambda_{p,\perp}-\lambda_{m,\perp}}\right)_{j+\frac{1}{2},k}+ \left(\frac{\lambda_{p,\perp}\lambda_{m,\perp}l}{\lambda_{p,\perp}-\lambda_{m,\perp}}\right)_{j,k+\frac{1}{2}}+ \left(\frac{\lambda_{p,\perp}\lambda_{m,\perp}l}{\lambda_{p,\perp}-\lambda_{m,\perp}}\right)_{j-\frac{1}{2},k}+ \left(\frac{\lambda_{p,\perp}\lambda_{m,\perp}l}{\lambda_{p,\perp}-\lambda_{m,\perp}}\right)_{j,k-\frac{1}{2}}
\end{equation}
\end{subequations}
The requirement of positivity of the coefficient matrix in Term 5 leads to the following limit on the global time step.
\begin{equation}
\Delta t\leq \Delta t_{p}=min_{j,k}\left[\frac{A_{j,k}}{max_{r}\left\{l_{0} \ eig_{r}(\textbf{A}_{0})- b\right\}_{j,k}}\right]
\label{eq:2d_dt_3}
\end{equation}

\subsubsection{Time step restriction based on stability}
We extend the simplification made in the 1D stability analysis to 2D by considering the simpler scalar numerical diffusion model (with the numerical diffusion coefficient $\lambda_{\perp}$). The equilibrium distributions for this model are
\begin{subequations}
\label{eq:2d_dt_4}
\begin{equation}
\textbf{f}^{eq}_{1}=\frac{\textbf{U}}{2}+ \frac{\textbf{G}_{\perp}}{2\lambda_{\perp}} 
\end{equation}
\begin{equation}
\textbf{f}^{eq}_{2}= \frac{\textbf{U}}{4}- \frac{\lambda_{\parallel}n_{1}+2\lambda_{\perp}n_{2}}{4\lambda_{\perp}\lambda_{\parallel}}\textbf{G}_{1}- \frac{\lambda_{\parallel}n_{2}-2\lambda_{\perp}n_{1}}{4\lambda_{\perp}\lambda_{\parallel}}\textbf{G}_{2}
\end{equation}
\begin{equation}
\textbf{f}^{eq}_{3}= \frac{\textbf{U}}{4}- \frac{\lambda_{\parallel}n_{1}-2\lambda_{\perp}n_{2}}{4\lambda_{\perp}\lambda_{\parallel}}\textbf{G}_{1}- \frac{\lambda_{\parallel}n_{2}+2\lambda_{\perp}n_{1}}{4\lambda_{\perp}\lambda_{\parallel}}\textbf{G}_{2}
\end{equation}
\end{subequations}
Then, as per Bouchut's stability criterion 
\begin{subequations}
\begin{equation}
eig\left(\frac{\partial \textbf{f}^{eq}_{1}}{\partial \textbf{U}}\right) \subset\left[0,\infty\right)
\label{eq:2d_dt_5a}
\end{equation}
\begin{equation}
eig\left(\frac{\partial \textbf{f}^{eq}_{2}}{\partial \textbf{U}}\right) \subset\left[0,\infty\right)
\label{eq:2d_dt_5b}
\end{equation}
\begin{equation}
eig\left(\frac{\partial \textbf{f}^{eq}_{3}}{\partial \textbf{U}}\right) \subset\left[0,\infty\right)
\label{eq:2d_dt_5c}
\end{equation}
\end{subequations}
Now, $\textbf{f}^{eq}_{2}+\textbf{f}^{eq}_{3}=\frac{\textbf{U}}{2}- \frac{\textbf{G}_{\perp}}{2\lambda_{\perp}}$, which is independent of $\lambda_{\parallel}$. So, we make an approximation by replacing \eqref{eq:2d_dt_5b} and \eqref{eq:2d_dt_5c} by the condition
\begin{equation}
eig\left(\frac{\partial \left(\textbf{f}^{eq}_{2}+ \textbf{f}^{eq}_{3}\right)}{\partial \textbf{U}}\right) \subset\left[0,\infty\right)
\label{eq:2d_dt_6}
\end{equation}
The criteria \eqref{eq:2d_dt_5a} and \eqref{eq:2d_dt_6} then lead to
	\begin{equation}
	\lambda_{\perp} \geq \lambda_{max}= max\left(\left|u_{\perp}-a\right|, \left|u_{\perp}\right|, \left|u_{\perp}+a\right|\right)
	\label{eq:2d_dt_7}
	\end{equation}
	An estimate of the global time step based on stability criterion is then given by
	\begin{equation}
	\Delta t \leq \Delta t _{s}= min_{j,k}\left[\frac{A_{j,k}}{(\lambda_{max})_{\xi}l_{\xi}+ (\lambda_{max})_{\eta}l_{\eta}}\right]
	\label{eq:2d_dt_8}
	\end{equation}
	\textbf{Global time step }: Finally, we compute the global time step as follows.
	\begin{equation}
	\Delta t= \sigma \ min(\Delta t_{p}, \Delta t_{s}), \ 0<\sigma\leq1
	\label{eq:2d_dt_9}
	\end{equation}

\subsubsection{Extension to second order accuracy}
We follow the same strategy that we used in 1D to extend our basic scheme in 2D to second order accuracy. Flux limited approach is used to define the kinetic normal flux at $(j+\frac{1}{2},k)^{^{th}}$ interface as follows.
 \begin{equation}
(\textbf{h}_{\perp i})_{j+\frac{1}{2},k, 2O}= (\textbf{h}_{\perp i})_{j+\frac{1}{2},k}+ \frac{1}{2}\Phi\left( (\Delta \textbf{h}_{\perp i}^{+})_{j+ \frac{1}{2},k}, (\Delta \textbf{h}_{\perp i}^{+})_{j- \frac{1}{2},k}\right)- \frac{1}{2}\Phi\left((\Delta \textbf{h}_{\perp i}^{-})_{j+ \frac{1}{2},k},(\Delta \textbf{h}_{\perp i}^{-})_{j+ \frac{3}{2},k}\right)
\label{eq:2d_2O_1}
\end{equation}
Here $(\textbf{h}_{\perp i})_{j+\frac{1}{2},k}$ is the first order flux. Equation \eqref{eq:2d_2O_1} can be rewritten as
\begin{eqnarray}
\begin{bmatrix} h_{\perp,1i} \\ h_{\perp,2i} \\ h_{\perp,3i} \end{bmatrix}_{j+ \frac{1}{2},k, 2O}= \begin{bmatrix} h_{\perp,1i} \\ h_{\perp,2i} \\ h_{\perp,3i} \end{bmatrix}_{j+ \frac{1}{2},k}+&& \frac{1}{2} \begin{bmatrix}\phi\left\{\left(\lambda_{p,\perp}\Delta f^{eq}_{1i}\right)_{j+\frac{1}{2},k},\left(\lambda_{p,n}\Delta f^{eq}_{1i}\right)_{j-\frac{1}{2},k}\right\} \\0 \\0 \end{bmatrix}\nonumber\\
-&& \frac{1}{2} \begin{bmatrix}0 \\ \phi\left\{\left(\lambda_{m,\perp}\Delta f^{eq}_{2i}\right)_{j+\frac{1}{2},k},\left(\lambda_{m,\perp}\Delta f^{eq}_{2i}\right)_{j+\frac{3}{2},k}\right\} \\ \phi\left\{\left(\lambda_{m,\perp}\Delta f^{eq}_{3i}\right)_{j+\frac{1}{2},k},\left(\lambda_{m,\perp}\Delta f^{eq}_{3i}\right)_{j+\frac{3}{2},k}\right\} \end{bmatrix}
\label{eq:2d_2O_2}
\end{eqnarray}
The macroscopic normal flux at the cell-interface is obtained by taking moment of the kinetic flux, and is given by
\begin{equation}
\begin{split}
(G_{\perp i})_{j+\frac{1}{2},k, 2O}&= \textbf{P}_{i}(\textbf{h}_{\perp i})_{j+\frac{1}{2},k}\\
&=(G_{\perp i})_{j+\frac{1}{2},k} + \frac{1}{2}\phi\left\{\left(\lambda_{p,\perp}\Delta f^{eq}_{1i}\right)_{j+\frac{1}{2},k},\left(\lambda_{p,\perp}\Delta f^{eq}_{1i}\right)_{j-\frac{1}{2},k}\right\}\\
&-\frac{1}{2}\phi\left\{\left(\lambda_{m,\perp}\Delta f^{eq}_{2i}\right)_{j+\frac{1}{2},k},\left(\lambda_{m,\perp}\Delta f^{eq}_{2i}\right)_{j+\frac{3}{2},k}\right\} -  \frac{1}{2}\phi\left\{\left(\lambda_{m,\perp}\Delta f^{eq}_{3i}\right)_{j+\frac{1}{2},k},\left(\lambda_{m,\perp}\Delta f^{eq}_{3i}\right)_{j+\frac{3}{2},k}\right\}
\end{split}
\label{eq:2d_2O_3}
\end{equation}  
At this point, we make an approximation. It has its basis in the work of Kumar \& Dass \cite{kumar}, who have, while working in continuous molecular velocity space, approximated the integral (w.r.t. molecular velocity) of the limiter function of two variables by the limiter function of integral of the two variables. In our framework, integrals are replaced by summations. Thus, we approximate \eqref{eq:2d_2O_3} by the following expression.
\begin{eqnarray}
(G_{\perp i})_{j+\frac{1}{2},k, 2O}= (G_{\perp i})_{j+\frac{1}{2},k} &&+ \frac{1}{2}\phi\left\{\left(\lambda_{p,\perp}\Delta f^{eq}_{1i}\right)_{j+\frac{1}{2},k},\left(\lambda_{p,\perp}\Delta f^{eq}_{1i}\right)_{j-\frac{1}{2},k}\right\} \nonumber \\
&&- \frac{1}{2}\phi\left\{\left(\lambda_{m,\perp}\Delta (f^{eq}_{2i}+ f^{eq}_{3i})\right)_{j+\frac{1}{2},k},\left(\lambda_{m,\perp}\Delta (f^{eq}_{2i}+ f^{eq}_{3i})\right)_{j+\frac{3}{2},k}\right\}
\label{eq:2d_2O_4}
\end{eqnarray}
For our model, $\textbf{f}^{eq}_{1}$(=$\frac{1}{\lambda_{p,\perp}-\lambda_{m,\perp}}\left(-\lambda_{m,\perp}\textbf{U}+ \textbf{G}_{\perp}\right)$) and $\textbf{f}^{eq}_{2}+\textbf{f}^{eq}_{3}$(=$\frac{1}{\lambda_{p,\perp}-\lambda_{m,\perp}}\left(\lambda_{p,\perp}\textbf{U}- \textbf{G}_{\perp}\right)$) are independent of $\lambda_{\parallel}$. Thus, as a consequence of the approximation in \eqref{eq:2d_2O_4}, our flux for second order accuracy simplifies and becomes independent of $\lambda_{\parallel}$, thus becoming locally one-dimensional. Finally, the temporal derivative is approximated using SSPRK method, as described in \eqref{eq:1d_2O_6}.

\section{Kinetic model for viscous flows}
We consider a first order approximation to f, {\em i.e.}, $f= f^{CE}$, where $f^{CE}$ is the Chapman-Enskog distribution function. 
 For a first order approximation to $f$, the moments of variable velocity Boltzmann equation give us the macroscopic Navier-Stokes equations, given by 
\begin{equation}
\frac{\partial \textbf{U}}{\partial t}+ \frac{\partial \textbf{G}_{i}}{\partial x_{i}}= \frac{\partial \textbf{G}_{vis, i}}{\partial x_{i}}
\label{eq:vis_1}
\end{equation}
with
\begin{equation}
\textbf{U}= \begin{bmatrix} \rho \\ \rho u_{j} \\ \rho E \end{bmatrix}, \textbf{G}_{i}= \begin{bmatrix} \rho u_{i} \\ \rho u_{i}u_{j}+ p \delta_{ij} \\ (\rho E+ p)u_{i} \end{bmatrix}, \textbf{G}_{vis, i}= \begin{bmatrix} 0 \\ \tau_{ij} \\ \tau_{ij}u_{j} -q_{i} \end{bmatrix}
\label{eq:vis_2}
\end{equation}
and
\begin{equation}
\tau_{ij}= \mu\left(\frac{\partial u_{i}}{\partial x_{j}}+ \frac{\partial u_{j}}{\partial x_{i}}\right) -\frac{2 \mu}{3}\frac{\partial u_{k}}{\partial x_{k}}\delta_{ij}, \ q_{i} =-K \frac{\partial T}{\partial x_{i}}
\label{eq:vis_3}
\end{equation}
 Here, the moment relations are

\begin{equation}
\int_{\mathbb{R}^{N}}d\textbf{v} \int_{\mathbb{R}^{+}}dI \ \bm{\Psi} f^{CE} = \textbf{U}, \ \int_{\mathbb{R}^{N}}v_{i}d\textbf{v} \int_{\mathbb{R}^{+}}dI \ \bm{\Psi} f^{CE}= \textbf{G}_{i}- \textbf{G}_{vis,i}= \textbf{G}_{net,i}
\label{eq:vis_4}
\end{equation}

In 1D, we define $f^{CE}_{i}$ in the same fashion as we did $f^{eq}_{i}$, but satisfying the moment relations \eqref{eq:vis_4}, {\em i.e.}, 
\begin{subequations}
\label{eq:vis_5}
\begin{equation}
\left\langle f^{CE}_{i} \right\rangle= f^{CE}_{1i}+ f^{CE}_{2i}= U_{i}
\end{equation}
\begin{equation}
\left\langle vf^{CE}_{i} \right\rangle= \lambda_{p} f^{CE}_{1i}+  \lambda_{m} f^{CE}_{2i} = G_{i} - G_{vis, i}
\end{equation}
\end{subequations}
The moment relations \eqref{eq:vis_5} give us
\begin{subequations}
\label{eq:vis_6}
\begin{equation}
f^{CE}_{1i}= \frac{-\lambda_{m}}{\lambda_{p}- \lambda_{m}}U_{i} + \frac{1}{\lambda_{p}- \lambda_{m}}\left(G_{i}- G_{vis, i}\right)= f^{eq}_{1i}- \frac{1}{\lambda_{p}- \lambda_{m}}G_{vis, i}
\end{equation}
\begin{equation}
f^{CE}_{2i}= \frac{\lambda_{p}}{\lambda_{p}- \lambda_{m}}U_{i} - \frac{1}{\lambda_{p}- \lambda_{m}}\left(G_{i}- G_{vis, i}\right)= f^{eq}_{2i}+ \frac{1}{\lambda_{p}- \lambda_{m}}G_{vis, i}
\end{equation}
\end{subequations}
We use the operator-splitting strategy to solve a Flexible Velocity Boltzmann Equation, with the only difference being that we now relax the distribution function to the Chapman-Enskog distribution function in the collision step. The net kinetic numerical flux is evaluated as follows.
\begin{subequations}
\label{eq:vis_7}
\begin{equation}
(\textbf{h}_{net,i})_{j+\frac{1}{2}} = \frac{1}{2}\left\{(\textbf{h}_{net,i})_{j}+ (\textbf{h}_{net,i})_{j+1}\right\}-\frac{1}{2}\left\{(\Delta \textbf{h}_{net,i}^{+})_{j+\frac{1}{2}}- (\Delta \textbf{h}_{net,i}^{-})_{j+\frac{1}{2}}\right\}, \ \textbf{h}_{net,i}=\Lambda\textbf{f}_{i}^{CE}
\end{equation}
\begin{equation}
(\Delta \textbf{h}_{net,i}^{\pm})_{j+\frac{1}{2}}= \left(\Lambda^{\pm}\Delta \textbf{f}_{i}^{CE}\right)_{j+ \frac{1}{2}}
\end{equation}
\end{subequations}

The macroscopic flux at the interface is then evaluated to be 
\begin{eqnarray}
(G_{net,i})_{j+\frac{1}{2}} =&& \textbf{P}_{i} (\textbf{h}_{net,i})_{j+1/2}\nonumber\\
=&& (G_{i})_{j+\frac{1}{2}} - \left[\left(\frac{\lambda_{p}}{\lambda_{p}- \lambda_{m}}\right)_{j+1/2}\left(G_{vis, i}\right)_{j}- \left(\frac{\lambda_{m}}{\lambda_{p}- \lambda_{m}}\right)_{j+1/2} \left(G_{vis, i}\right)_{j+1}\right]
\label{eq:vis_8}
\end{eqnarray}
Here, $(G_{i})_{j+\frac{1}{2}}$ is the inviscid flux for first order accuracy (Equation \eqref{eq:1d_euler_15}). At this point, we simplify our flux by approximating $\frac{\lambda_{p}}{\lambda_{p}- \lambda_{m}}$ and -$\frac{\lambda_{m}}{\lambda_{p}- \lambda_{m}}$ by the fraction $\frac{1}{2}$, which holds true for the scalar numerical diffusion model. This approximation prevents the velocities $\lambda_{p}$ and $\lambda_{m}$, which are determined on the basis of inviscid considerations, from influencing viscous fluxes. Further, to ensure that our scheme is second order accurate, the inviscid anti-diffusion terms are added as well. Thus, our final flux is given by
\begin{equation}
(G_{net,i})_{j+\frac{1}{2}}= (G_{i})_{j+\frac{1}{2}, 2O}- \frac{1}{2}\left[\left(G_{vis, i}\right)_{j}+ \left(G_{vis, i}\right)_{j+1}\right]
\label{eq:vis_9}
\end{equation}
Here, $(G_{i})_{j+\frac{1}{2}, 2O}$ is the inviscid flux for second order accuracy, given by Equation \eqref{eq:1d_2O_5}. Finally, the temporal derivative is discretized using SSPRK method (see \eqref{eq:1d_2O_6}).

\subsection{Time step restriction based on stability}
We have attempted to obtain a stability criterion for the viscous case by imposing Bouchut's stability criteria on our Chapman-Enskog type distribution, $\textbf{f}_{CE}$. For stability analysis, we consider the simpler symmetrical kinetic model, for which these distributions in vector form are given by
	\begin{equation}
	\textbf{f}^{CE}_{1}= \frac{\textbf{U}}{2}+ \frac{\textbf{G}-\textbf{G}_{vis}}{2\lambda}, \  \textbf{f}^{CE}_{2}= \frac{\textbf{U}}{2}- \frac{\textbf{G}-\textbf{G}_{vis}}{2\lambda}
	\label{eq:res2_2a}
	\end{equation}
	We then impose the following criteria on these distributions, based on Bouchut's condition.
	\begin{equation}
	eig\left(\frac{\partial \textbf{f}^{CE}_{1,2}}{\partial \textbf{U}}\right) \subset\left[0,\infty\right)
	\label{eq:res2_2b}
	\end{equation}
	Here, $eig$ refers to the eigenspectrum. Condition \eqref{eq:res2_2b} gives us
	\begin{equation}
	\lambda \geq \rho\left(\frac{\partial}{\partial \textbf{U}}(\textbf{G}-\textbf{G}_{vis})\right)
	\label{eq:res2_2c}
	\end{equation}
	Here $\rho$ is the spectral radius. Using the definitions $\lambda_{max}=\rho\left(\frac{\partial \textbf{G}}{\partial \textbf{U}}\right)$, $\lambda_{vis\_max}=\rho\left(\frac{\partial \textbf{G}_{vis}}{\partial \textbf{U}}\right)$, we make the following estimate of the lower bound on $\lambda$, based on \eqref{eq:res2_2c}.
	\begin{equation}
	\lambda \geq \rho\left(\frac{\partial \textbf{G}}{\partial \textbf{U}}\right)+ \rho\left(\frac{\partial \textbf{G}_{vis}}{\partial \textbf{U}}\right)= \lambda_{max}+ \lambda_{vis\_max}
	\label{eq:res2_2d}
	\end{equation}
	An estimate of the time step based on stability is then given by
	\begin{equation}
	\Delta t = \sigma \ min_{j}\left(\frac{\Delta x}{\lambda_{max,j}+ \lambda_{vis\_max,j}}\right), \ \sigma\text{= CFL no}
	\label{eq:res2_2e}
	\end{equation}
In 2D, the time step is estimated using a similar analysis as
\begin{equation}
\Delta t =\sigma \ min_{j,k}\left[\frac{A}{\lambda_{max}l+ \lambda_{vis\_max}l}\right]_{j,k}
\label{eq:res2_2f}
\end{equation}
where, $\lambda_{max}l= (\lambda_{max})_{\xi}l_{\xi}+ (\lambda_{max})_{\eta}l_{\eta}$. The term $\lambda_{vis\_max}l$ is approximated by assuming orthogonality of $\xi$ and $\eta$ (refer \cite{1987PhDT_Martinelli}) and by ignoring cross terms, and is given by
\begin{equation}
\left(\lambda_{vis\_max}l\right)_{j,k}= \frac{\gamma}{Re\ Pr}\left[\frac{\mu}{\rho A}\left(l^{2}_{\xi}+l^{2}_{\eta}\right)\right]_{j,k}
\label{eq:res2_2g}
\end{equation}

\section{Results and Discussion}

\subsection{Experimental Order of Convergence}
To determine the Experimental Order of Convergence (EOC) of our first and second order schemes, we have solved a simple 1D Euler test case with the following initial conditions.
\begin{subequations}
\label{eq:EOC_1}
\begin{equation}
\rho(x,0)= \rho_{0}(x)= 1+ 0.2 sin(\pi x), \ x\in [0,2]
\end{equation}
\begin{equation}
u(x,0)= 0.1, \ p(x,0)= 0.5
\end{equation}
\end{subequations}
The pressure and velocity are thus initially constant, whereas initial density is perturbed with a sinusoidal variation in space. Periodic boundary conditions are applicable at the two ends. The exact solution for this test case is known and is given by
\begin{subequations}
\label{eq:EOC_2}
\begin{equation}
\rho(x,t)= \rho_{0}(x- ut)= 1+ 0.2 sin\left\{\pi (x- 0.1 t)\right\}
\end{equation}
\begin{equation}
u(x,t)= 0.1, \ p(x,t)= 0.5
\end{equation}
\end{subequations}
We have solved this problem numerically and considered its solution at time t= 0.5. The numerical solution is obtained for varying grid sizes, {\em i.e.}, $Nx$ (=$\frac{2}{\Delta x}$)= 40, 80, 160, etc. Then, the $L_{1}$ and $L_{2}$ errors in solution are computed as follows.  
\begin{subequations}
\label{eq:EOC_3}
\begin{equation}
\left\|\varepsilon_{Nx}\right\|_{L_{1}}= \Delta x \sum^{Nx}_{i=1}|\rho^{i}- \rho^{i}_{exact}|
\end{equation}
\begin{equation}
\left\|\varepsilon_{Nx}\right\|_{L_{2}}= \sqrt{\Delta x \sum^{Nx}_{i=1}(\rho^{i}- \rho^{i}_{exact})^{2}}
\end{equation}
\end{subequations}
Here, $\rho^{i}$ and $\rho^{i}_{exact}$ are the numerical and exact solutions for the $i^{th}$ cell. Now, for a $p^{th}$ order accurate scheme,
\begin{subequations}
\label{eq:EOC_4}
\begin{equation}
\left\|\varepsilon_{Nx}\right\|= C \Delta x^{p}+ O(\Delta x^{p+1})\text{. Similarly,}
\end{equation}
\begin{equation}
\left\|\varepsilon_{Nx/2}\right\|= C (2 \Delta x)^{p}+ O(\Delta x^{p+1})\text{, }(Nx\propto\frac{1}{\Delta x})
\end{equation}
\end{subequations}
Thus,
\begin{equation}
\frac{\left\|\varepsilon_{Nx/2}\right\|}{\left\|\varepsilon_{Nx}\right\|}= 2^{p}+ O(\Delta x) \Rightarrow log_{2}\left(\frac{\left\|\varepsilon_{Nx/2}\right\|}{\left\|\varepsilon_{Nx}\right\|}\right)= p+ O(\Delta x)
\label{eq:EOC_5}
\end{equation}
The experimental order of convergence (EOC) of the scheme is then given by
\begin{equation}
\textrm{EOC} = log_{2}\left(\frac{\left\|\varepsilon_{Nx/2}\right\|}{\left\|\varepsilon_{Nx}\right\|}\right)
\label{eq:EOC_6}
\end{equation}

\begin{table}
\centering
\begin{tabular}{ |c|c|c|c|c|c| }
\hline
Nx& $\Delta$x& $L_{1}$ Error & EOC & $L_{2}$ Error & EOC\\
\hline
40	& 0.05 & 0.0125806130 & & 0.0099139194 & \\
80	& 0.025 & 0.0063876317 & 0.977849 & 0.0050279880 & 0.979474 \\
160	& 0.0125 & 0.0032325335 & 0.982616 & 0.0025455194 & 0.982021 \\
320	& 0.00625 & 0.0016215746 & 0.995270 & 0.0012771561 & 0.995025 \\
640	& 0.003125 & 0.0008132367 & 0.995648 & 0.0006405558 & 0.995539 \\
1280 & 0.0015625 & 0.0004072322 & 0.997824 & 0.0003207729 & 0.997772 \\
2560 & 0.00078125 & 0.0002036996 & 0.999408 & 0.0001604550 & 0.999384 \\
\hline
\end{tabular}
\centering
\caption{EOC using $L_{1}$ and $L_{2}$ error norms for first order accuracy}
\label{table:EOC_1}
\end{table}

\begin{table}
\centering
\begin{tabular}{ |c|c|c|c|c|c| }
\hline
Nx& $\Delta$x& $L_{1}$ Error & EOC & $L_{2}$ Error & EOC\\
\hline
40 & 0.05 & 0.0019706495 & & 0.0019530945 & \\
80 & 0.025 & 0.0005608529 & 1.812977 & 0.0006506868 & 1.585726 \\
160 & 0.0125 & 0.0001501011 & 1.901687 & 0.0002130568 & 1.610725 \\
320 & 0.00625 & 0.0000402887 & 1.897486 & 0.0000694866 & 1.616432 \\
640 & 0.003125 & 0.0000105659 & 1.930960 & 0.0000225218 & 1.625412 \\
1280 & 0.0015625 & 0.0000027395 & 1.947406 & 0.0000072668 & 1.631922 \\
2560 & 0.00078125 & 0.0000007073 & 1.953521 & 0.0000023369 & 1.636734 \\
\hline
\end{tabular}
\centering
\caption{EOC using $L_{1}$ and $L_{2}$ error norms for second order accuracy with minmod limiter}
\label{table:EOC_2}
\end{table}

\begin{table}
\centering
\begin{tabular}{ |c|c|c|c|c|c| }
\hline
Nx& $\Delta$x& $L_{1}$ Error & EOC & $L_{2}$ Error & EOC\\
\hline
40 & 0.05 & 0.0003650312 & & 0.0002856627 & \\
80 & 0.025 & 0.0000848129 & 2.105663 & 0.0000663599 & 2.105928 \\
160 & 0.0125 & 0.0000206898 & 2.035361 & 0.0000162291 & 2.031728 \\
320 & 0.00625 & 0.0000051456 & 2.007508 & 0.0000040400 & 2.006162 \\
640 & 0.003125 & 0.0000012848 & 2.001821 & 0.0000010090 & 2.001455 \\
1280 & 0.0015625 & 0.0000003206 & 2.002553 & 0.0000002518 & 2.002460 \\
2560 & 0.00078125 & 0.0000000797 & 2.007936 & 0.0000000626 & 2.007912 \\
\hline
\end{tabular}
\centering
\caption{EOC using $L_{1}$ and $L_{2}$ error norms for second order accuracy, unlimited ($\phi$= 1)}
\label{table:EOC_3}
\end{table}

\begin{figure}
\centering
\begin{tabular}{cc}
\includegraphics[width=0.45\textwidth]{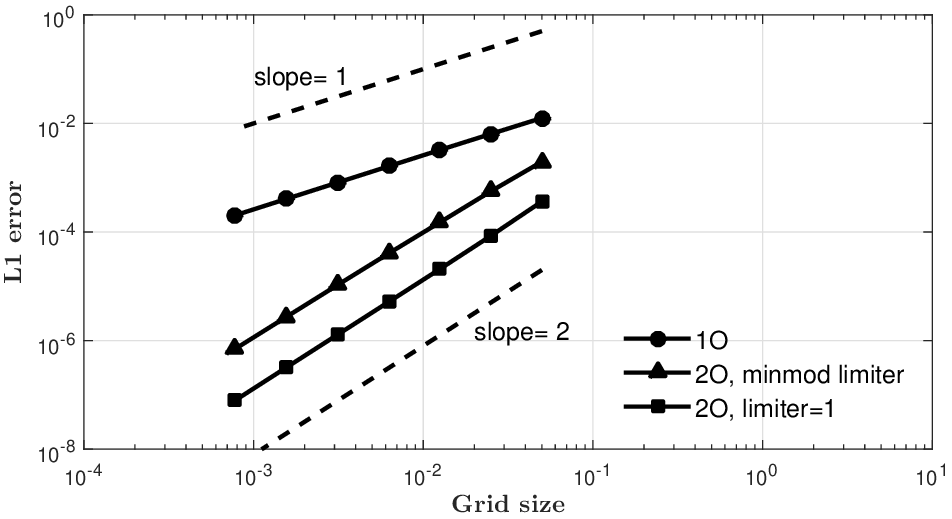} & \includegraphics[width=0.45\textwidth]{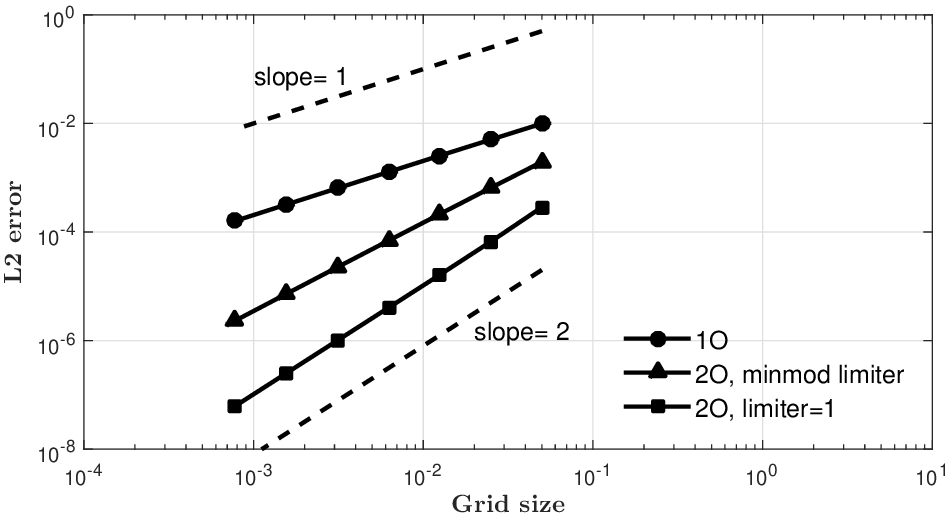}
\end{tabular}
\caption{Left: $L_{1}$ error norm vs grid size, Right: $L_{2}$ error norm vs grid size}
\label{fig:EOC_1}
\end{figure}
The $L_{1}$ and $L_{2}$ errors of the present scheme for I order accuracy are tabulated in Table \ref{table:EOC_1}. The II order results are tabulated in Table \ref{table:EOC_2} (limited) and Table \ref{table:EOC_3} (unlimited, {\em i.e.}, $\phi$= 1) respectively. The log-log plots comparing the EOC with slopes 1 and 2 are shown in Figure \ref{fig:EOC_1}.

\subsection{1D Euler tests}

\begin{table}
\centering
\resizebox{\textwidth}{!}{%
\begin{tabular}{ |c|c|c|c|c|c|c|c|c|c| }
\hline
$x_{1}$& $x_{2}$ &$x_{0}$& $\rho_{L}$& $u_{L}$& $p_{L}$& $\rho_{R}$& $u_{R}$& $p_{R}$& $t_{final}$\\
\hline
0& 1& 0.5& 1.0& 1.0& $\frac{1}{\gamma M(=2)^{2}}$& $\frac{\frac{\gamma +1}{\gamma -1}\frac{p_{R}}{p_{l}}+1}{\frac{\gamma +1}{\gamma -1}+ \frac{p_{R}}{p_{L}}}$& $\sqrt{\frac{\gamma (2+ (\gamma -1)M^{2})p_{R}}{(2\gamma M^{2}+1-\gamma)\rho_{R}}}$& $p_{L}\frac{2\gamma M^{2}- (\gamma -1)}{\gamma +1}$ &1.5\\
0& 1& 0.5& 1.4& 0& 1.0 &1.0 &0.0 &1.0 &2.0 \\
0& 1& 0.5& 3.86& -0.81& 10.33& 1.0& -3.44& 1.0& 1.0\\
0& 1& 0.5& 1.4& 0.1& 1.0& 1.0& 0.1& 1.0& 1.0\\
0 & 1 & 0.3 & 1 & 0.75 & 1 & 0.125 & 0 & 0.1 & 0.2 \\ 
0& 1& 0.5& 1.0& -2.0& 0.4& 1.0& 2.0& 0.4& 0.15 \\
0& 1& 0.5& 1.0& 0.0& 1000.0& 1.0& 0.0& 0.01& 0.012\\
0& 1& 0.4& 5.99924& 19.5975& 460.894& 5.99242& -6.19633& 46.0950& 0.035\\
0& 1& 0.8& 1.0& -19.59745& 1000.0& 1.0& -19.59745& 0.01& 0.012\\
-1& 1& -0.8& 3.857143& 2.629369& 10.3333& 1+ 0.2$sin(5\pi x)$ & 0.0& 1.0& 0.47\\
\hline
\end{tabular}}
\caption{Initial condition for 1D test cases}
\label{table:1d_euler}
\end{table}

\begin{figure}
\centering
\resizebox{\textwidth}{!}{%
\includegraphics[width=15cm]{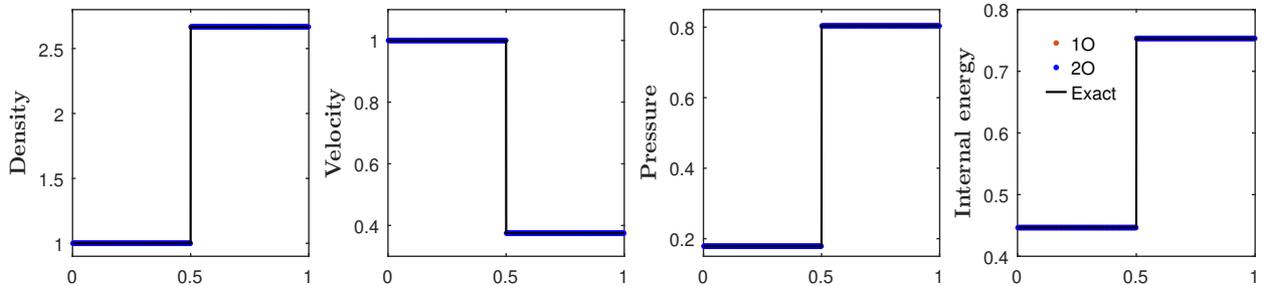}}
\caption{\label{fig:1d_euler_1} Test case 1: Steady shock}
\end{figure}

\begin{figure}
\centering
\resizebox{\textwidth}{!}{%
\includegraphics[width=15cm]{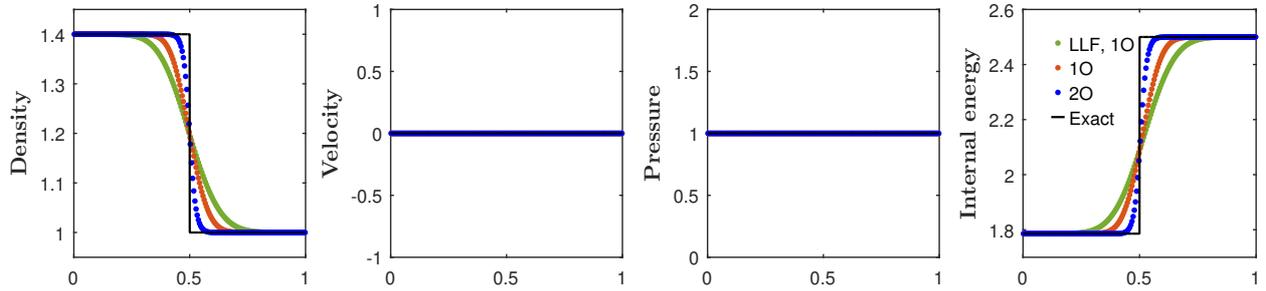}}
\caption{\label{fig:1d_euler_2} Test case 2: Steady contact-discontinuity}
\end{figure}

\begin{figure}
\centering
\resizebox{\textwidth}{!}{%
\includegraphics[width=15cm]{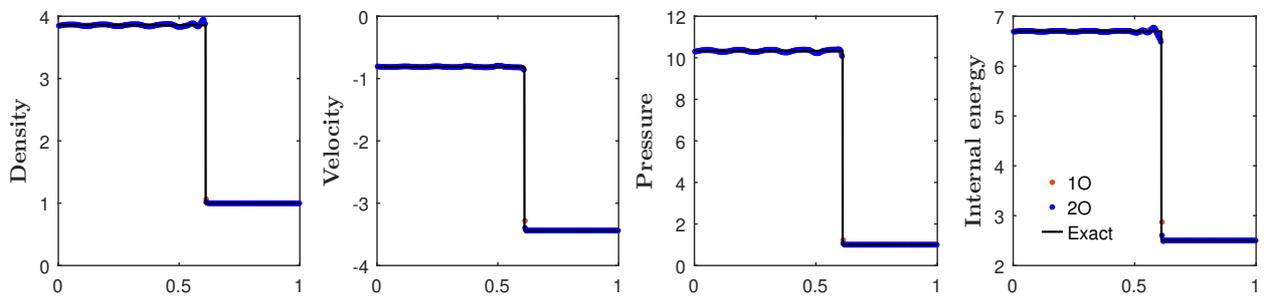}}
\caption{\label{fig:1d_euler_3} Test case 3: Slowly moving shock}
\end{figure}

\begin{figure}
\centering
\resizebox{\textwidth}{!}{%
\includegraphics[width=15cm]{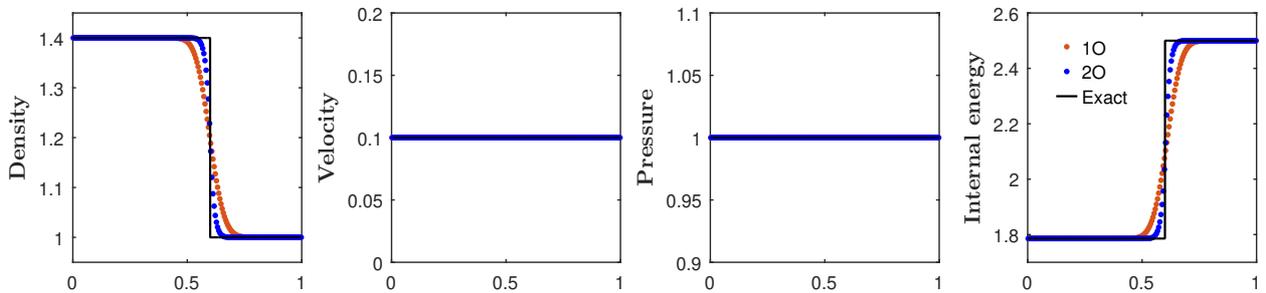}}
\caption{\label{fig:1d_euler_4} Test case 4: Slowly moving contact-discontinuity}
\end{figure}

\begin{figure}
\centering
\resizebox{\textwidth}{!}{%
\includegraphics[width=15cm]{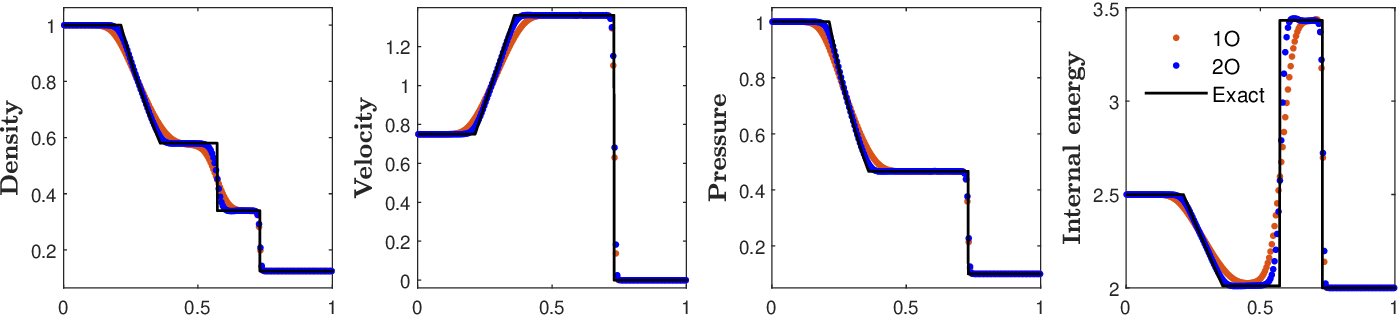}}
\caption{\label{fig:1d_euler_5} Test case 5: Sod's shock tube problem}
\end{figure}

\begin{figure}
\centering
\resizebox{\textwidth}{!}{%
\includegraphics[width=15cm]{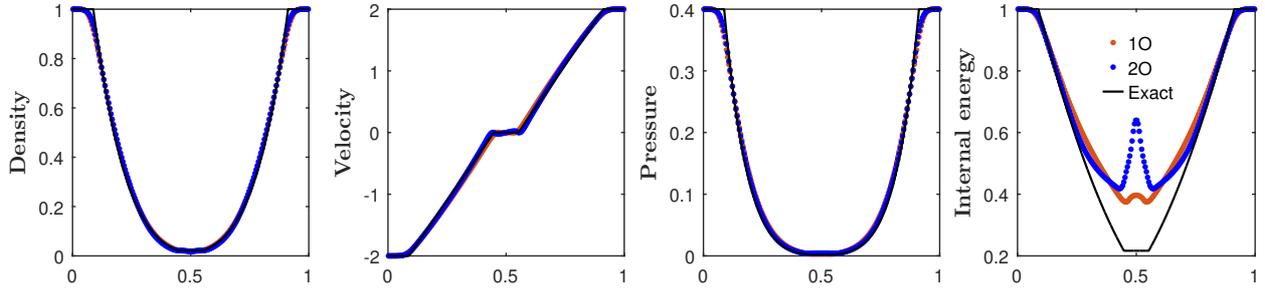}}
\caption{\label{fig:1d_euler_6} Test case 6: Overheating problem}
\end{figure}

\begin{figure}
\centering
\resizebox{\textwidth}{!}{%
\includegraphics[width=15cm]{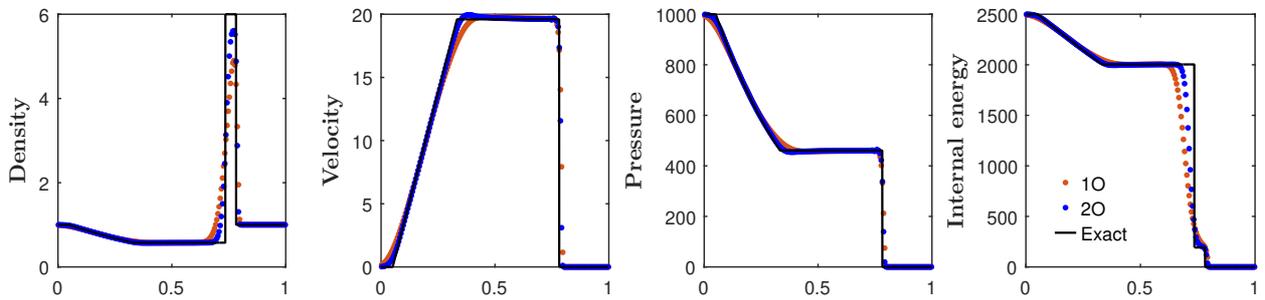}}
\caption{\label{fig:1d_euler_7} Test case 7: Left half portion of Woodward and Colella problem}
\end{figure}

\begin{figure}
\centering
\resizebox{\textwidth}{!}{%
\includegraphics[width=15cm]{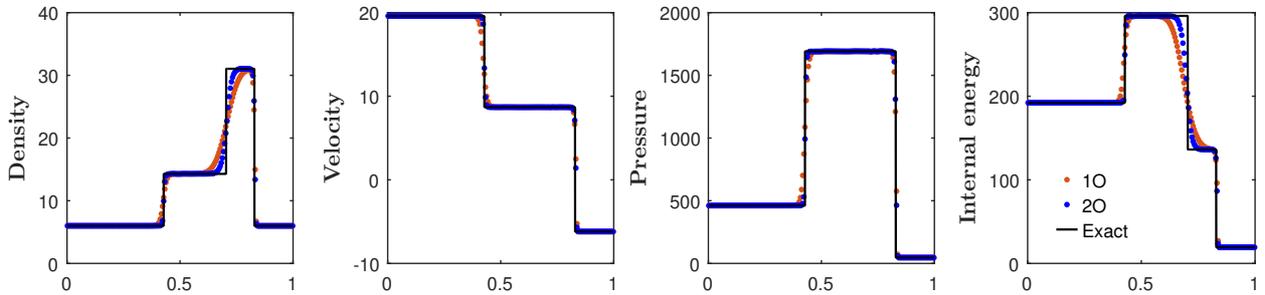}}
\caption{\label{fig:1d_euler_8} Test case 8: Colliding strong shocks}
\end{figure}

\begin{figure}
\centering
\resizebox{\textwidth}{!}{%
\includegraphics[width=15cm]{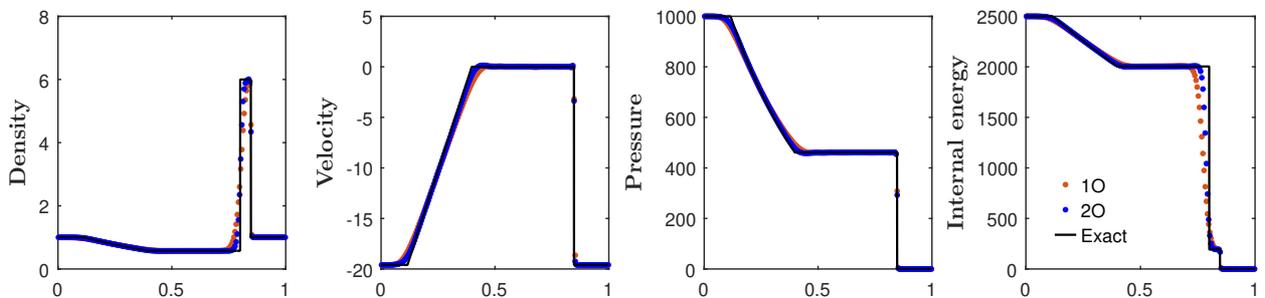}}
\caption{\label{fig:1d_euler_9} Test case 9}
\end{figure}

\begin{figure}
\centering
\resizebox{\textwidth}{!}{%
\includegraphics[width=15cm]{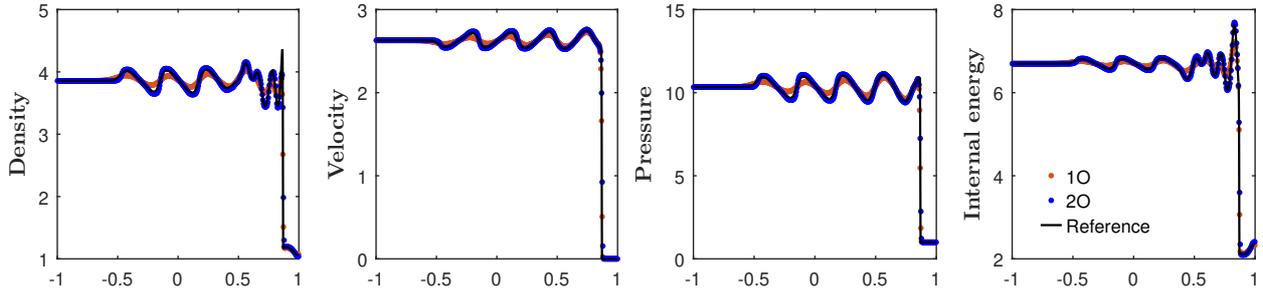}}
\caption{\label{fig:1d_euler_10} Test case 10: Shock-entropy wave collision}
\end{figure}

We have solved an extensive set of benchmark 1D Euler problems to test the accuracy and robustness of our numerical scheme. The initial conditions for these test cases have been tabulated in  Table \ref{table:1d_euler}. In the table, $x_{1}$ and $x_{2}$ denote the two ends of the domain, whereas $x_{0}$ is the position of initial discontinuity. For all the test cases, the cell size $\Delta x$= 0.005, CFL number $\sigma$= 0.8, and Neumann boundary conditions are applied at the two ends. Test case 1 has a steady shock, introduced as an initial discontinuity, with a freestream Mach 2 flow. The results in Figure \ref{fig:1d_euler_1} show that our numerical scheme captures the steady shock exactly for both first and second order accuracy. Test case 2 comprises a steady contact-discontinuity as the initial condition. The results in Figure \ref{fig:1d_euler_2} show that our numerical scheme captures the discontinuity sharply and with much less numerical diffusion than the Local Lax-Friedrichs (LLF) scheme. Test case 3 consists of a slowly moving shockwave. The results in Figure \ref{fig:1d_euler_3} show the shock captured over a few cells, with minor oscillations. The slowly moving contact-discontinuity in Test case 4 (Figure \ref{fig:1d_euler_4} ) is also captured with reasonable accuracy, without any oscillations. Test cases 5 to 9 are taken from Toro (\cite{toro2013riemann}). Test case 5 is the Sod's shock tube problem. Its solution consists of a shock wave and a contact-discontinuity traveling to the right and an expansion wave (containing a sonic point) going to the left. The results in Figure \ref{fig:1d_euler_5} show that no entropy-violating expansion shock is formed. Test case 6 is the well-known overheating problem for which several Riemann solvers fail. Its solution consists of two strong symmetric expansions to the left and right, with a contact-discontinuity of vanishing strength in the middle. For this test case, pressure at the center reaches near vacuum, making it a suitable test case for assessing the performance of a scheme at low densities. The results in Figure \ref{fig:1d_euler_6} show that our scheme does not fail for this test case, and captures the the expansions accurately. However, there is an increase in internal energy at the center, which is observed for many numerical methods due to numerical overheating. Test cases 7 to 9 test the robustness of a scheme in handling large gradients. Test case 7 is the left half of the blast wave problem of Woodward and Colella. Its solution consists of a strong shock to the right, a contact-discontinuity in the middle, and an expansion fan to the left. Test case 8 involves the collision of two strong shocks; its solution consists of a left facing shock (traveling very slowly to the right), a right traveling contact-discontinuity and a right traveling shock. Test case 9 consists of a left rarefaction wave, a right traveling shock wave, and a stationary contact-discontinuity. The results for these test cases are shown in Figures \ref{fig:1d_euler_7} to \ref{fig:1d_euler_9}. The results are reasonably accurate. Test case 10 is the shock-entropy wave interaction problem. It comprises a Mach 3 shock traveling right and interacting with a stationary medium with an initial sinusoidal perturbation in density. This initial disturbance gives rise to a continuous interaction of smooth flow with discontinuities. Similar interactions occur in compressible turbulence simulations. This makes it a suitable problem to test the ability of a numerical scheme to resolve complex interactions, which can be used in turbulent computations. The results in Figure \ref{fig:1d_euler_10} show that our second order accurate solution matches well with the reference (fine grid) solution.

\subsection{2D Euler tests}
Some standard 2D inviscid test cases are solved to showcase the accuracy, robustness, as well as positivity preserving properties of our proposed numerical method. The boundary conditions used for our kinetic model are derived in \ref{appendix:a2}. For all test cases, CFL no $\sigma$= 0.8 unless otherwise specified. For steady state results, our strategy is to evolve the solution in time until a minimum density residual of $10^{-10}$ or maximum time steps of 50000 is reached, whichever occurs earlier.

\subsubsection{Oblique shock reflection}
\begin{figure}[h!] 
\centering
\begin{tabular}{cc}
\includegraphics[width=0.45\textwidth]{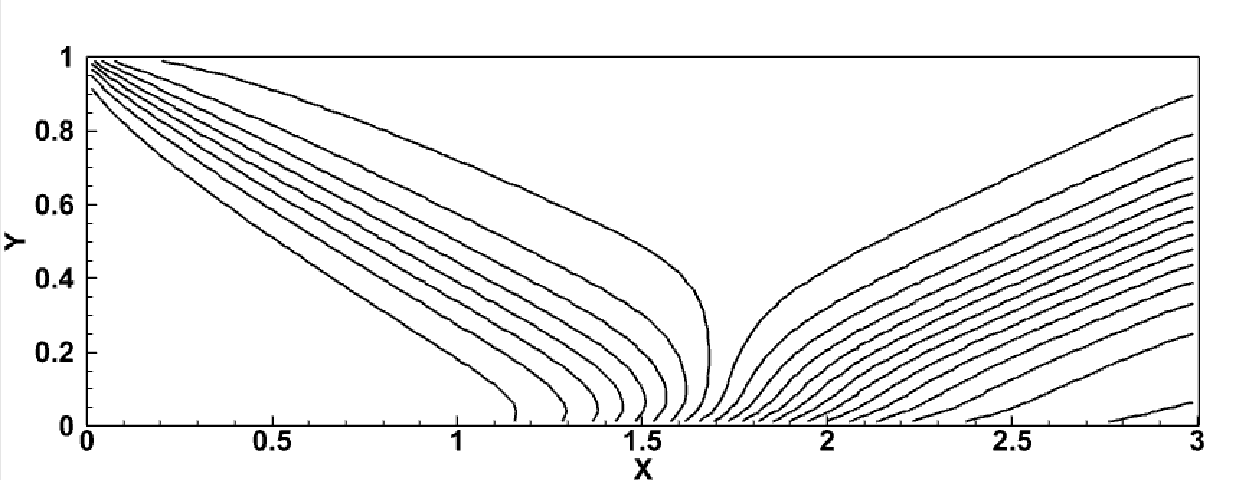} & \includegraphics[width=0.45\textwidth]{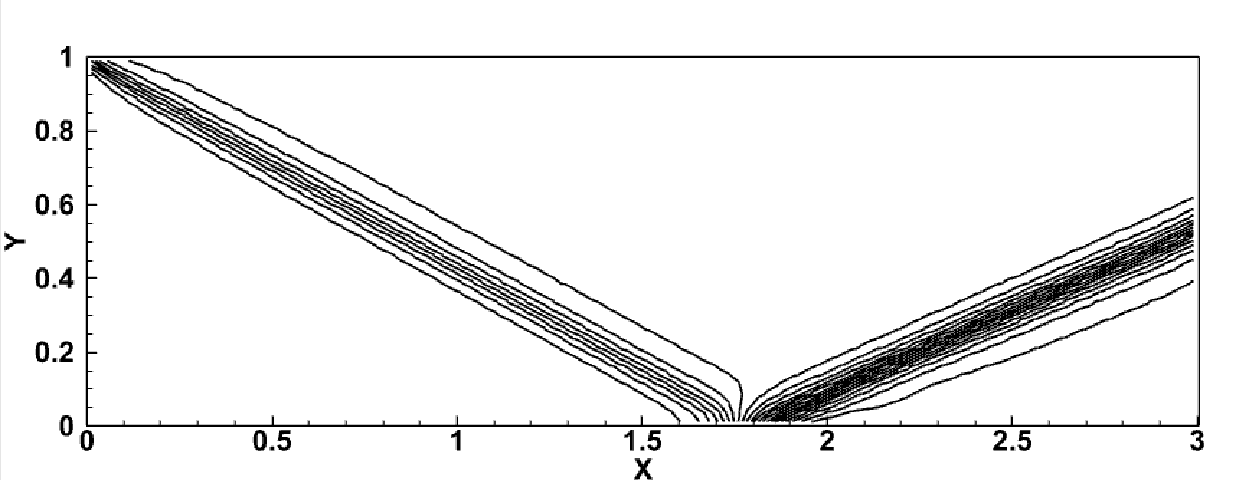}\\
\includegraphics[width=0.45\textwidth]{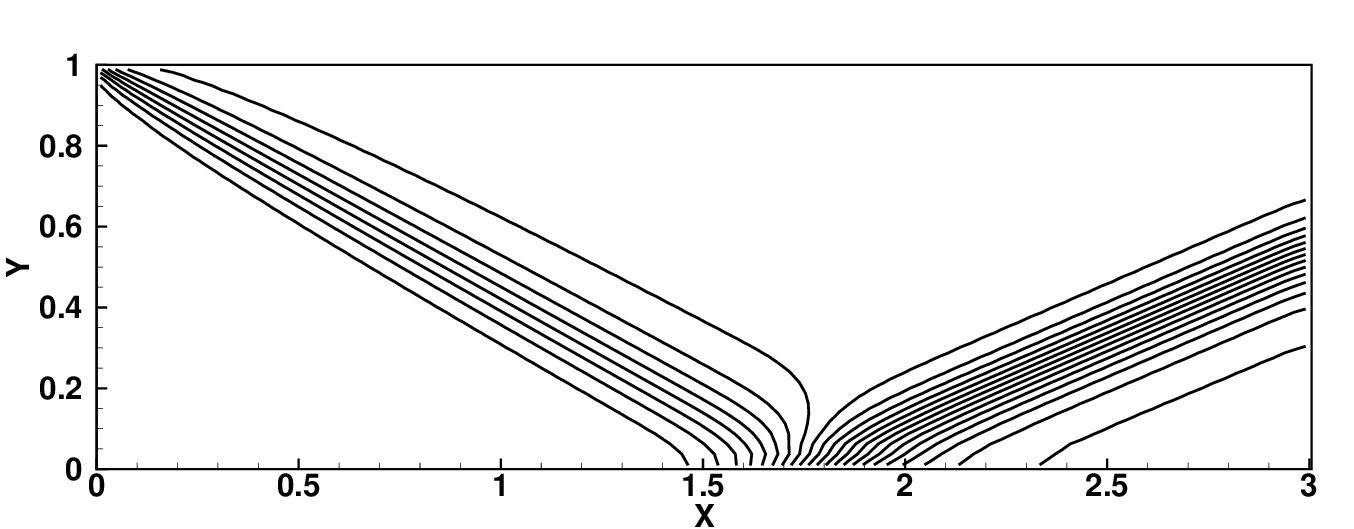} & \includegraphics[width=0.45\textwidth]{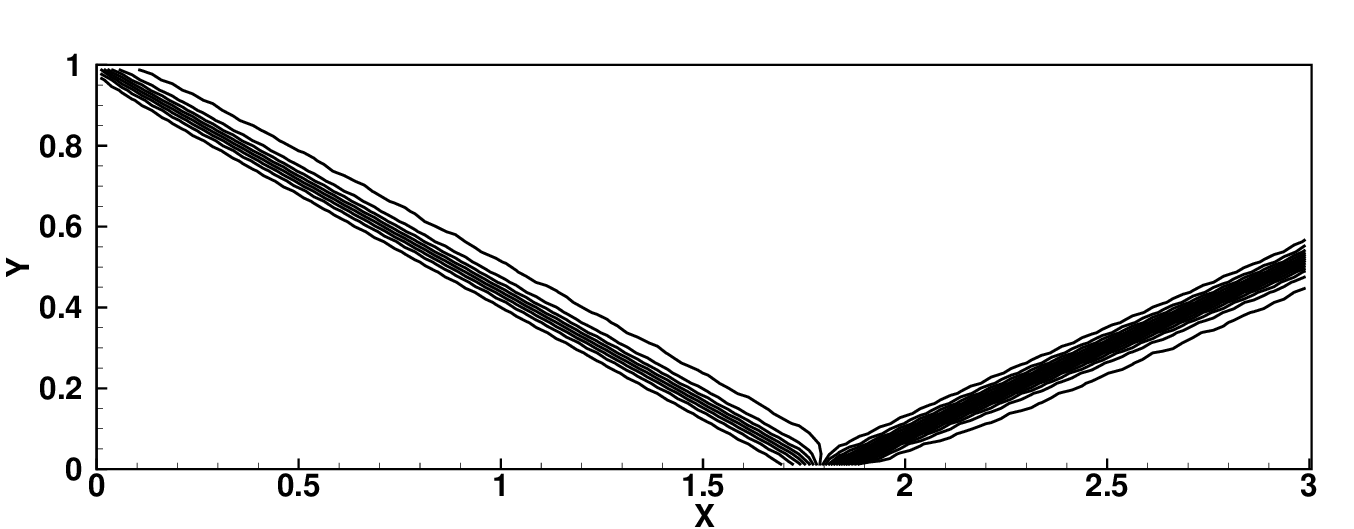}\\
\includegraphics[width=0.45\textwidth]{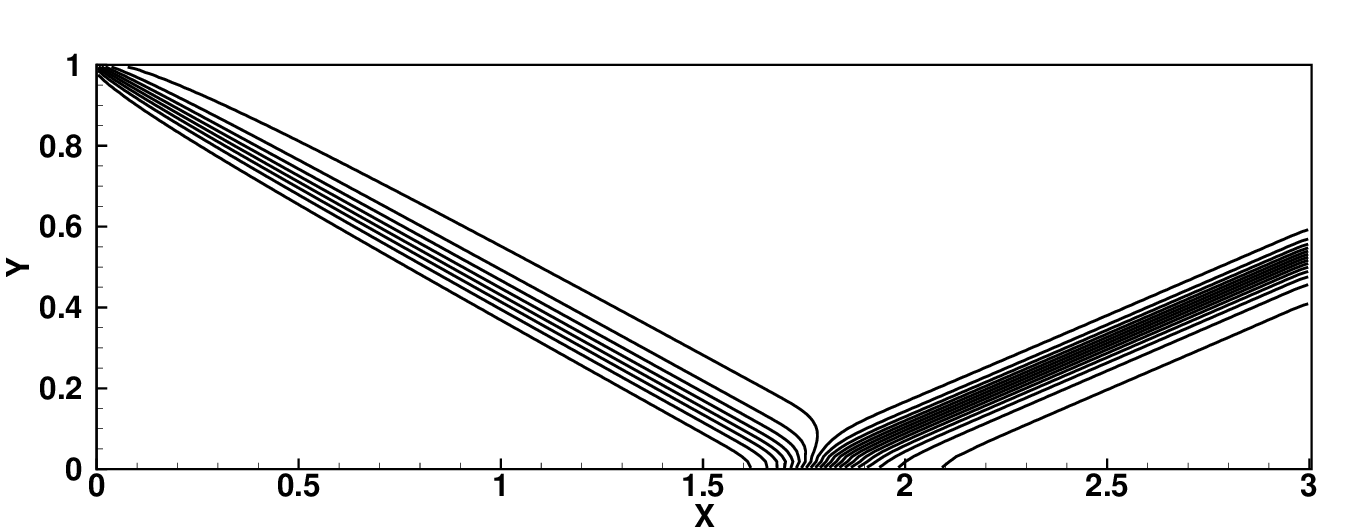} & \includegraphics[width=0.45\textwidth]{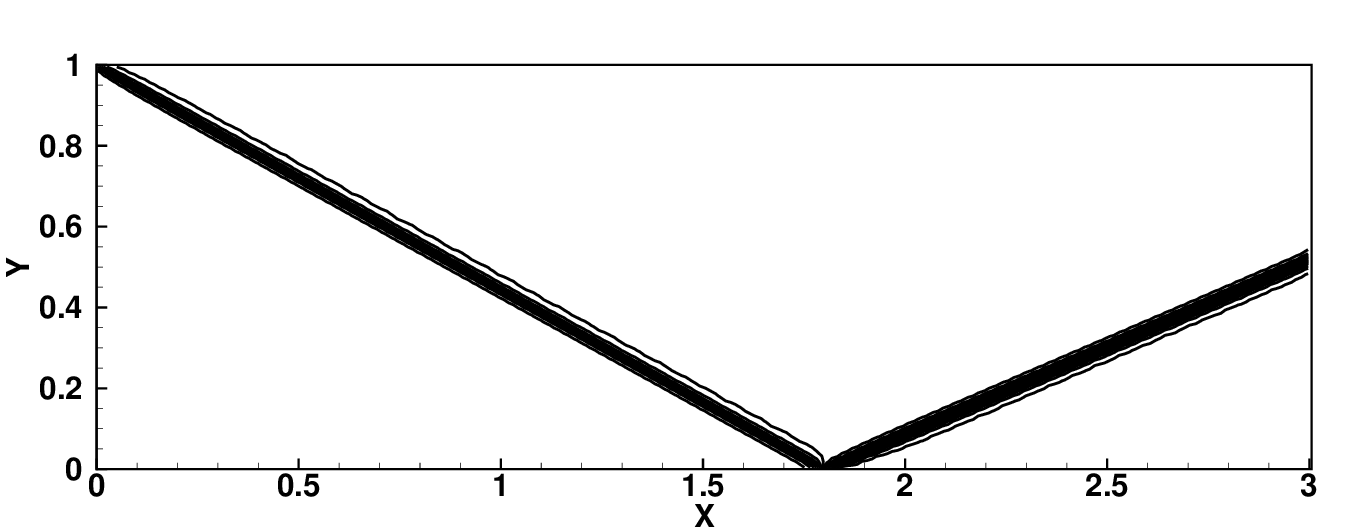} 
\end{tabular}
\caption{Oblique shock reflection with shock angle $29^\circ$, inflow Mach no. 2.9 - Pressure contours (0.7:0.1:2.9), Top row) I order and II order accurate results for D2Q4 model on $120 \times 40$ grid, Middle row) I order and II order accurate results for present scheme on $120 \times 40$ grid, Bottom row) I order and II order accurate results for present scheme on $240 \times 80$ grid} 
\label{fig:2d_euler_1}
\end{figure}
\begin{figure}[h!] 
\centering
\begin{tabular}{cc}
\includegraphics[width=0.45\textwidth]{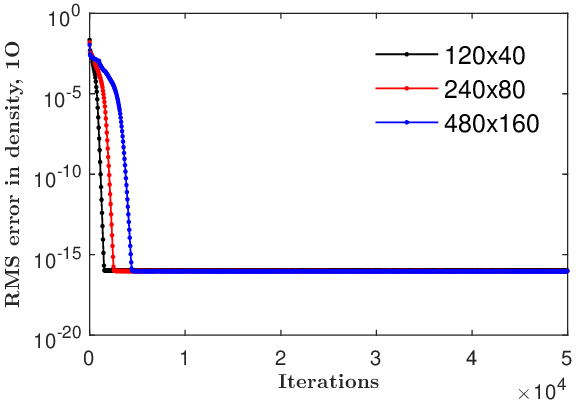} & \includegraphics[width=0.45\textwidth]{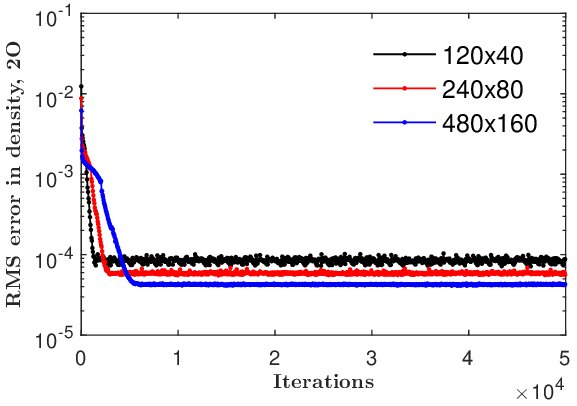} 
\end{tabular}
\caption{Oblique shock reflection test case: RMS error in density vs number of iterations for, Left) I order result, Right) II order result}
\label{fig:2d_euler_2}
\end{figure}
This test case comprises an oblique shock striking a solid wall at an incident shock angle of $29^\circ$ and reflecting from it \cite{yee1982high}. The freestream flow from left to right has a Mach no. of 2.9. The computational domain is $[0,3]\times[0,1]$. The left boundary has supersonic inflow; hence, freestream conditions are applied there. Post-shock conditions using compressible flow relations are applied at the top. Flow tangency (inviscid wall) conditions are applied at the bottom and supersonic outflow conditions are applied at the right boundary. Freestream initial conditions are used. The pressure contours for our first and second order accurate steady state results are shown in Figure \ref{fig:2d_euler_1}. For comparison, we have also added the results for the standard D2Q4 (2 dimension, 4 velocity) kinetic model, with their velocities set to the maximum wave speed in the domain. For the same grid size, the oblique shock has a sharper profile for the present scheme than for the D2Q4 model. Figure \ref{fig:2d_euler_2} shows the variation in root-mean-square (RMS) error in density with the number of time iterations for different cell sizes. The results are shown for both first and second order accuracy.

\subsubsection{Supersonic flow over a compression ramp}

\begin{figure}[h!] 
\centering
\begin{tabular}{cc}
\includegraphics[width=0.45\textwidth]{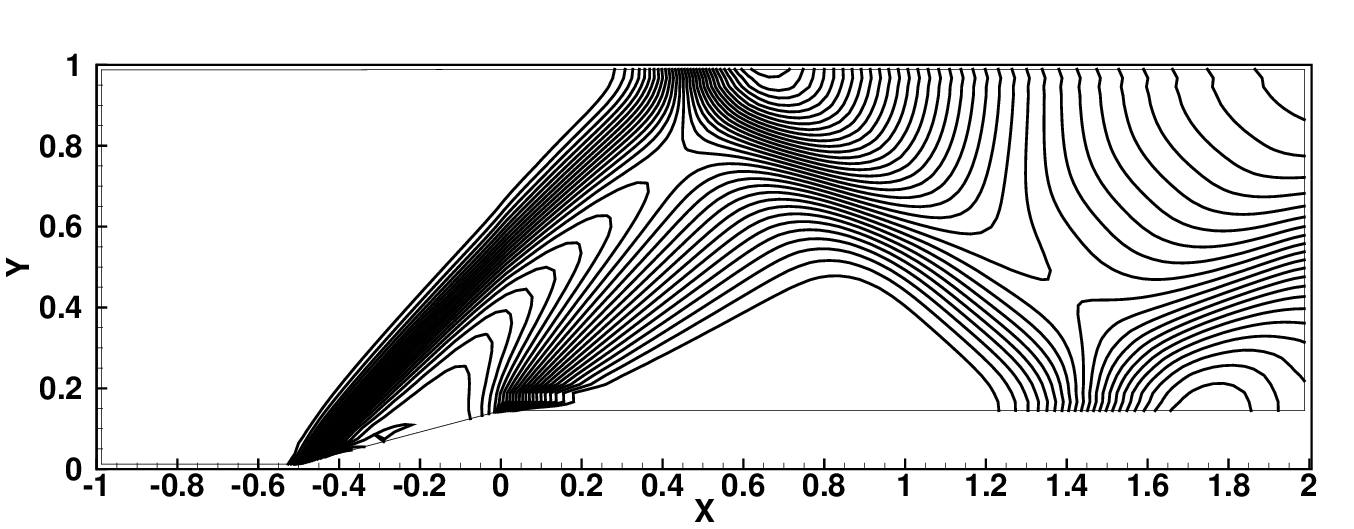} & \includegraphics[width=0.45\textwidth]{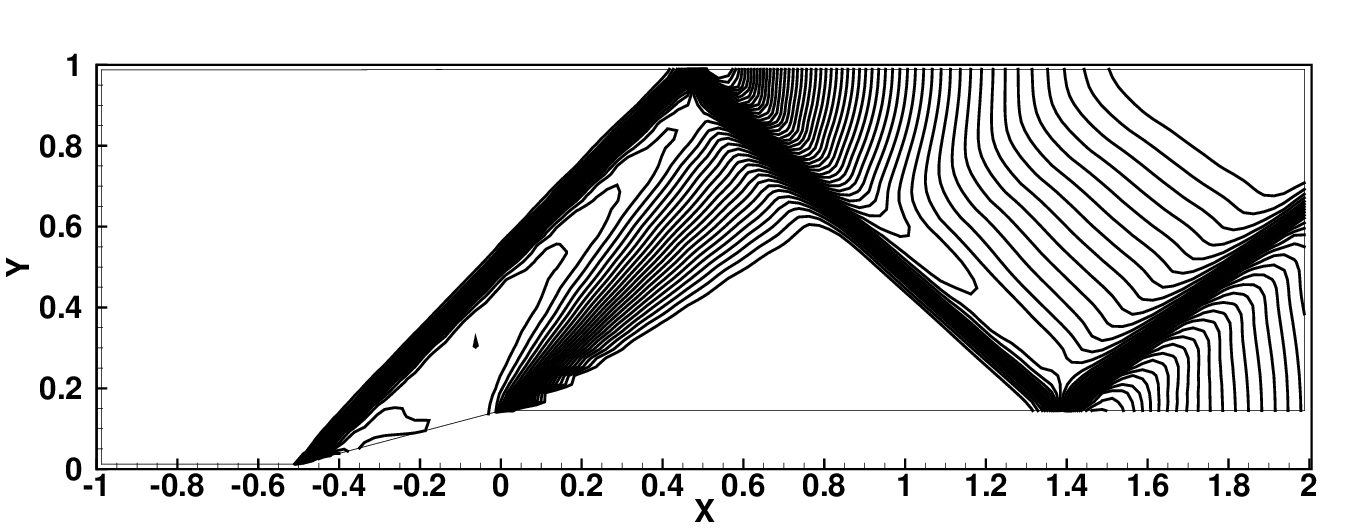}\\
\includegraphics[width=0.45\textwidth]{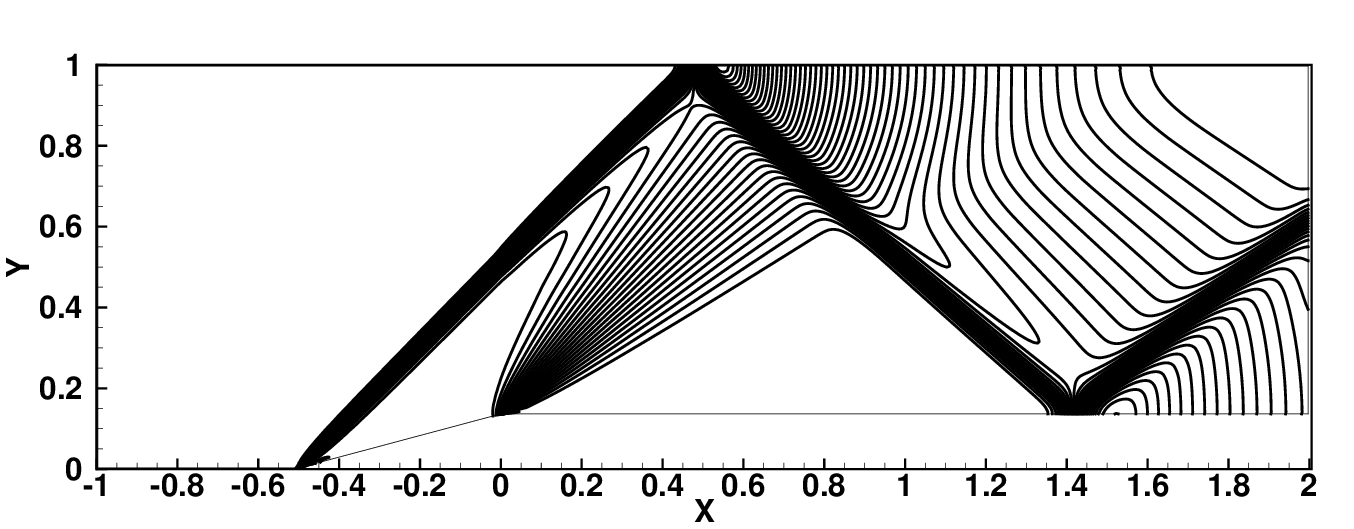} & \includegraphics[width=0.45\textwidth]{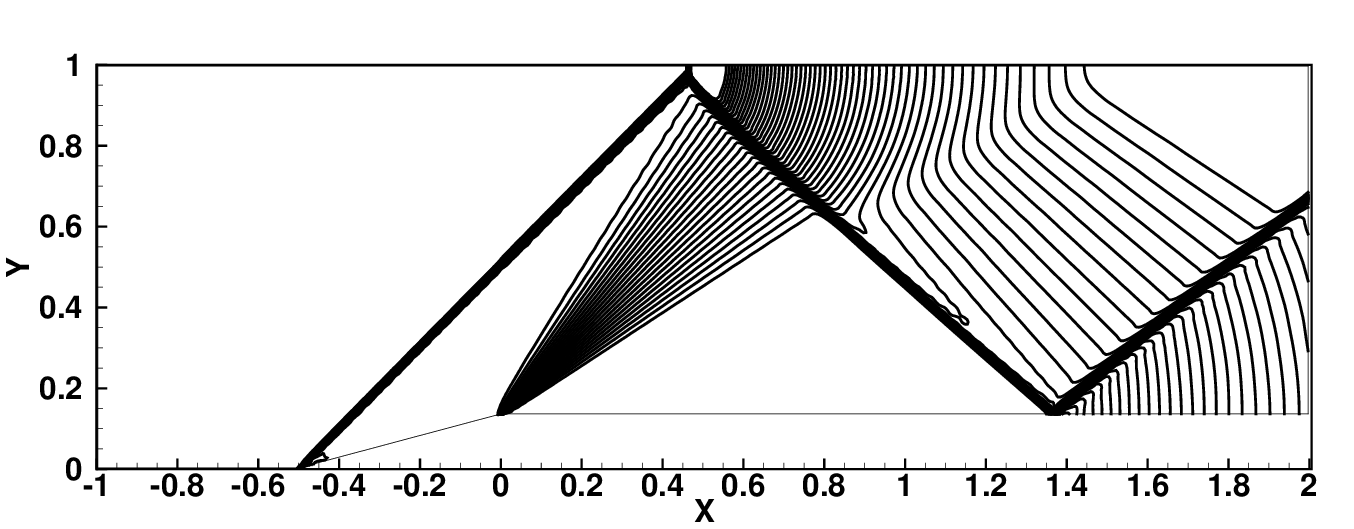} 
\end{tabular}
\caption{Mach 2 flow over a $15^\circ$ ramp - Pressure contours (1.1:0.05:3.8), Top) I order and II order accurate results on $120 \times 40$ grid, Bottom) I order and II order accurate results on $480 \times 160$ grid}
\label{fig:2d_euler_3}
\end{figure}
This test case consists of a Mach 2 flow over a $15^\circ$ compression ramp in a wind tunnel \cite{levy1993use}. The computational domain is $[1,2]\times[0,1]$ with a $15^\circ$ ramp at the bottom from x= -0.5 to x= 0. The applied boundary conditions are: supersonic inflow at the left boundary, flow tangency conditions at the top and bottom walls, and supersonic outflow conditions at the right boundary. Freestream initial conditions are used. The steady state results are shown in Figure \ref{fig:2d_euler_3}. The solution comprises an oblique shock originating at the concave corner and expansion fans starting from the convex corner. The oblique shock strikes and reflects from the top and bottom walls while also interacting with the expansion fan. As our contours show, no entropy-violating expansion shocks are formed.

\subsubsection{Grid-aligned oblique shock on a non-Cartesian grid}

\begin{figure}[h!] 
\centering
\begin{tabular}{cc}
\includegraphics[width=0.45\textwidth]{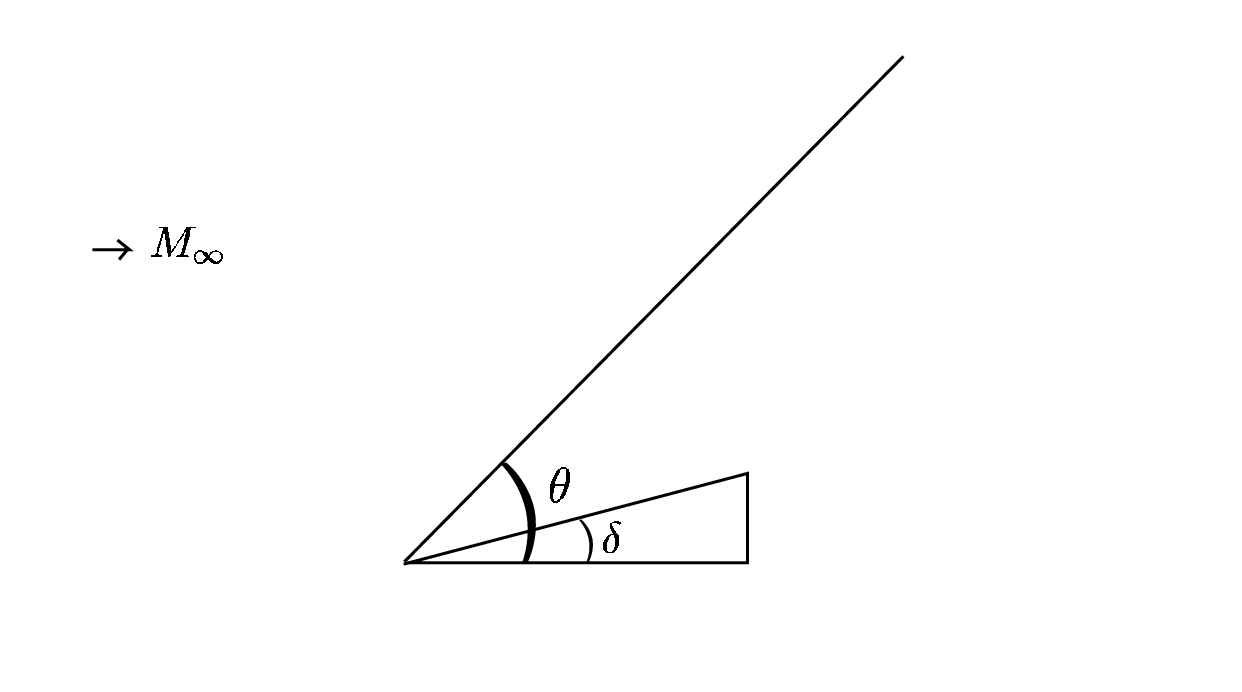} & \includegraphics[width=0.45\textwidth]{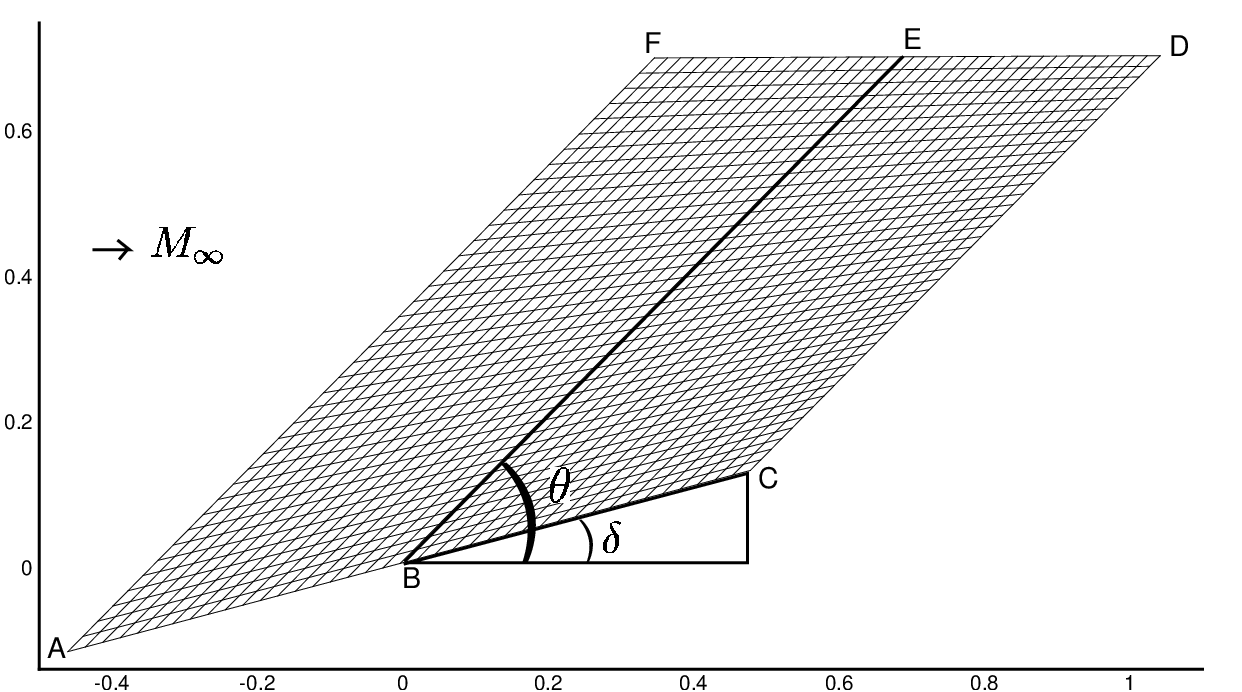}
\end{tabular}
\caption{Mach 2 flow over a $15^\circ$ wedge, Left) Setup showing the wedge and the oblique shock, Right) Computational domain and mesh}
\label{fig:2d_euler_3a}
\end{figure}

\begin{figure}[h!] 
\centering
\begin{tabular}{cc}
\includegraphics[width=0.45\textwidth]{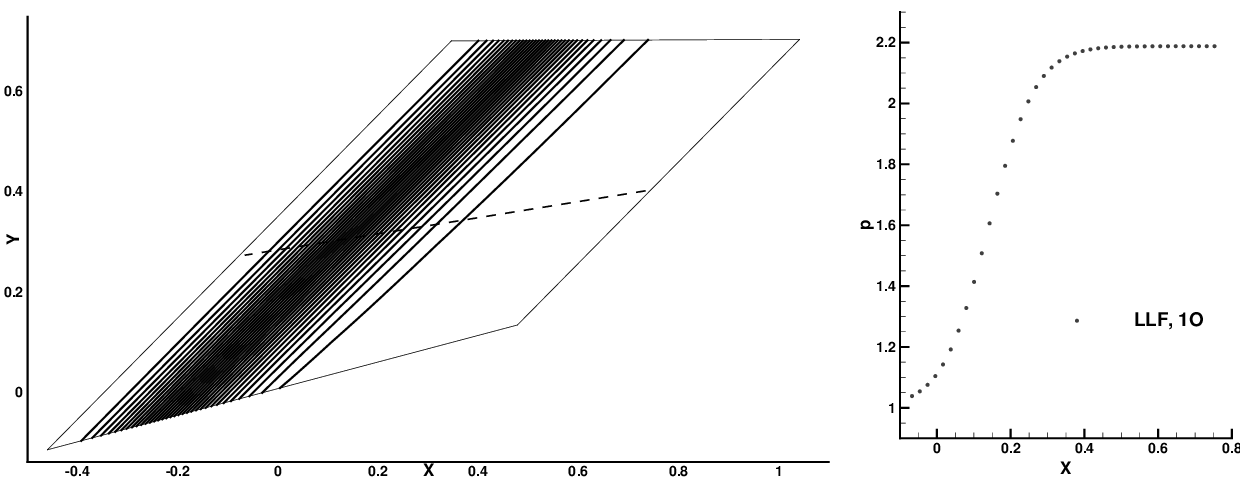} & \includegraphics[width=0.45\textwidth]{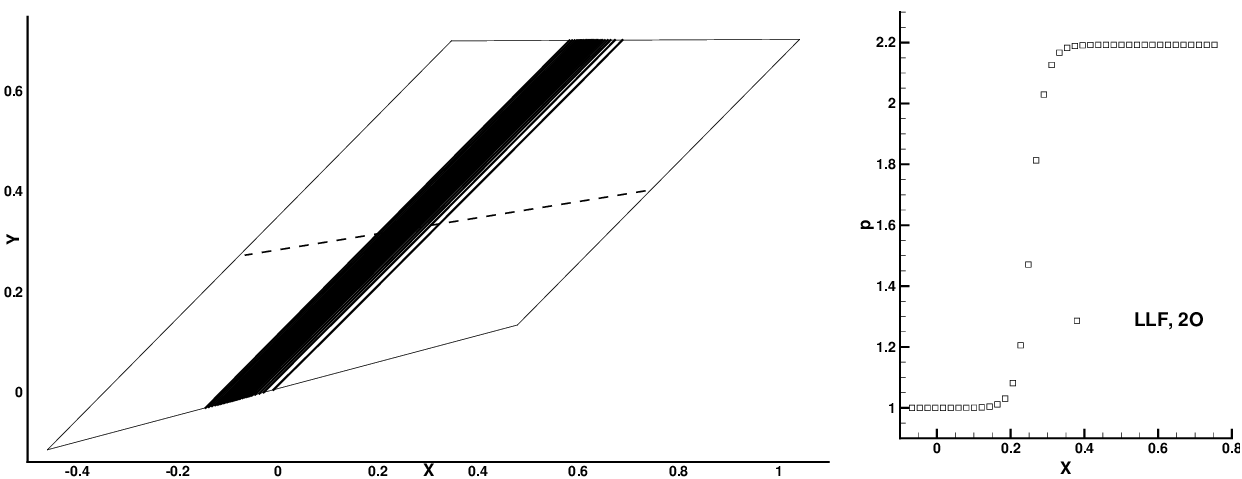}\\
\includegraphics[width=0.45\textwidth]{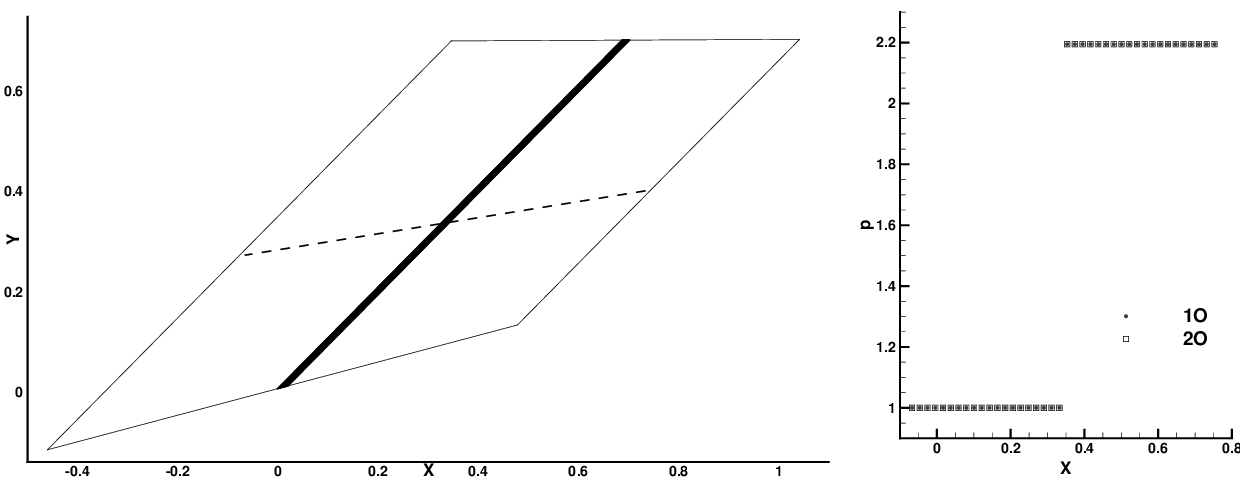} & \includegraphics[width=0.45\textwidth]{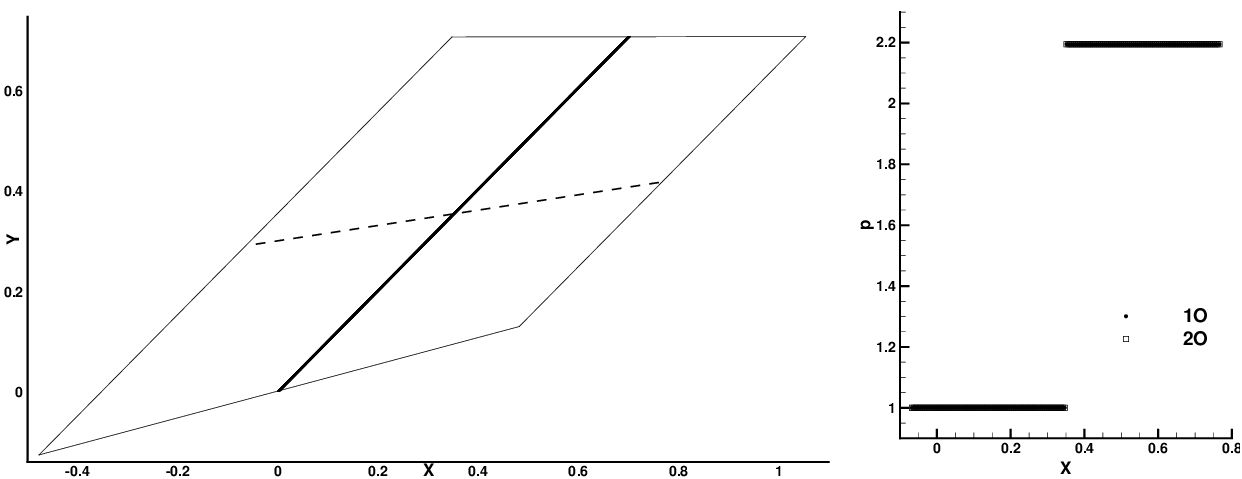}
\end{tabular}
\caption{Pressure contours (1.1:0.0333:2.2) and pressure variation along the dashed ($j$= Ny/2) line for I order and II order accuracy: Top) LLF scheme on a $40 \times 40$ grid, Bottom) Present scheme on  $40 \times 40$ and $160 \times 160$ grids.}%
\label{fig:2d_euler_3b}%
\end{figure}
This test case has been devised to test the ability of a numerical scheme to capture a grid-aligned steady shock on a non-Cartesian grid. The flow conditions have been taken from the previous test case, with a freestream Mach 2 flow from left to right across a flat surface inclined at a wedge angle $\delta= 15^\circ$ with the flow. For $\gamma$= 1.4 (assumed throughout this work) and $\delta= 15^\circ$, the oblique shock relations give us a shock wave angle of $\theta= 45.3436^\circ$. The computational domain ABCDEF along with the mesh is shown in Figure \ref{fig:2d_euler_3a}. The boundary AC is aligned at a wedge angle $\delta$ with the incoming freestream flow, with the wall starting at B ((0,0)) and ending at C. The boundaries AF and CD, as well as all lines between, are aligned at the shock wave angle $\theta$ with the incoming flow. Finally, FED is parallel to the freestream flow. The domain is initialized with freestream conditions. Supersonic inflow conditions are applied at AF, and supersonic outflow conditions are applied at the boundaries AB and CD. Flow tangency (inviscid wall) conditions are applied at BC. Through numerical experiments, we have observed that constant extrapolation at the boundary FED is effective in preventing any deformation of the shock near the top. Thus, we have applied constant extrapolation at the boundary FD. The steady state results for this problem are shown in Figure \ref{fig:2d_euler_3b}. We observe that the more diffusive Local Lax-Friedrichs (LLF) scheme diffuses the shock over many cells. Our scheme, on the other hand, captures the grid-aligned oblique shock exactly for first as well as second order accuracy.

\subsubsection{Hypersonic flow over a half-cylinder}
\begin{figure}
\centering
\begin{tabular}{cccc}
a)\includegraphics[width=0.2\textwidth]{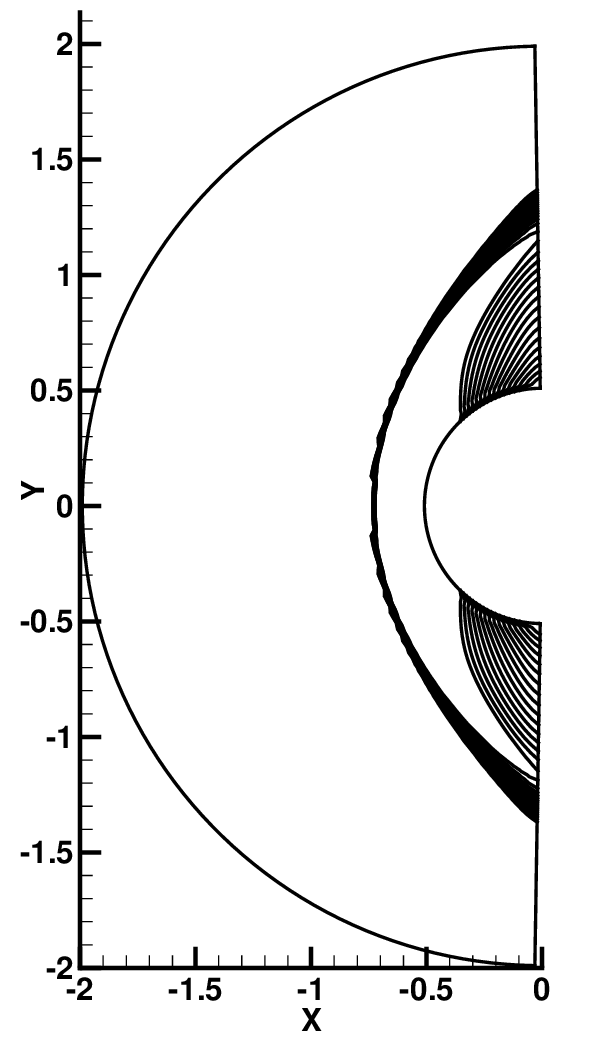} & b)\includegraphics[width=0.2\textwidth]{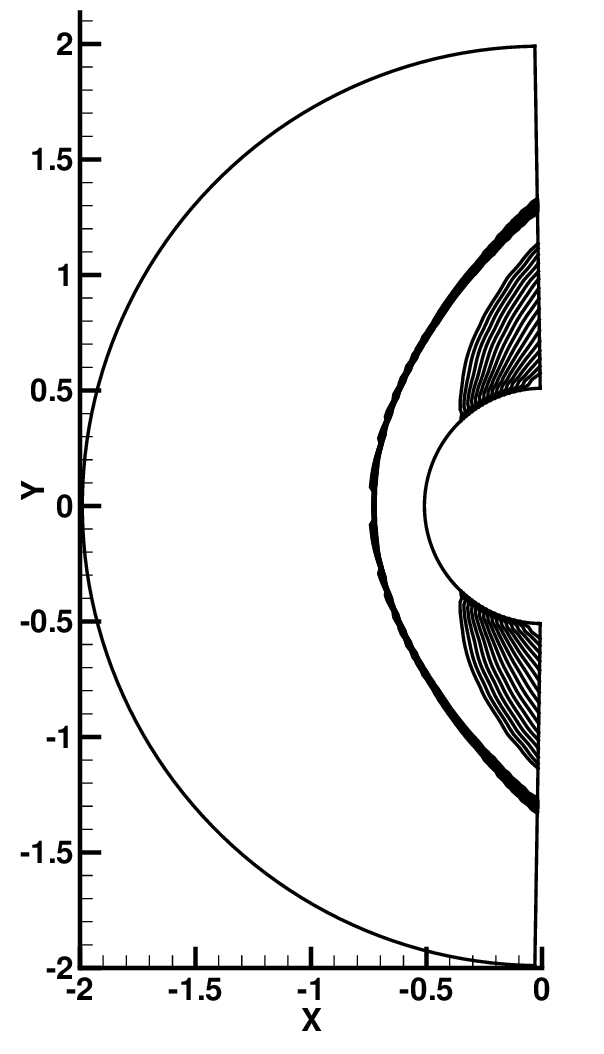} & c)\includegraphics[width=0.2\textwidth]{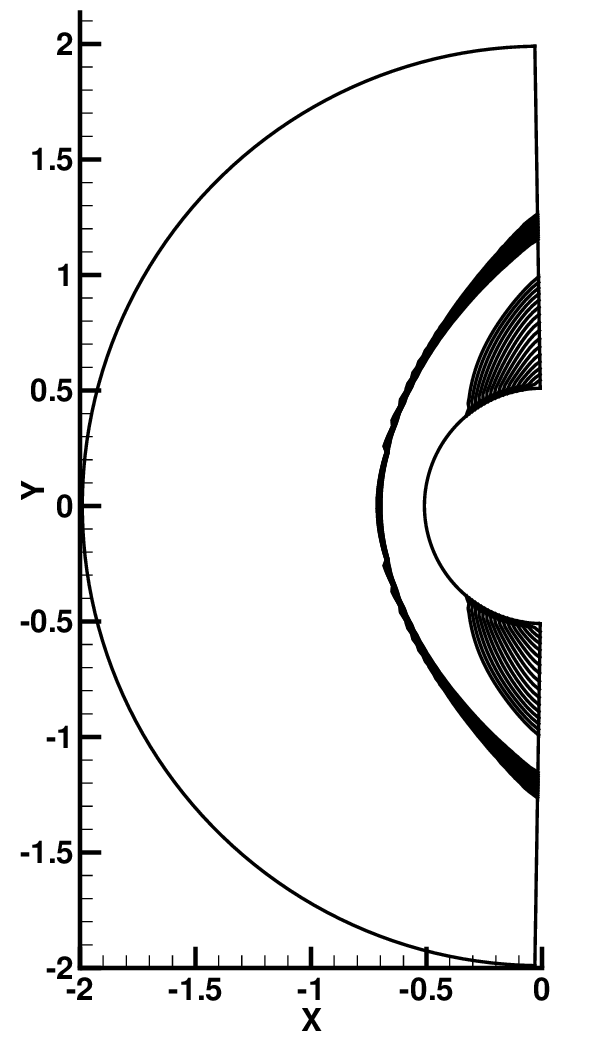} & d)\includegraphics[width=0.2\textwidth]{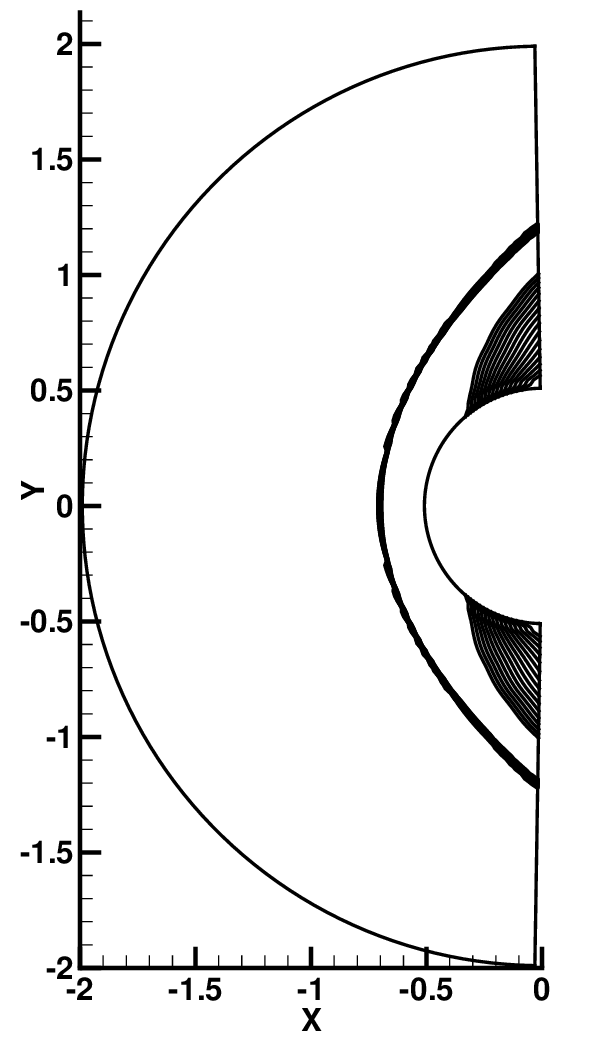}
\end{tabular}
\caption{Hypersonic flow past a half-cylinder, density contours (2:0.2:5) on $100 \times 80$ grid.(a) Mach 6 flow, I order (b) Mach 6 flow, II order, (c) Mach 20 flow, I order (d) Mach 20 flow, II order}
\label{fig:2d_euler_4}
\end{figure}
This test case assesses a scheme for a form of numerical shock instability called the carbuncle phenomenon. For this problem, we consider freestream flow at Mach 6 and Mach 20 over a half-cylinder. The computational domain taken is $(r,\theta) \in [0.5,2]\times[\frac{\pi}{2},\frac{3\pi}{2}]$, with constant grid spacing along $r$ and $\theta$ directions. Supersonic inflow conditions are applied at $r=2$ and flow tangency conditions are applied at $r=0.5$. Supersonic outflow conditions are applied at $\theta= \frac{\pi}{2}$ and $\theta= \frac{3\pi}{2}$ boundaries. Freestream initial conditions are used. For this test case, the steady state solution consists of a bow shock formed in front of and detached from the half-cylinder. Many numerical schemes, especially Riemann solvers, such as the Roe scheme, develop an unusual feature called a carbuncle shock, with the bow shock breaking on the stagnation line \cite{quirk1997contribution}. These carbuncle shocks are often considered spurious numerical artifacts, and significant research effort is spent to obtain carbuncle-free solutions. Our results are shown in Figure \ref{fig:2d_euler_4} and no carbuncle shock is observed in our solution.  

\subsubsection{Supersonic flow over a forward-facing step}
\begin{figure}[h!] 
\centering
\begin{tabular}{cc}
\includegraphics[width=0.45\textwidth]{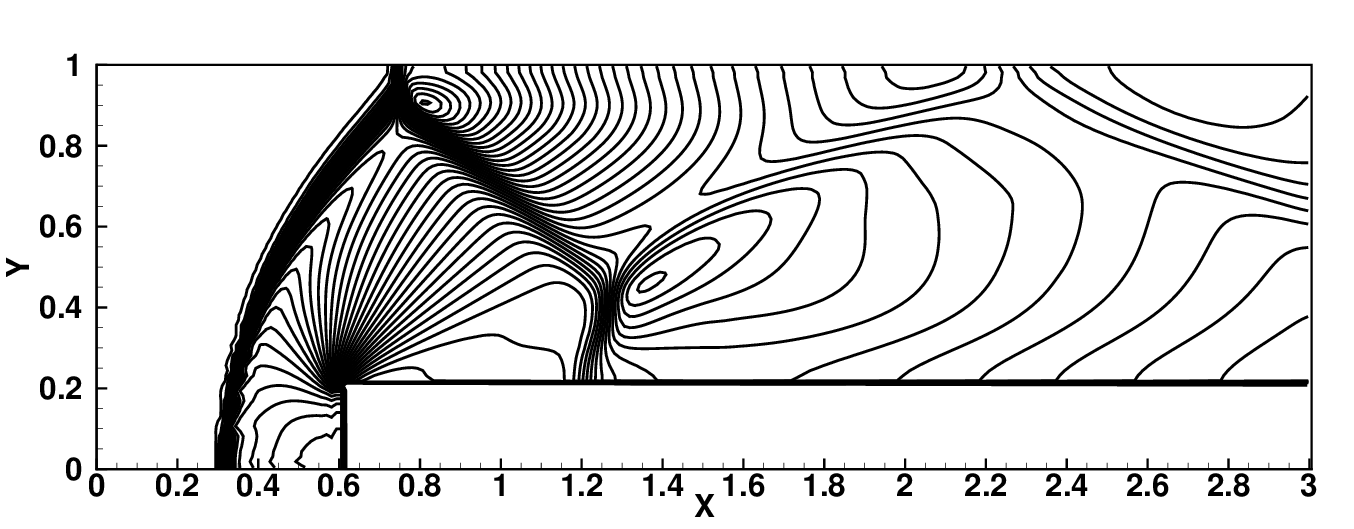} & \includegraphics[width=0.45\textwidth]{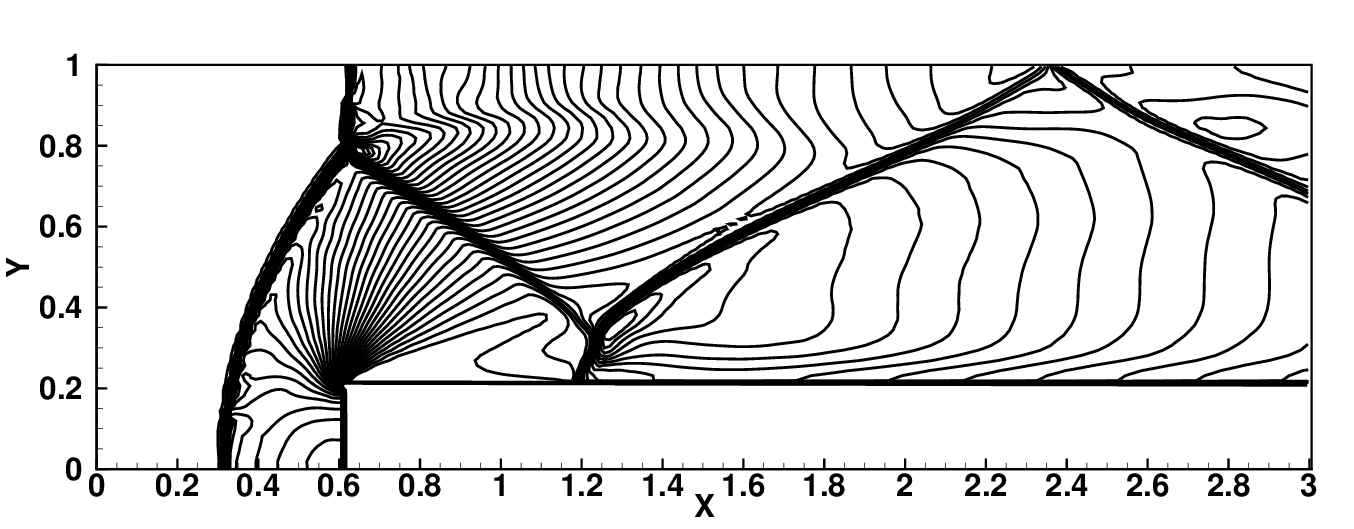}\\
\includegraphics[width=0.45\textwidth]{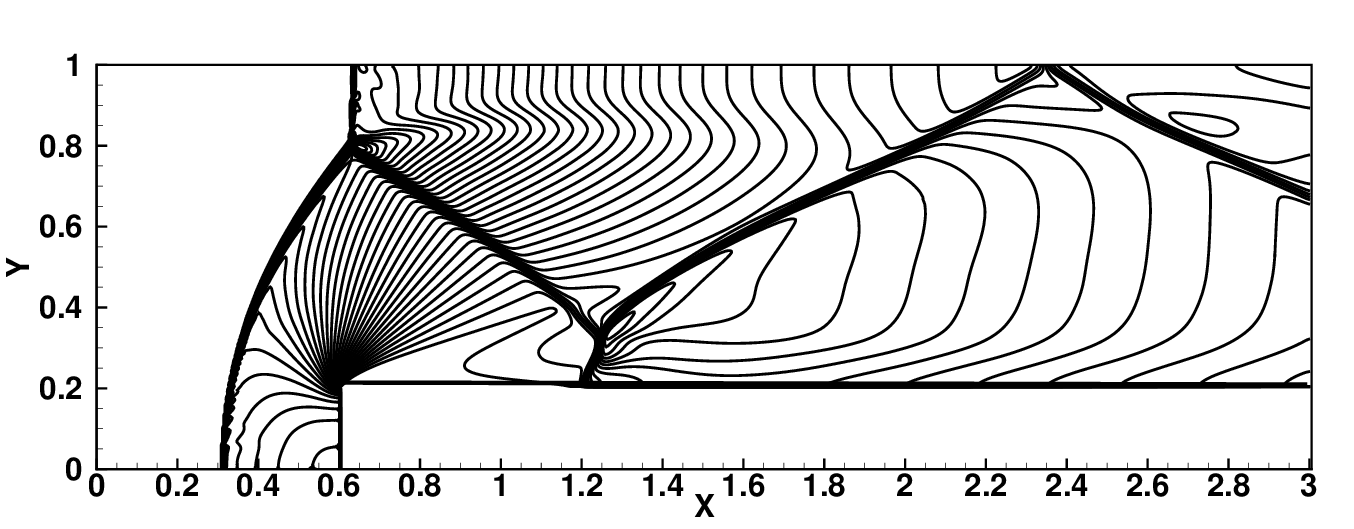} & \includegraphics[width=0.45\textwidth]{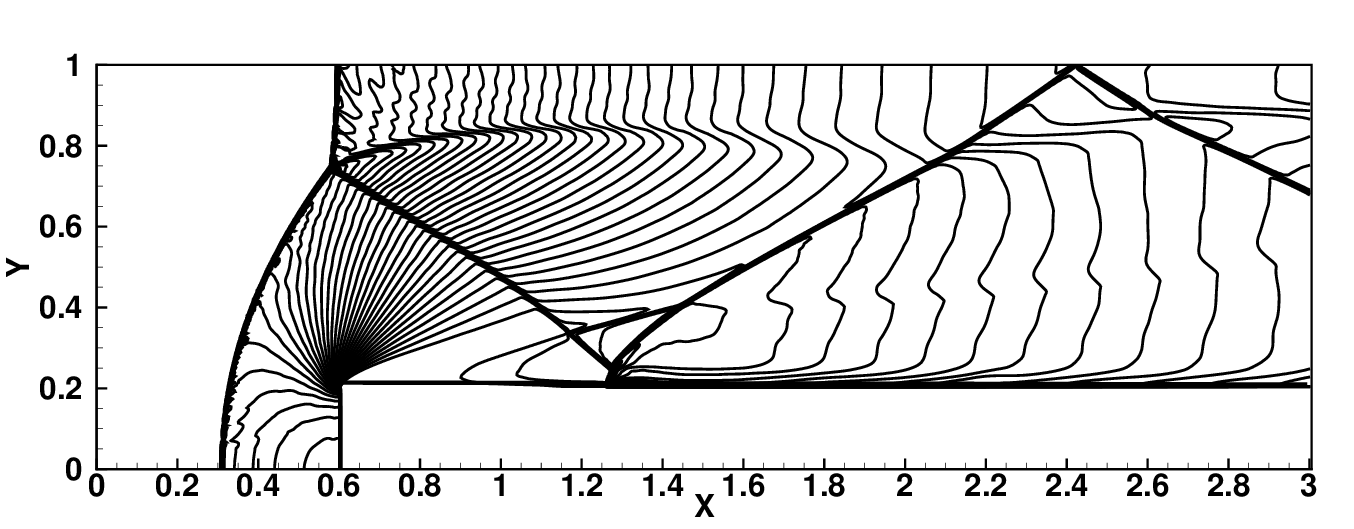} 
\end{tabular}
\caption{Mach 3 flow over a forward-facing step in wind tunnel, $t=4$, density contours (1:0.15:6.5); Top) I order and II order accurate results on $240 \times 80$ grid, Bottom) I order and II order accurate results on $960 \times 320$ grid}
\label{fig:2d_euler_5}
\end{figure}
For this unsteady problem, we consider a Mach 3 flow over a forward-facing step in a wind tunnel \cite{woodward1984numerical}. The dimensions of the wind tunnel are $[0,3]\times[0,1]$. The step is 0.2 unit high and is located at the bottom at a distance of 0.6 unit from the left end. The boundary conditions applied are: supersonic inflow at the left boundary, flow tangency conditions at the top and bottom walls (including the step), and supersonic outflow conditions at the right boundary. Freestream initial conditions are used. A lambda shock formed near the top boundary is visible at time $t=4$. A slip stream can be seen beyond the triple point, which can be captured well only by low diffusive schemes. The first and second order results for this test case are shown in Figure \ref{fig:2d_euler_5}.  

\subsubsection{Odd-even decoupling}
\begin{figure}[h!] 
\centering
\begin{tabular}{cc}
\includegraphics[width=0.5\textwidth]{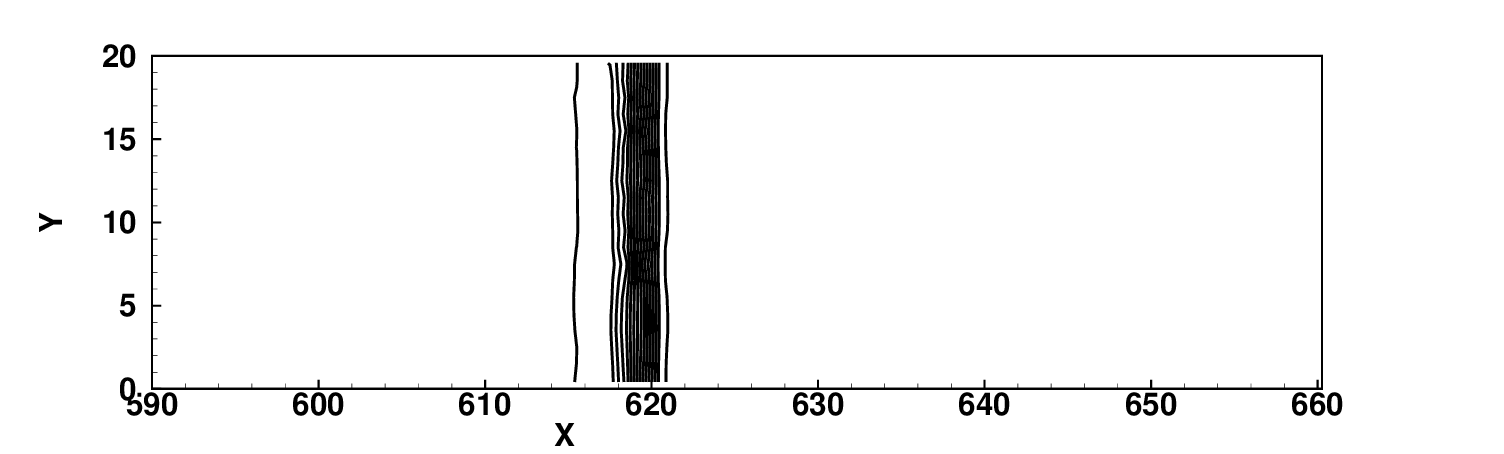} & \includegraphics[width=0.5\textwidth]{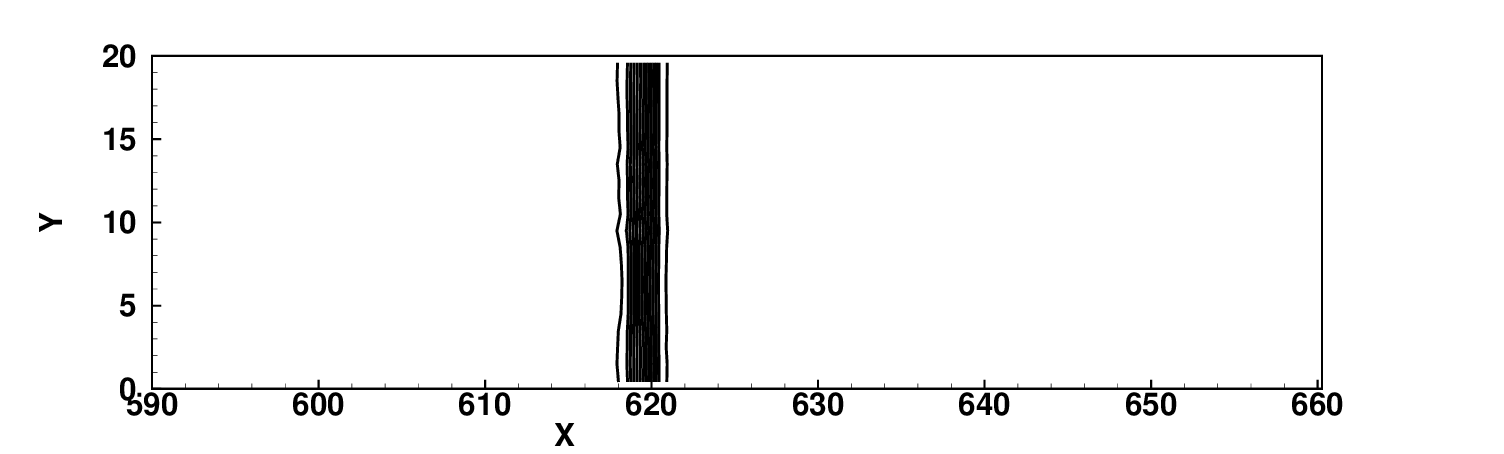} 
\end{tabular}
\caption{Mach 6 shock wave through a rectangular duct at $t=100$, $800 \times 20$ grid, density contours; Left) I order, Right) II order}
\label{fig:2d_euler_6}
\end{figure}
This test case consists of a planar Mach 6 shock that propagates through a stationary medium in a rectangular duct \cite{quirk1997contribution}. A Cartesian mesh of $800 \times 20$ square cells is used. The grid is perturbed along the center-line as follows.
\begin{equation}
(y_{i,j})_{mid}= \left\{\begin{array}{c}
(y_{i,j})_{mid}+ 10^{-3}, \ \text{if} \ i \ \text{is even}\\
(y_{i,j})_{mid}- 10^{-3}, \ \text{if}\ i \ \text{is odd} 
\end{array}\right\}
\label{eq:2d_euler_tc6}
\end{equation}
The right traveling shock is located at the interface between $20^{th}$ and $21^{st}$ cell at the initial time. The solution is sought at time $t=100$. For this problem, low-diffusive schemes like the Roe scheme develop oscillations due to a type of numerical instability called odd-even decoupling, which then destroys the solution. As our results in Figure \ref{fig:2d_euler_6} show, our scheme is free from this form of instability.

\subsubsection{Double Mach reflection}
\begin{figure}[h!] 
\centering
\begin{tabular}{cc}
\includegraphics[width=0.3\textwidth]{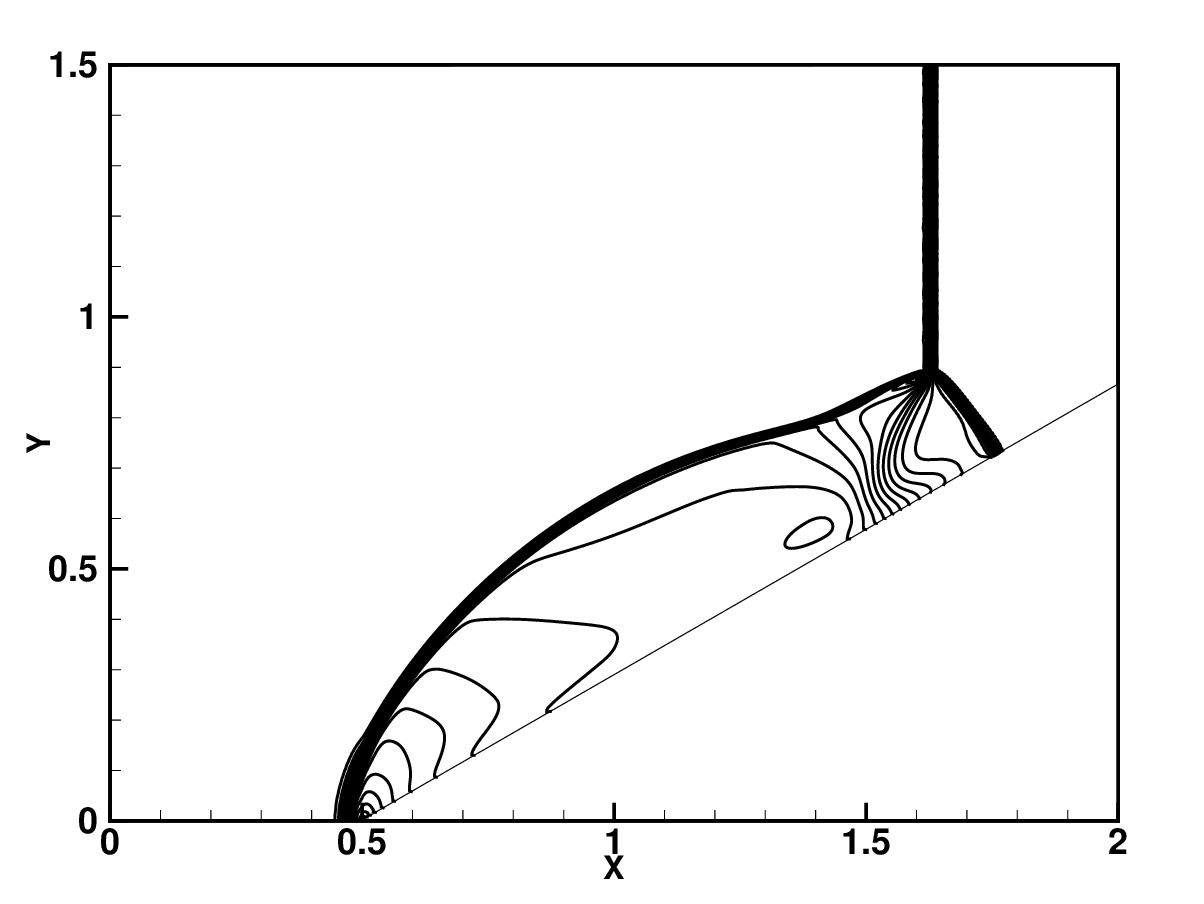} & \includegraphics[width=0.3\textwidth]{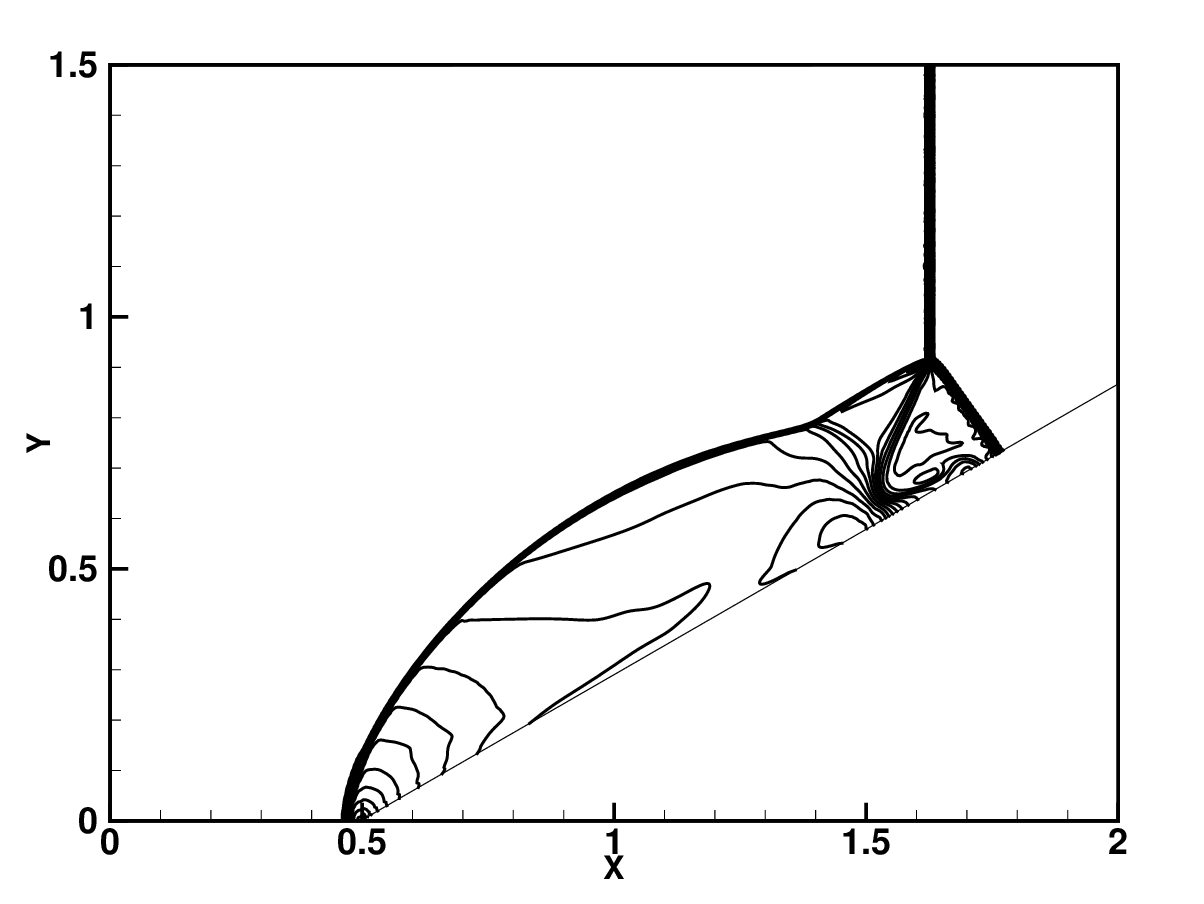}\\
\includegraphics[width=0.3\textwidth]{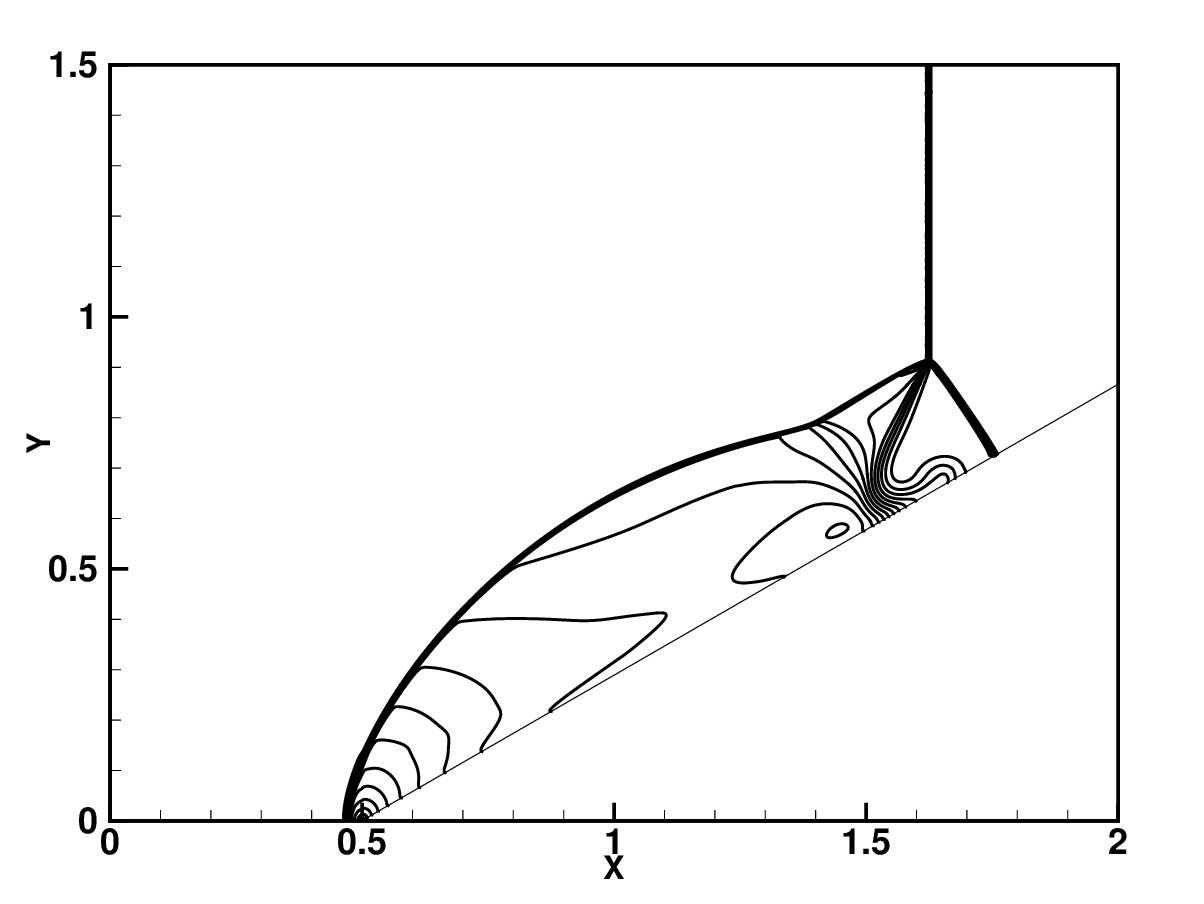} & \includegraphics[width=0.3\textwidth]{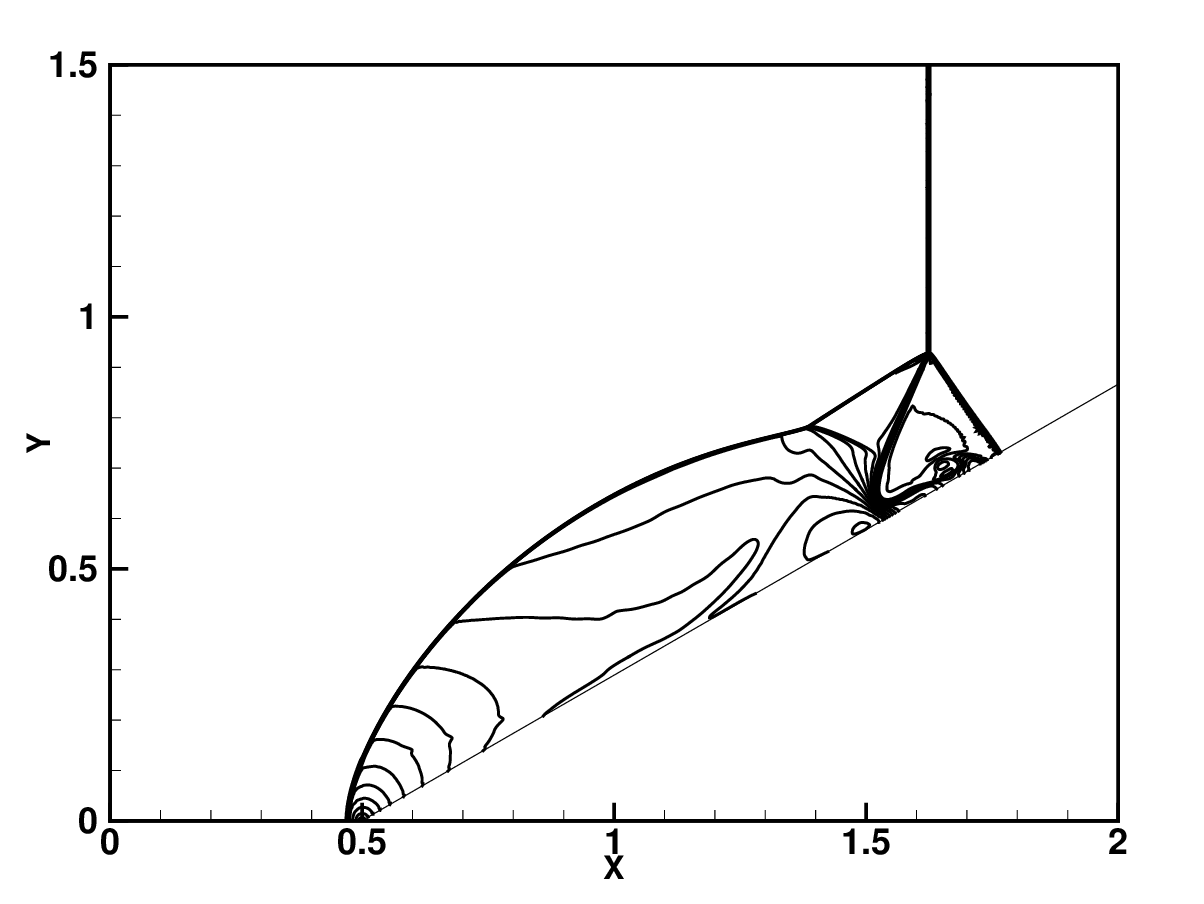} 
\end{tabular}
\caption{Double Mach reflection for Mach 5.5 shock across a $30^\circ$ wedge, $t=0.25$, density contours (1.5:0.5:19); Top) I order and II order accurate results on $400 \times 400$ grid, Bottom) I order and II order accurate results on $1200 \times 1200$ grid}
\label{fig:2d_euler_7}
\end{figure}
In this test case, a planar Mach 5.5 shock travels across a $30^\circ$ wedge \cite{quirk1997contribution}. The computational domain taken is $[0,2] \times [0,1.5]$ with a $30^\circ$ wedge at the bottom starting at $x= 0.5$. At the initial time, the right-traveling shock is placed at x= 0.25, with stationary medium to its right. Moving shock relations are used to determine the initial flow state to the left. The following boundary conditions are applied: initial left flow state at the left boundary, flow tangency conditions at the top and bottom boundaries, and constant extrapolation at the right boundary. When the moving shock collides with the wedge, it reflects over the surface as Mach reflection. The wave configuration consists of the initial shock, the reflected shock, one Mach stem, and one slip stream, all of which meet at a single triple point. Figure \ref{fig:2d_euler_7} shows the numerical solution of this unsteady problem at $t=0.25$. Some low-diffusive schemes produce an unphysical kinked Mach stem. No such kinked Mach stem is seen in our results.

\subsubsection{Shock diffraction}
\begin{figure}[h!] 
\centering
\begin{tabular}{cc}
\includegraphics[width=0.3\textwidth]{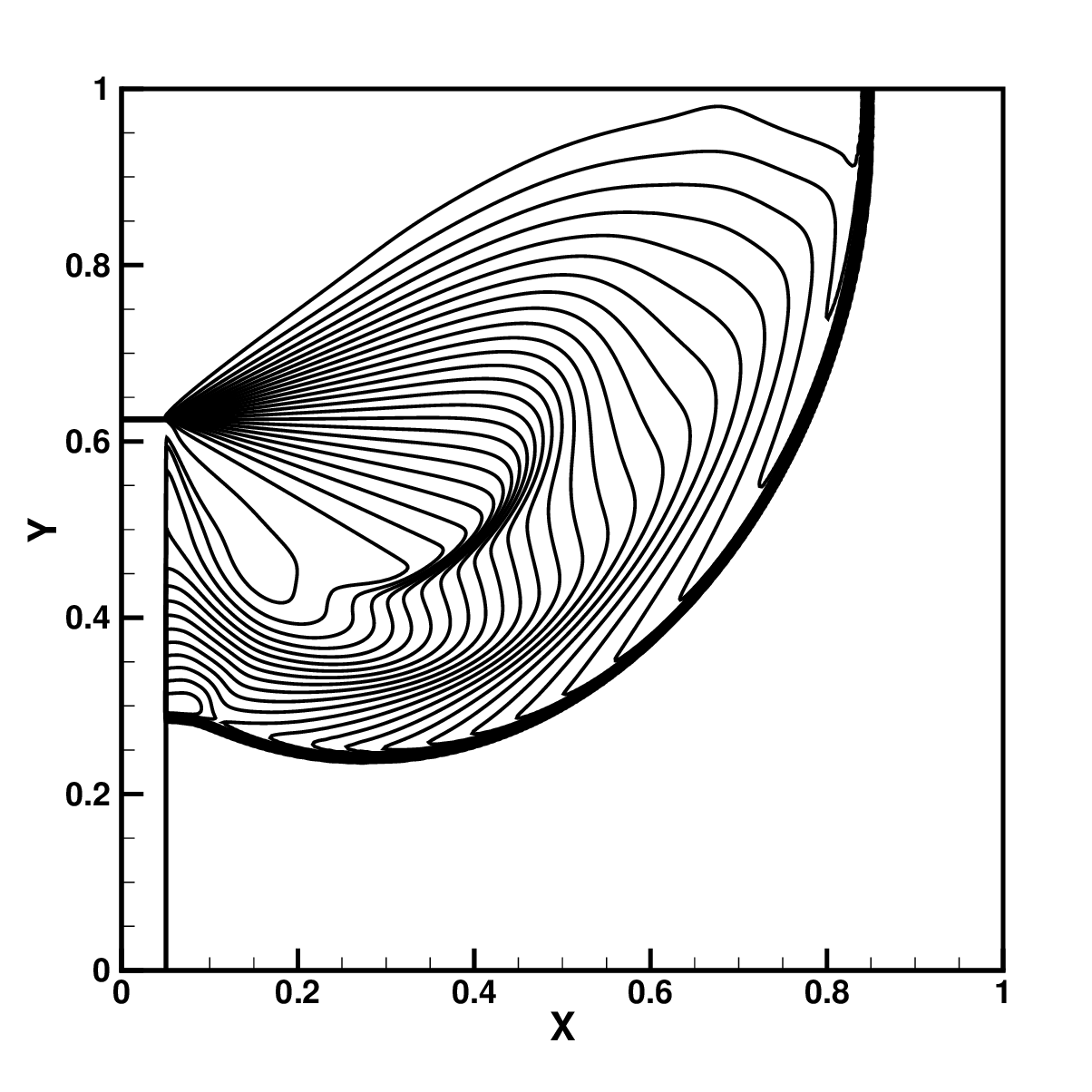} & \includegraphics[width=0.3\textwidth]{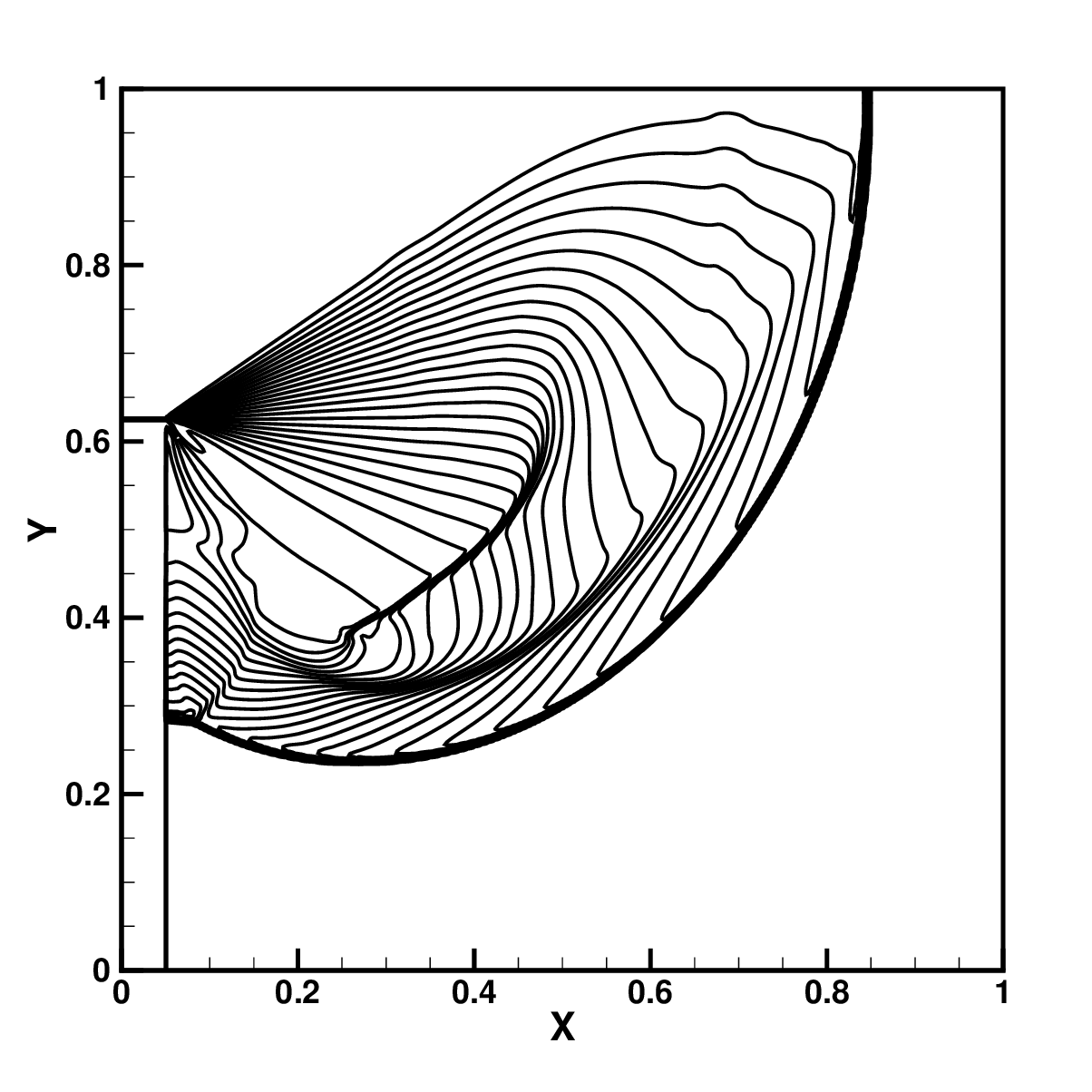}\\
\includegraphics[width=0.3\textwidth]{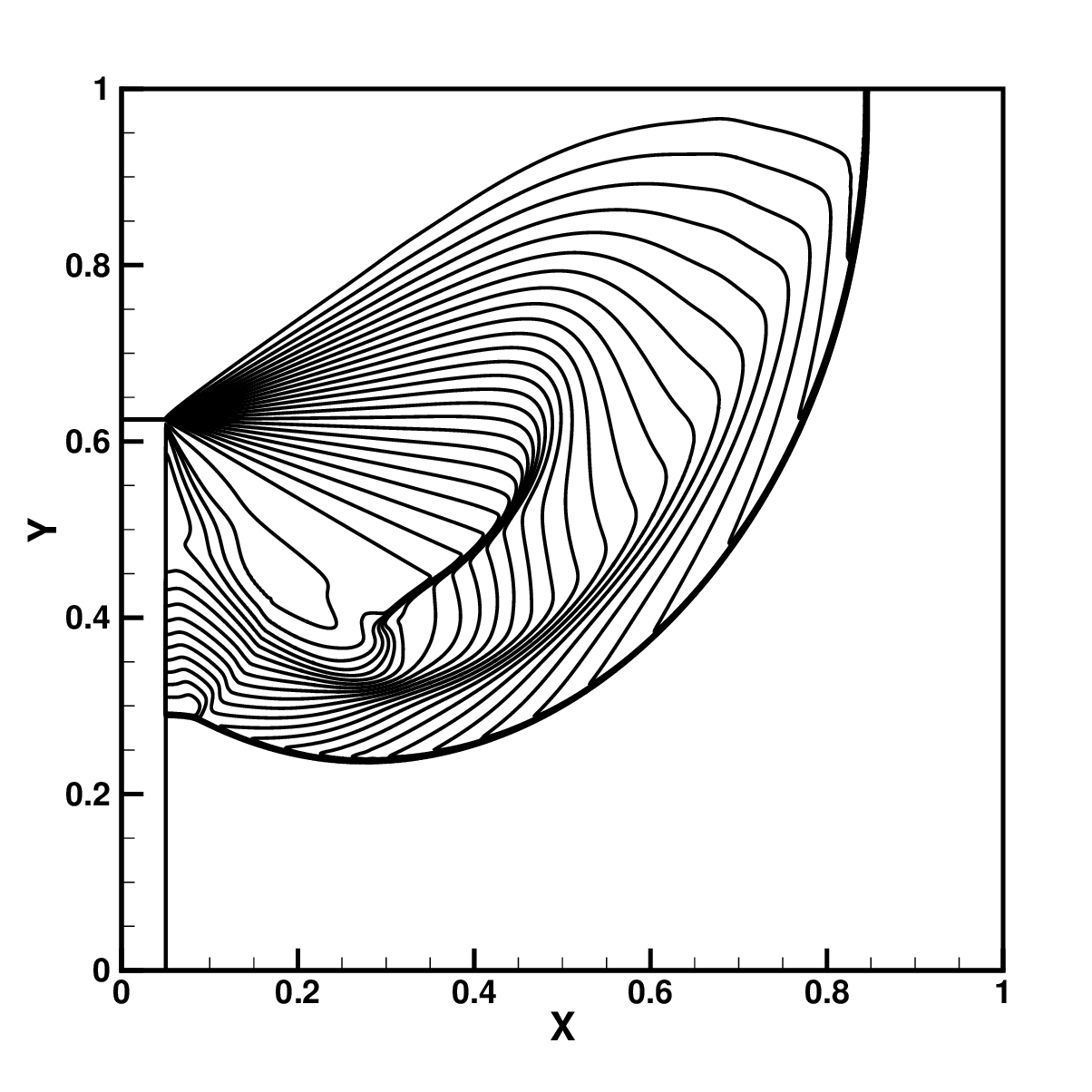} & \includegraphics[width=0.3\textwidth]{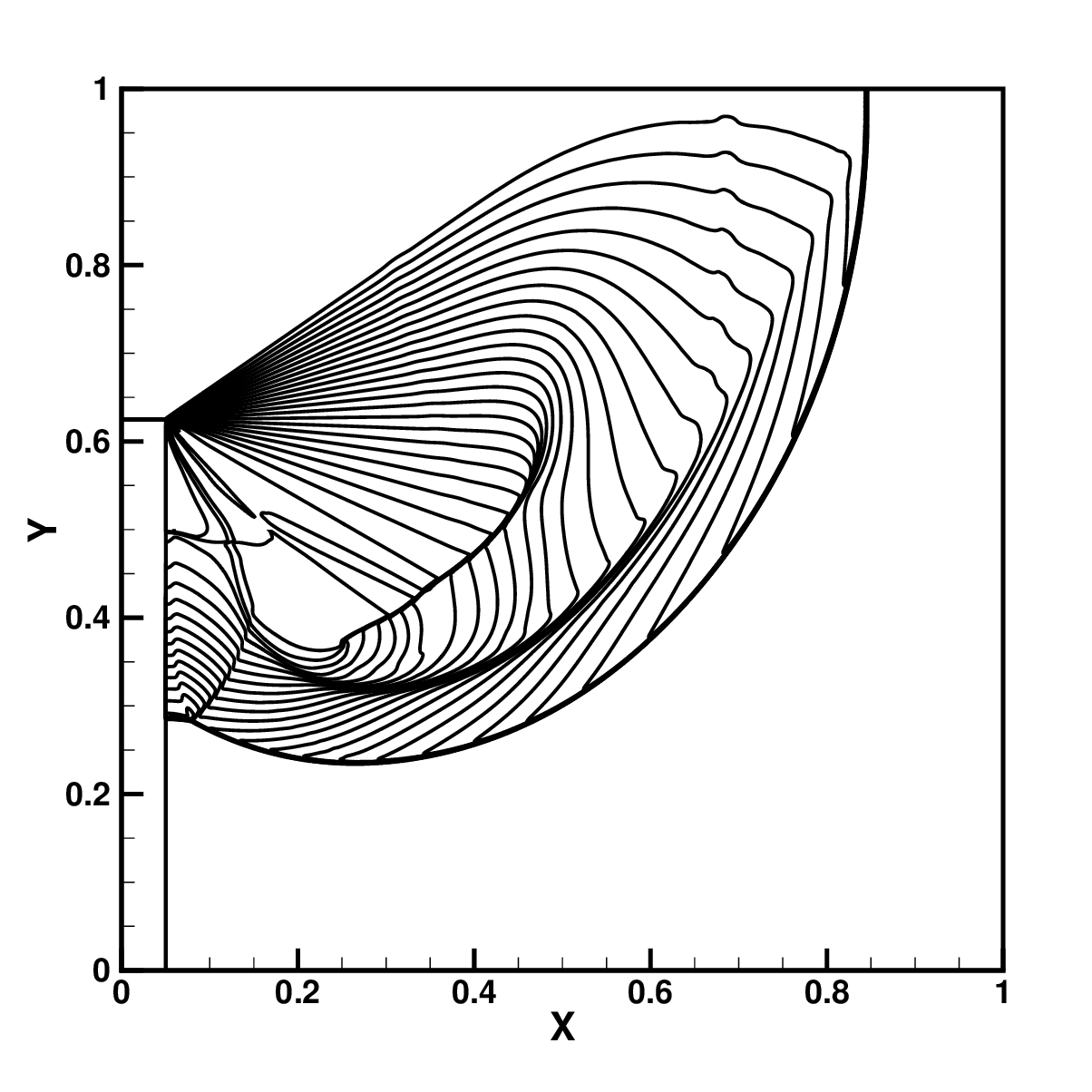} 
\end{tabular}
\caption{Shock diffracting around a $90^\circ$ corner, $t=0.1561$, density contours (0.5:0.25:6.75); Top) I order and II order accurate results on $400 \times 400$ grid, Bottom) I order and II order accurate results on $1200 \times 1200$ grid}
\label{fig:2d_euler_8}
\end{figure}
In this test case, a planar Mach 5.09 shock diffracts around a $90^\circ$ corner \cite{quirk1997contribution}. The computational domain taken is $[0,1] \times [0,1]$, with a corner at the bottom left end of width 0.05 unit and height 0.625 unit respectively. Initially, the planar shock is located at x= 0.05, traveling rightward into stationary medium. The boundary conditions are as follows: initial left flow state at the left boundary, flow tangency conditions at the top and for the corner, and constant extrapolation at the right and bottom boundaries. Figure \ref{fig:2d_euler_8} shows our numerical solution for this unsteady problem at time t= 0.1561. The solution has a complex wave structure that consists of the incident planar shock, the diffracted shock, a strong expansion fan, and a slip stream. Without an entropy fix, several low-diffusive schemes give rise to unphysical expansion shocks. Some Riemann solvers also produce oscillations near the location where the shock hits the top boundary. Our results are free of expansion shocks and oscillations. All flow features are captured well.  

\subsubsection{Positivity test cases}
\begin{figure}[h!] 
\centering
\begin{tabular}{cc}
\includegraphics[width=0.45\textwidth]{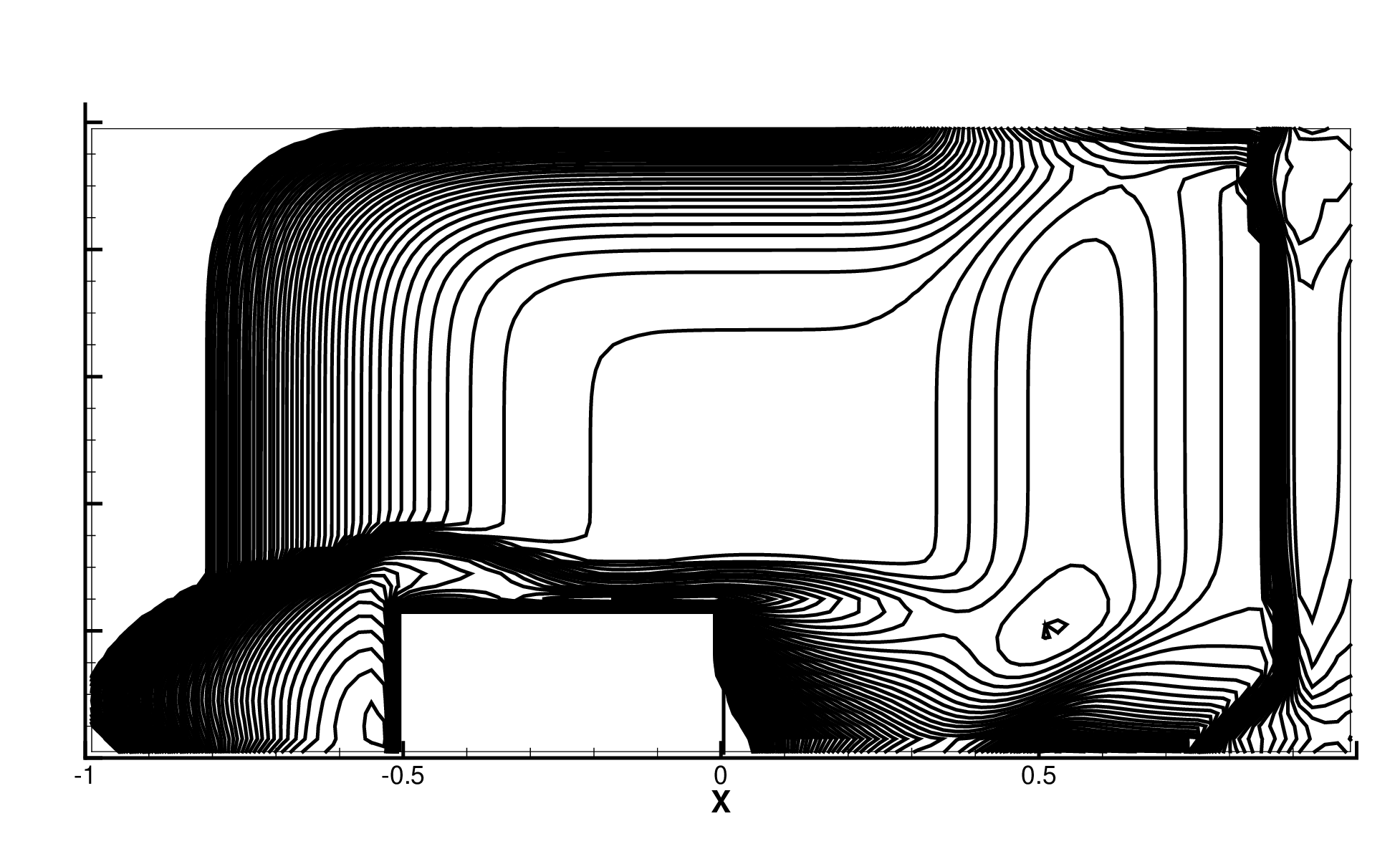} & \includegraphics[width=0.45\textwidth]{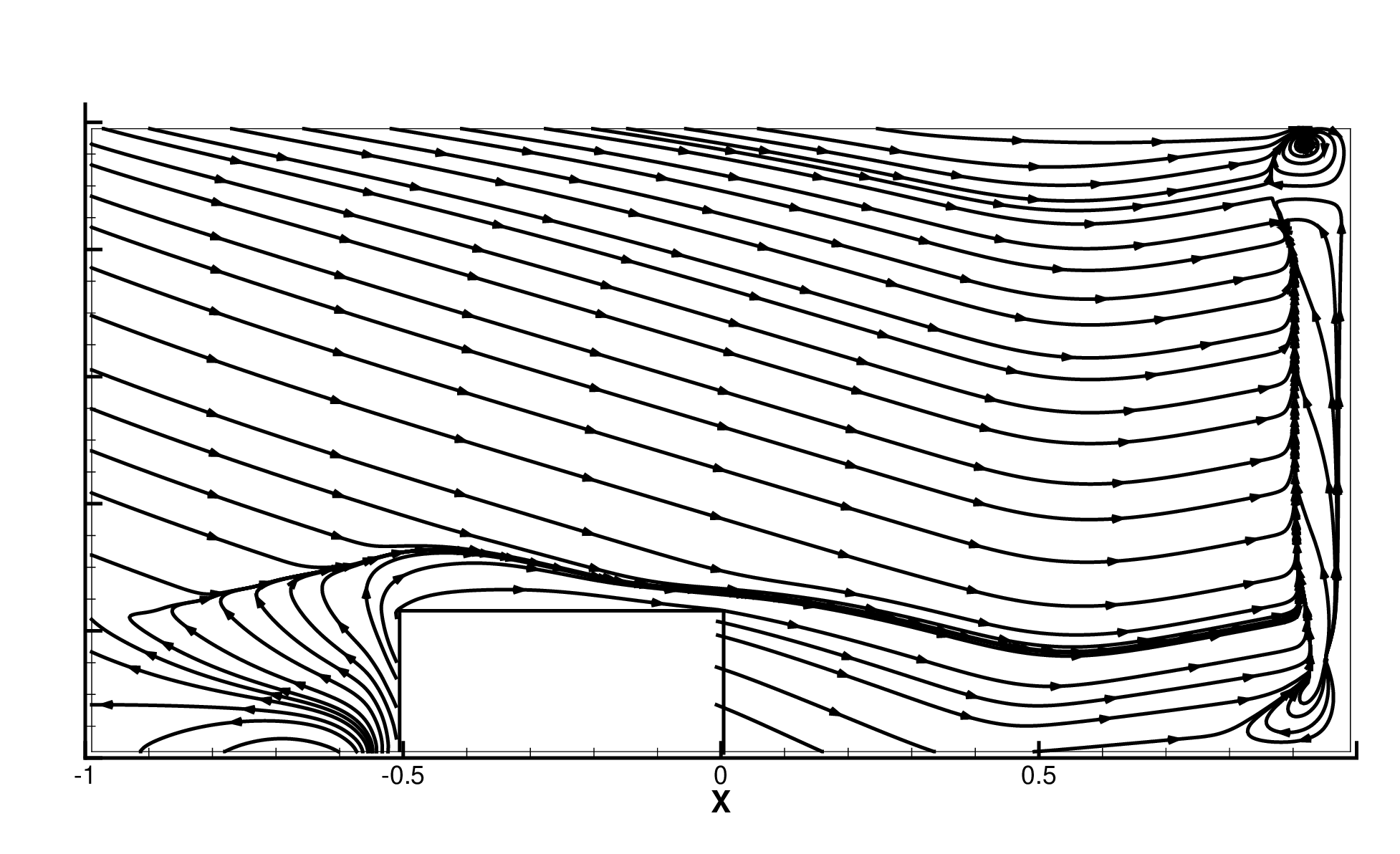}\\
\includegraphics[width=0.45\textwidth]{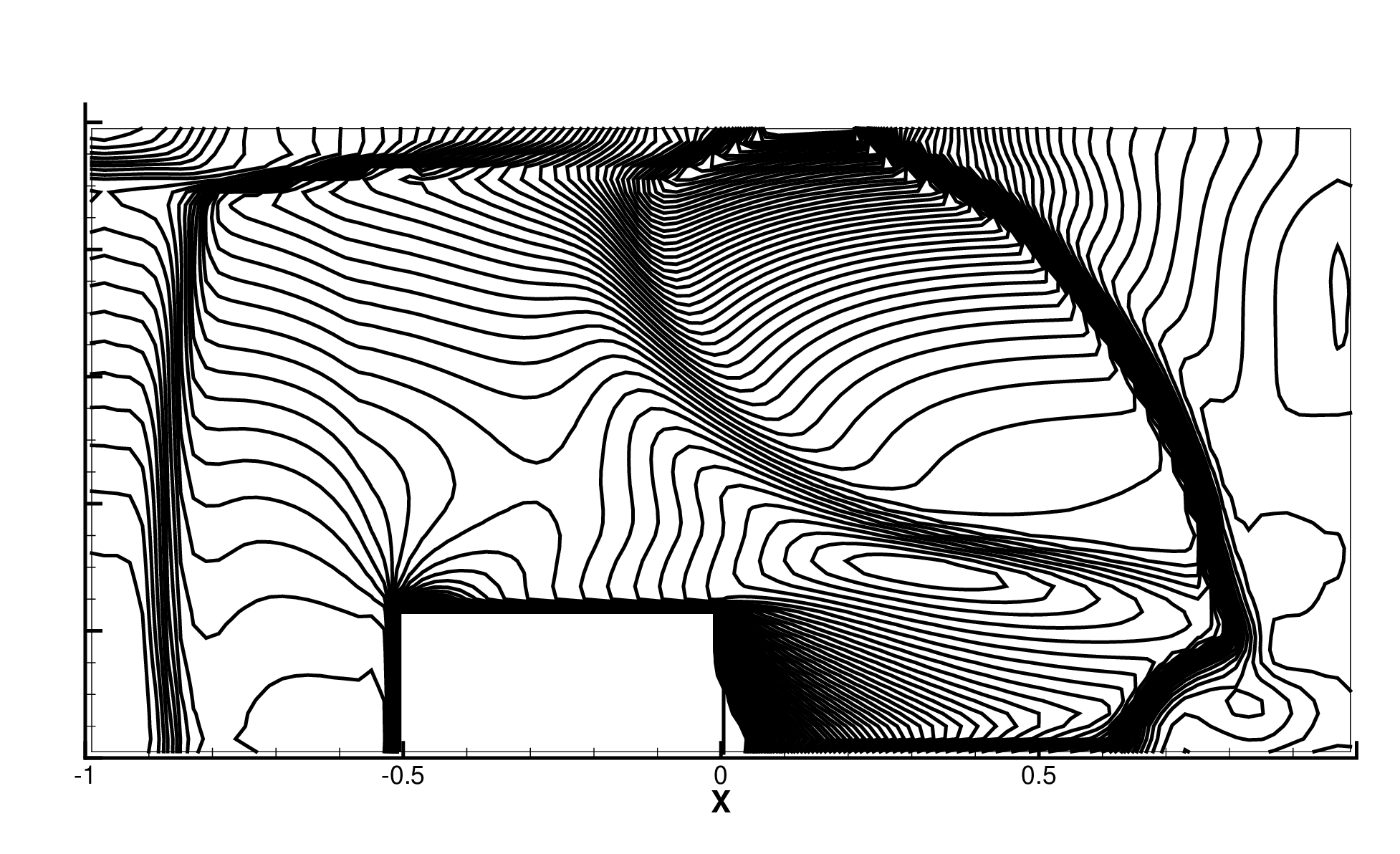} & \includegraphics[width=0.45\textwidth]{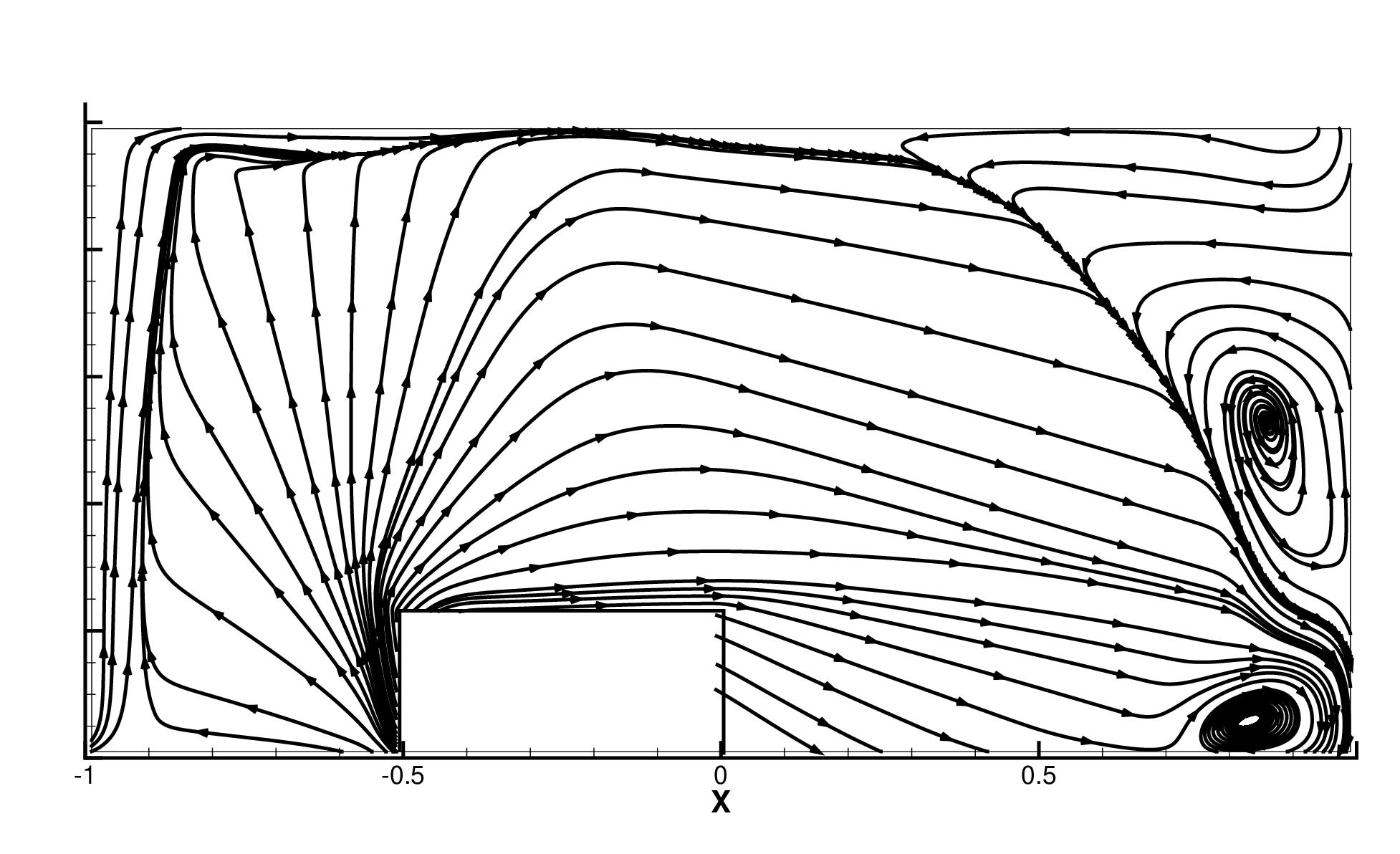} 
\end{tabular}
\caption{Parent Test case 8 (2D enclosure), I order results; Top) Pressure contours and streamlines at t= 0.00047, Bottom) Pressure contours and streamlines at t= 0.000955}
\label{fig:2d_euler_9}
\end{figure}

\begin{figure}[h!] 
\centering
\begin{tabular}{c}
\includegraphics[width=0.7\textwidth]{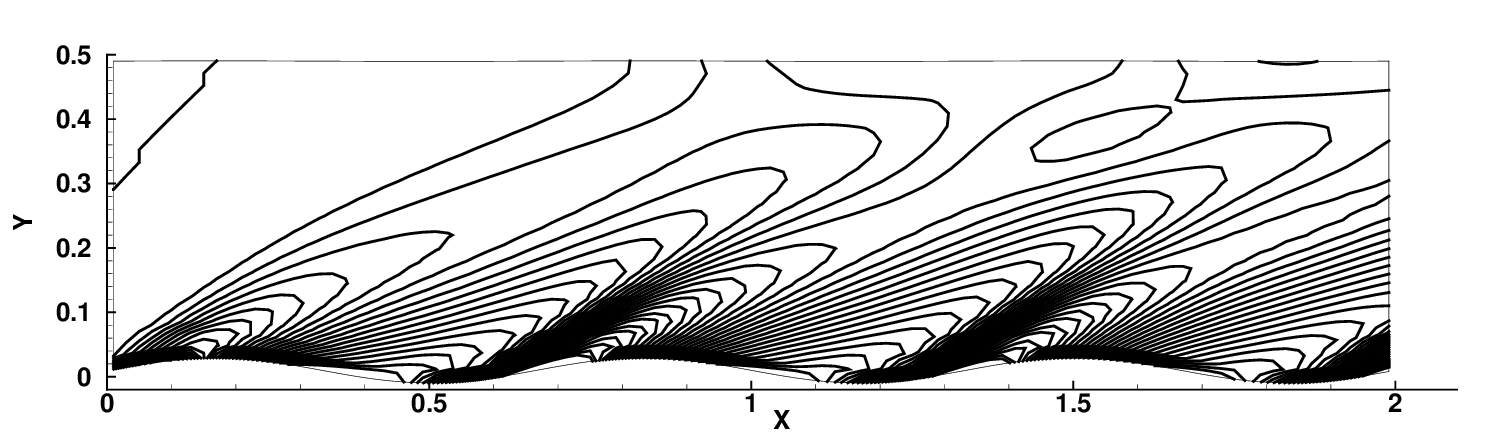} \\ \includegraphics[width=0.7\textwidth]{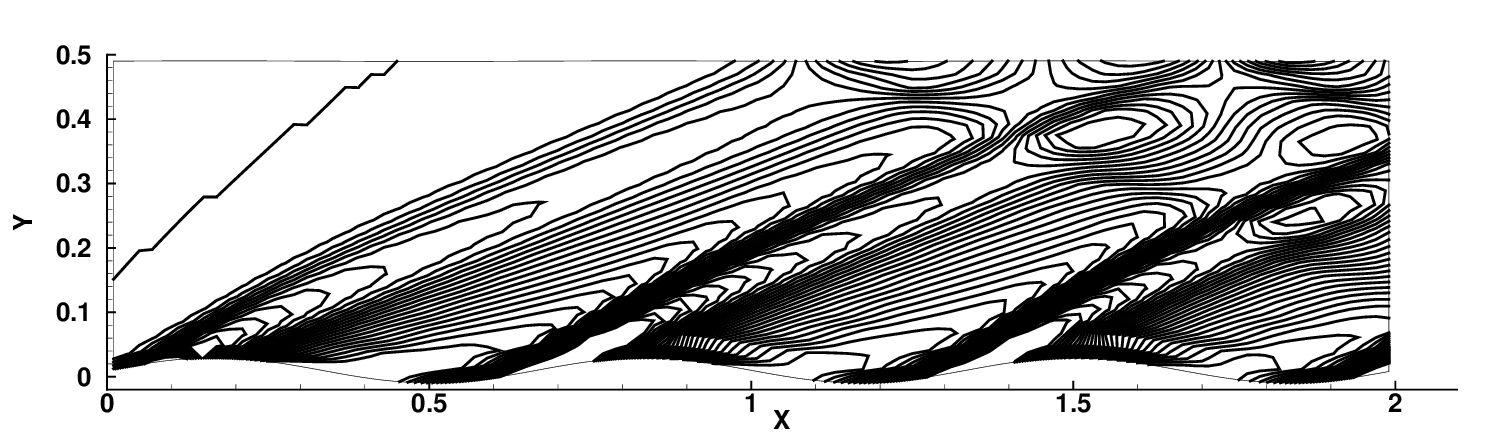} 
\end{tabular}
\caption{Parent Test case 11 (2D channel with wavy wall); pressure contours; Top) I order, Bottom) II order}
\label{fig:2d_euler_10}
\end{figure}

In this section, we have solved test cases taken from Parent \cite{PARENT2012173} which test a numerical scheme for its positivity preservation property. These test cases have large initial Mach numbers that create strong expansions (and thus low pressures) within the first few iterations, making them difficult to solve numerically. Since our proposed scheme for first order accuracy is positivity preserving, it is expected not to fail for these test cases and the results confirm this expectation. The first test case that we have solved is the 2D inclosure problem (Parent test case 8). It consists of a rectangular domain of dimension $[-1,1]\times[0,1]$ with a cutout at the bottom from x=-0.52 to x=0, and a height of 0.24 unit. A rectangular grid with a fixed cell size of 0.02 unit is chosen along both dimensions. CFL no. taken is 0.25. Since the flow is enclosed, flow tangency conditions are applied at all boundaries. The initial conditions are as follows.
\begin{eqnarray}
\text{Left state, x$\leq$ 0: }&& \ T= 35 \ K, \ M_{x}= 10, \ M_{y}= -3, \ p= 0.1 \ bar\nonumber\\
\text{Right state, x$>$ 0: }&& \ T= 35 \ K, \ M_{x}= 10, \ M_{y}= 2, \ p= 0.1 \ bar
\label{eq:2d_euler_tc_pos1}
\end{eqnarray}
This initial flow strikes and reflects from the walls and the cutout, creating complex flow features having strong pressure gradients. Our first order accurate results in Figure \ref{fig:2d_euler_9} show the pressure contours and streamlines at times t= 0.00047 and t= 0.000955. Our first order scheme thus successfully captures these strong gradients. Our second order method fails for this test case; it is not unexpected since our second order scheme is not necessarily positivity preserving. The next test case that we have solved consists of Mach 3 flow in a 2D channel with a wavy wall (Parent test case 11). The computational domain is $[0,2]\times[0,0.5]$, with a wavy wall at the bottom, given by y= $\frac{1}{50}sin(3\pi x)$. The grid is composed of 100 $\times$ 25 uniformly spaced cells. At initial time, flow throughout the domain is initialized with the following conditions:
\begin{equation}
T= 300 \ K, \ M_{x}= 3, \ M_{y}= 0, \ p= 0.102 \ bar
\label{eq:2d_euler_tc_pos2}
\end{equation}
Supersonic inflow and outflow conditions are applied at the left and right boundaries respectively, whereas flow tangency conditions are applied at the top and bottom. The wavy wall at the bottom alternately compresses and expands the supersonic flow. As a result, we get alternating oblique shocks and expansion fans emanating from the wavy wall. Our first and second order accurate steady state results (Figure \ref{fig:2d_euler_10}) capture these flow features accurately.

\subsubsection{NACA0012 airfoil test cases}
We have solved some benchmark test cases for the symmetric NACA0012 airfoil \cite{jones1985reference,dervieux1989numerical}. For these tests, an O-type structured grid with dimensions of 25 times the chord length is used around the airfoil. Farfield conditions described in \ref{appendix:a2} are applied at the outer boundary, whereas flow tangency conditions are applied at the surface of the airfoil. Periodic conditions are applied along the $\eta$-direction where the first and last grids meet. Freestream initial conditions are used, and steady state solutions are sought. Numerical tests are performed for the following supersonic, transonic, and subsonic test cases.
\begin{enumerate}
	\item $M_{\infty}=$ 1.2, A.O.A. (Angle of attack) = $0^\circ$
	\item $M_{\infty}=$ 1.2, A.O.A. = $7^\circ$
	\item $M_{\infty}=$ 0.8, A.O.A. = $1.25^\circ$
	\item $M_{\infty}=$ 0.85, A.O.A. = $1^\circ$
	\item $M_{\infty}=$ 0.63, A.O.A. = $2^\circ$
\end{enumerate}

\begin{figure}[h!] 
\begin{tabular}{cccc}
\centering
\includegraphics[width=0.22\textwidth]{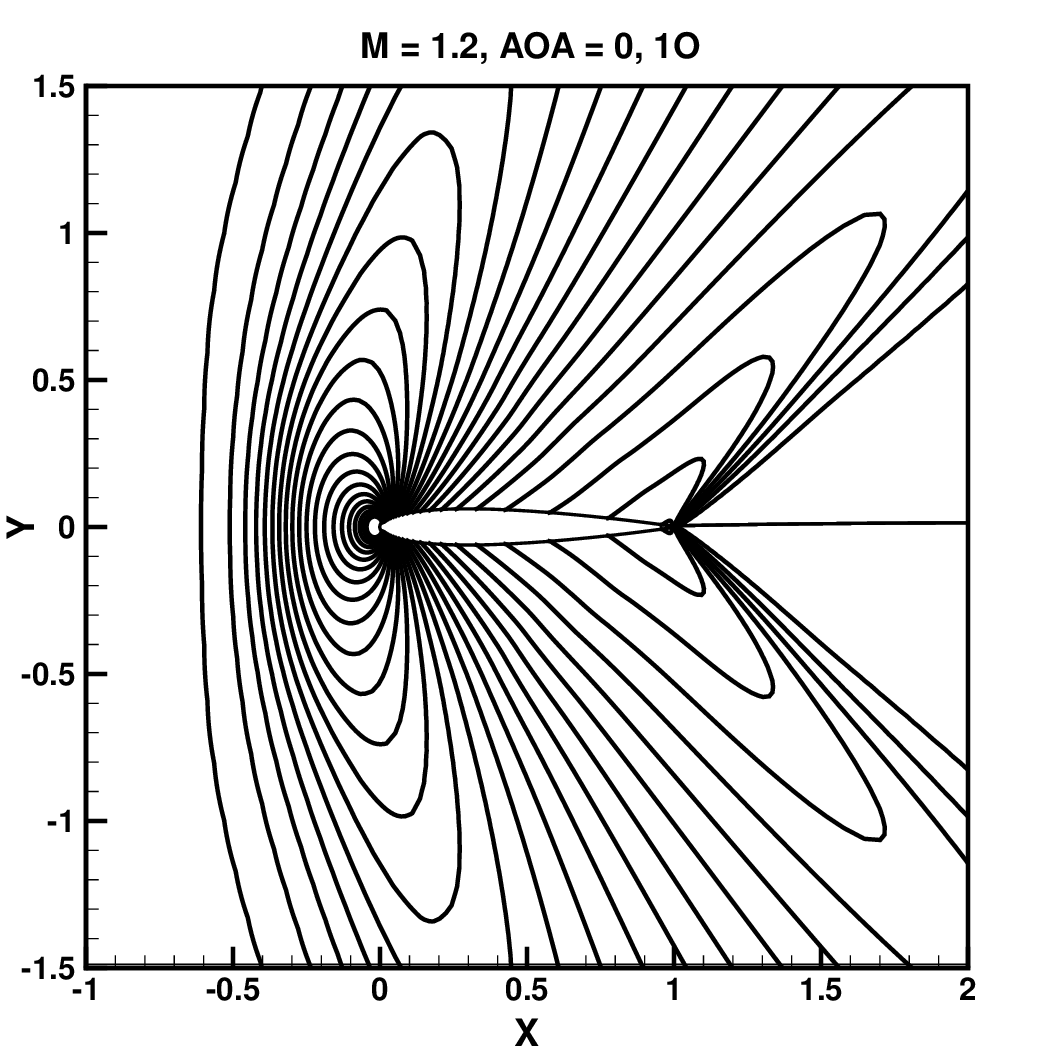} & \includegraphics[width=0.22\textwidth]{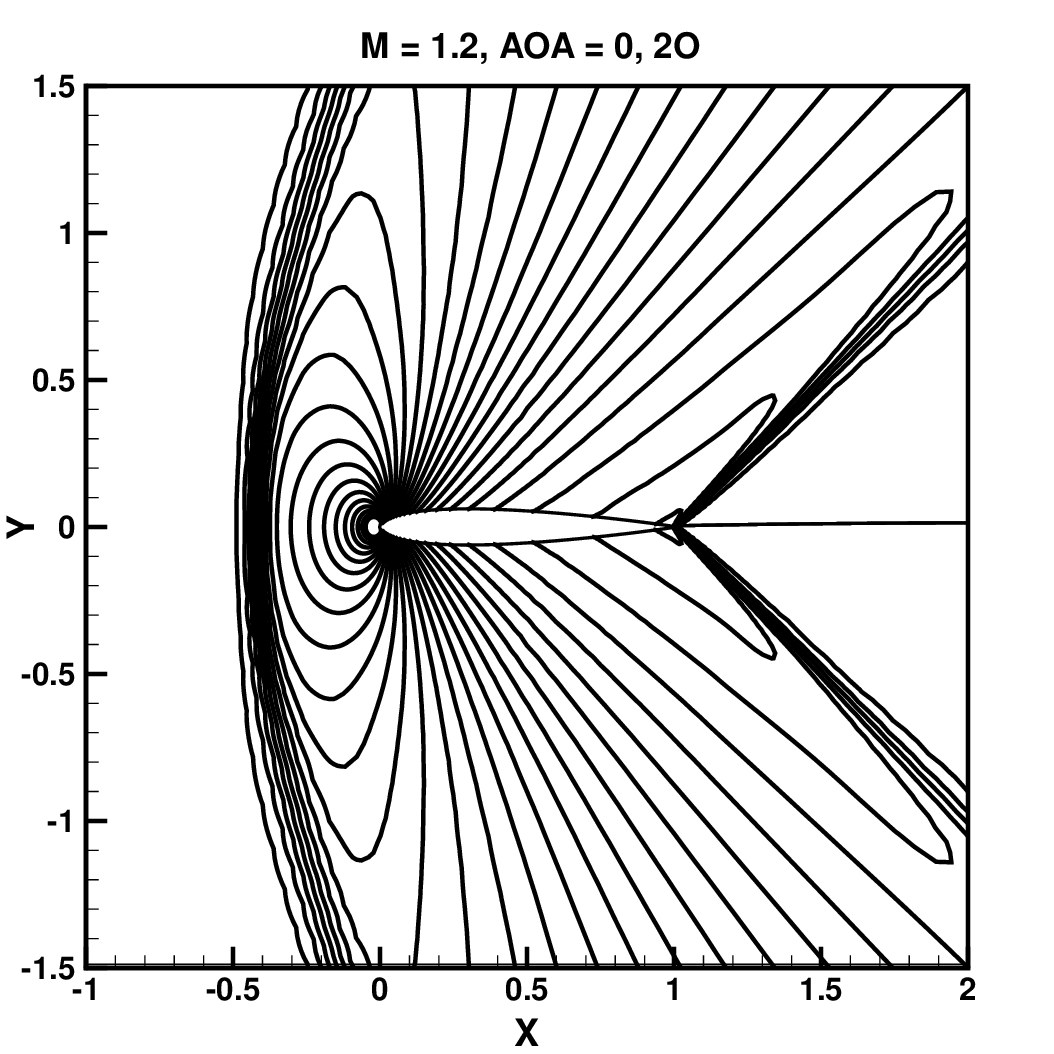} &
\includegraphics[width=0.23\textwidth]{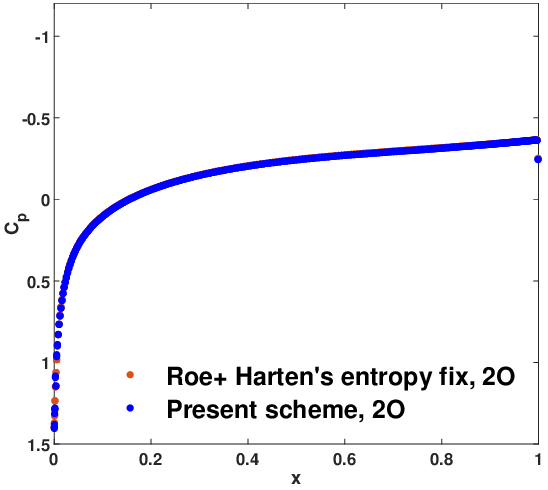} & \includegraphics[width=0.23\textwidth]{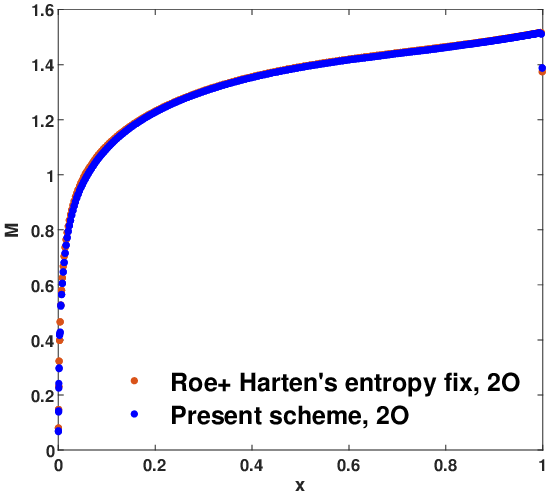} 
\end{tabular}
\caption{NACA0012, $M_{\infty}=$ 1.2, A.O.A$= 0^\circ$, $438 \times 107$ grid: a) I order, b) II order,  pressure contours (0.4:0.05:2.0), c)~$C_{p}$~vs~$x$, d) $M$ vs $x$ along the top and bottom surfaces of the airfoil.}
\label{fig:2d_euler_11}
\end{figure}
\begin{figure}[h!] 
\begin{tabular}{cccc}
\centering
\includegraphics[width=0.22\textwidth]{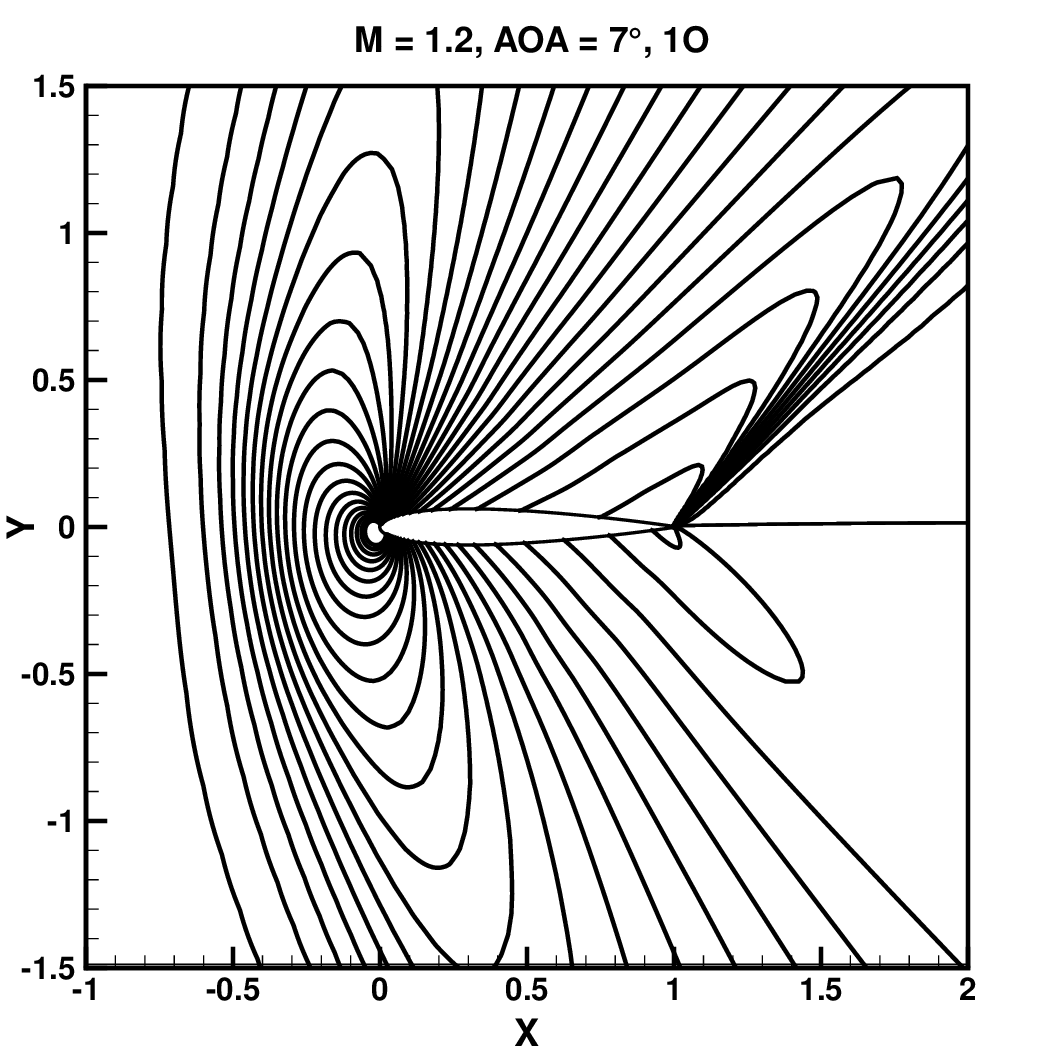} & \includegraphics[width=0.22\textwidth]{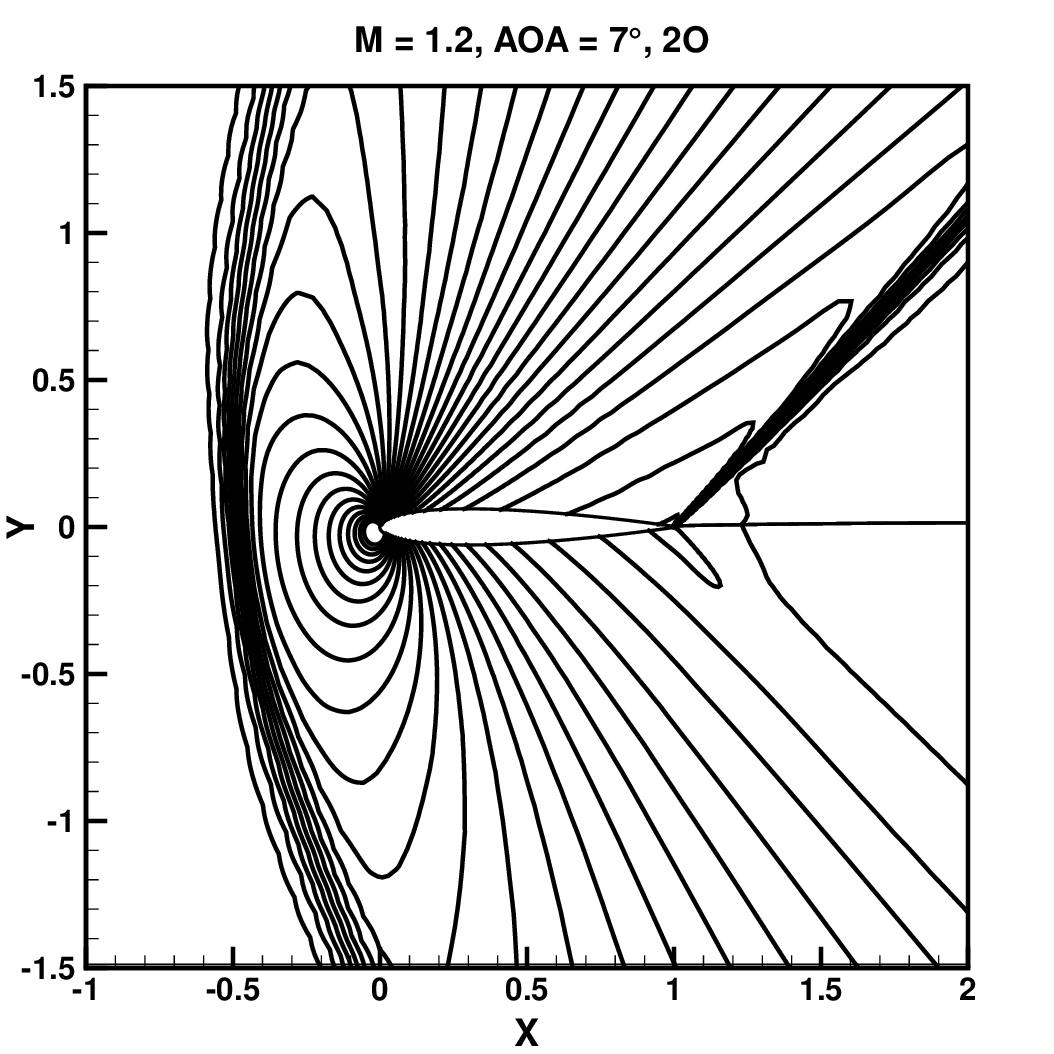} &
\includegraphics[width=0.23\textwidth]{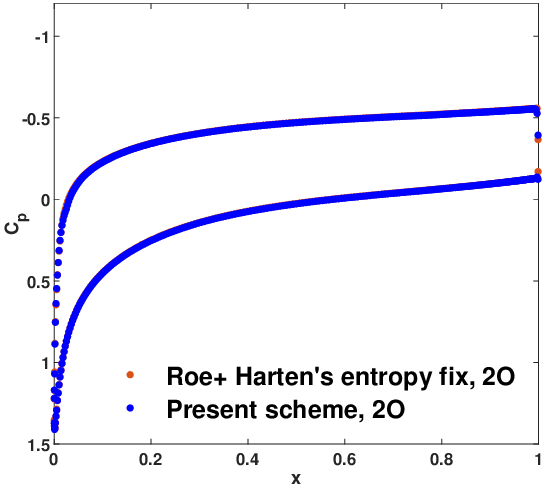} & \includegraphics[width=0.23\textwidth]{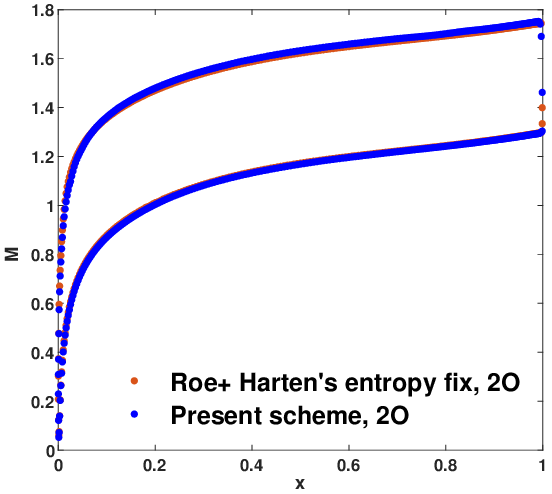} 
\end{tabular}
\caption{NACA0012, $M_{\infty}=$ 1.2, A.O.A$= 7^\circ$, $438 \times 107$ grid: a) I order, b) II order,  pressure contours (0.4:0.05:2.0), c)~$C_{p}$~vs~$x$, d) $M$ vs $x$ along the top and bottom surfaces of the airfoil.}
\label{fig:2d_euler_12}
\end{figure}
\begin{figure}[h!] 
\begin{tabular}{cccc}
\centering
\includegraphics[width=0.22\textwidth]{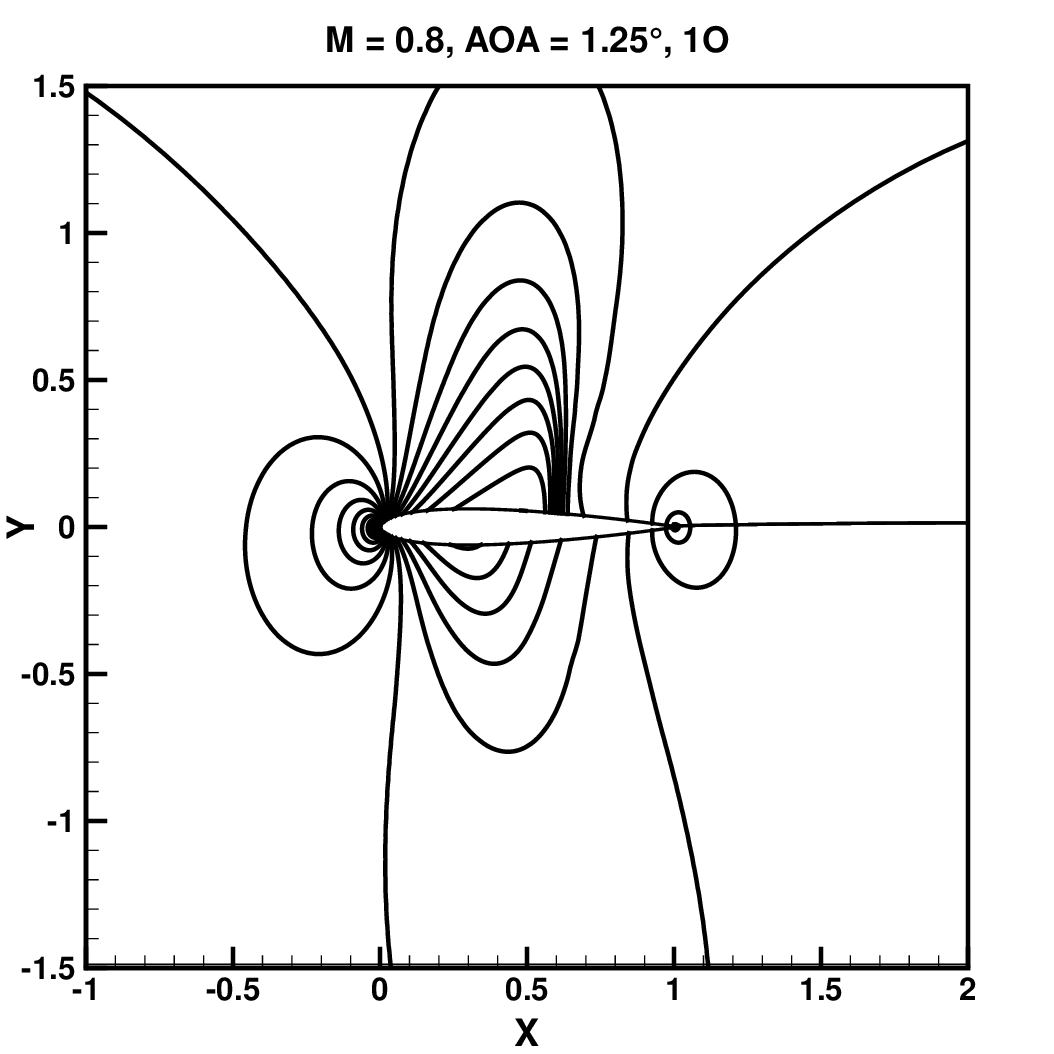} & \includegraphics[width=0.22\textwidth]{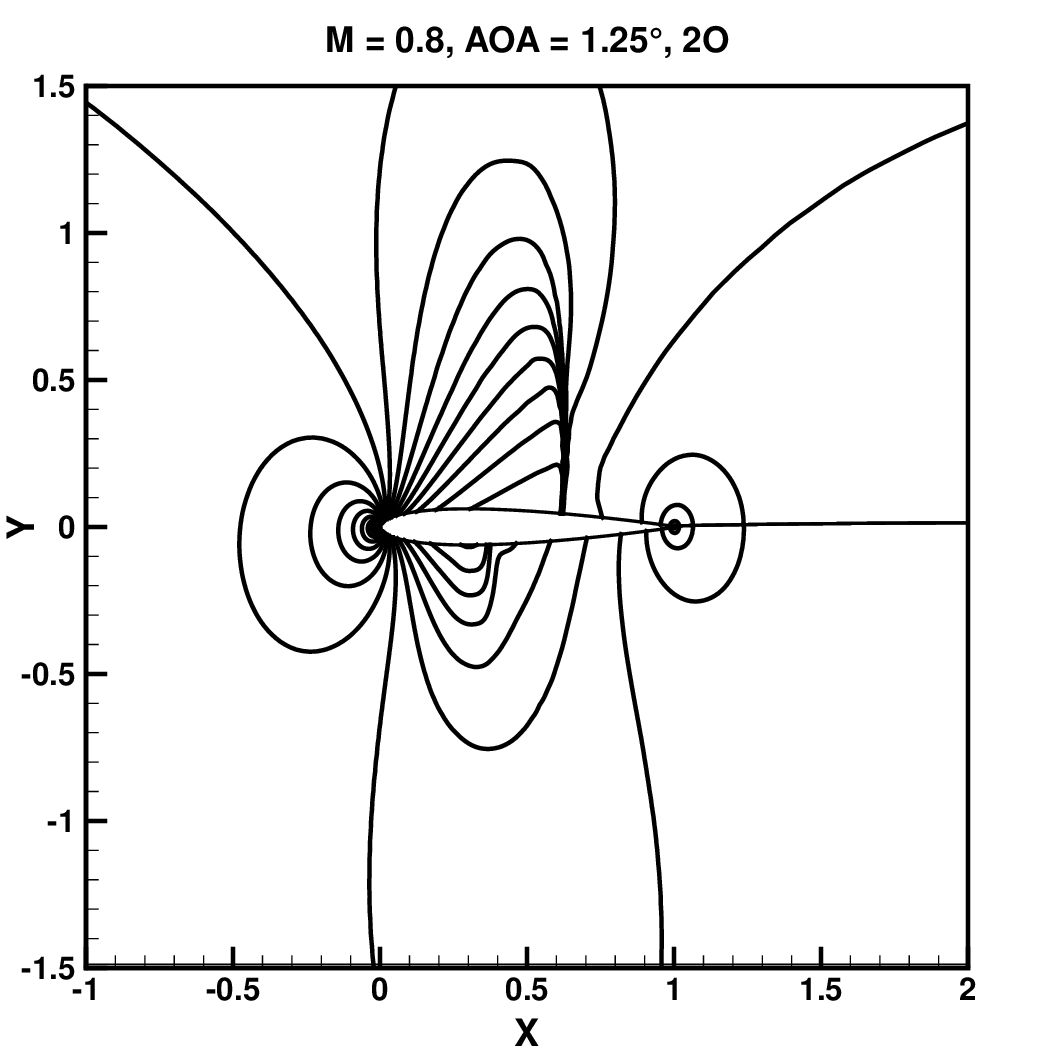} &
\includegraphics[width=0.23\textwidth]{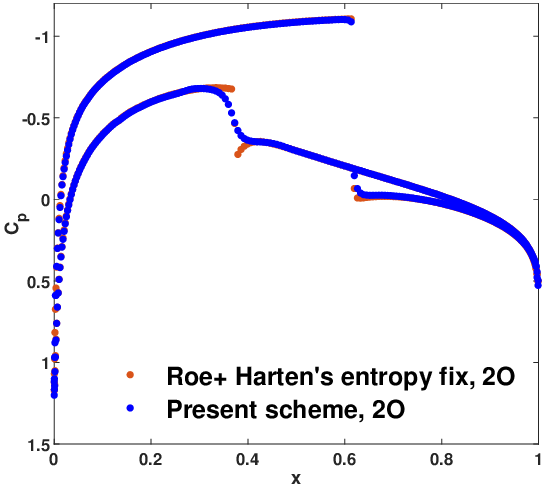} & \includegraphics[width=0.23\textwidth]{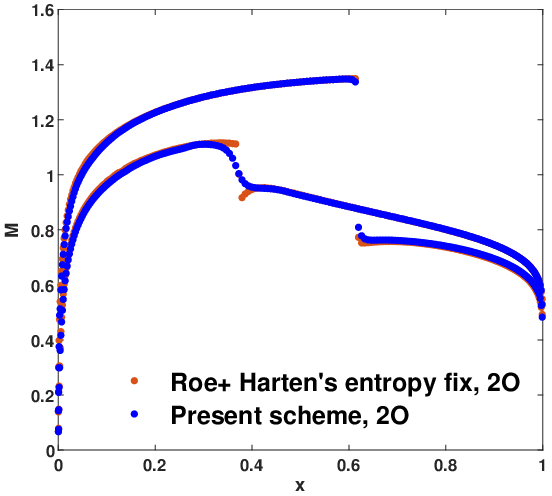} 
\end{tabular}
\caption{NACA0012, $M_{\infty}=$ 0.8, A.O.A$= 1.25^\circ$, $438 \times 107$ grid: a) I order, b) II order,  pressure contours (0.4:0.05:2.0), c)~$C_{p}$~vs~$x$, d) $M$ vs $x$ along the top and bottom surfaces of the airfoil.}
\label{fig:2d_euler_13}
\end{figure}
\begin{figure}[h!] 
\begin{tabular}{cccc}
\centering
\includegraphics[width=0.22\textwidth]{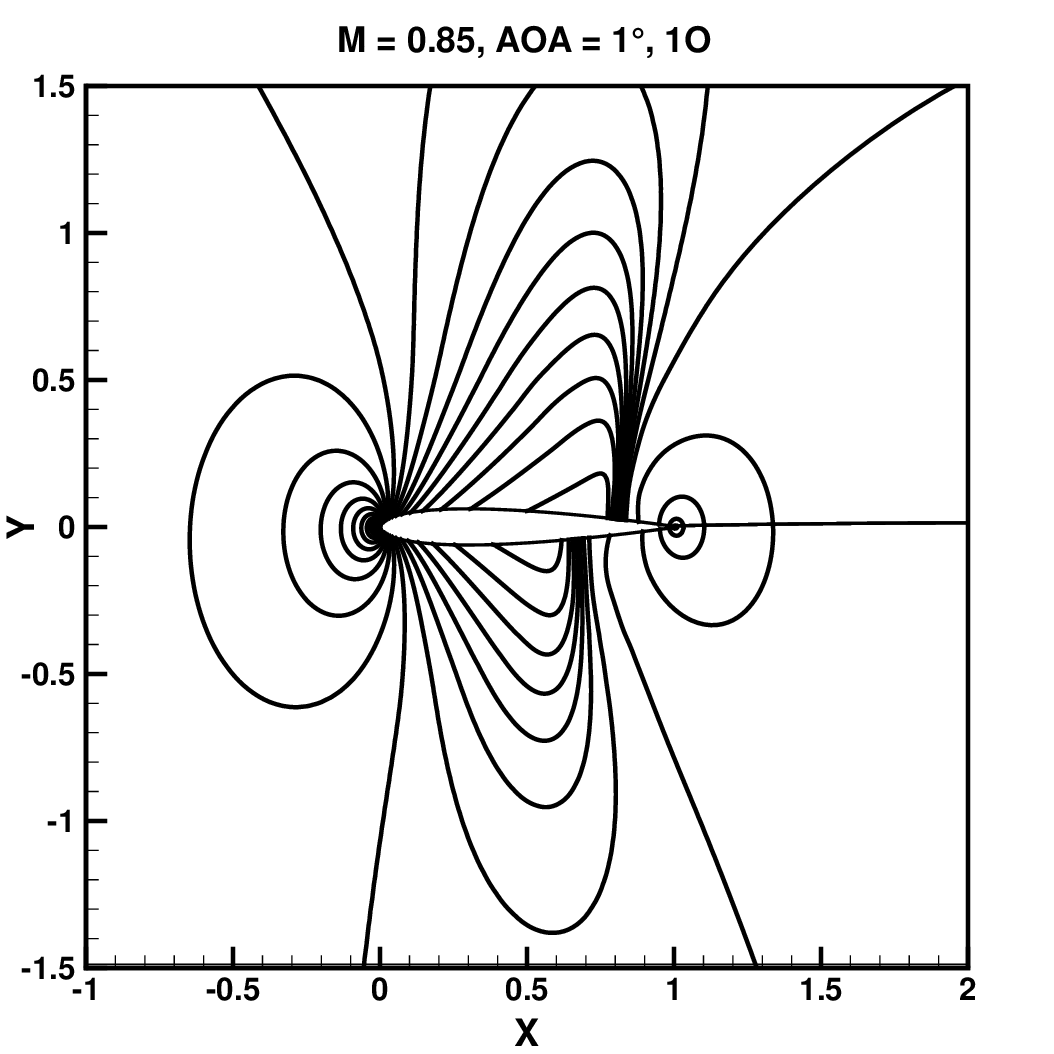} & \includegraphics[width=0.22\textwidth]{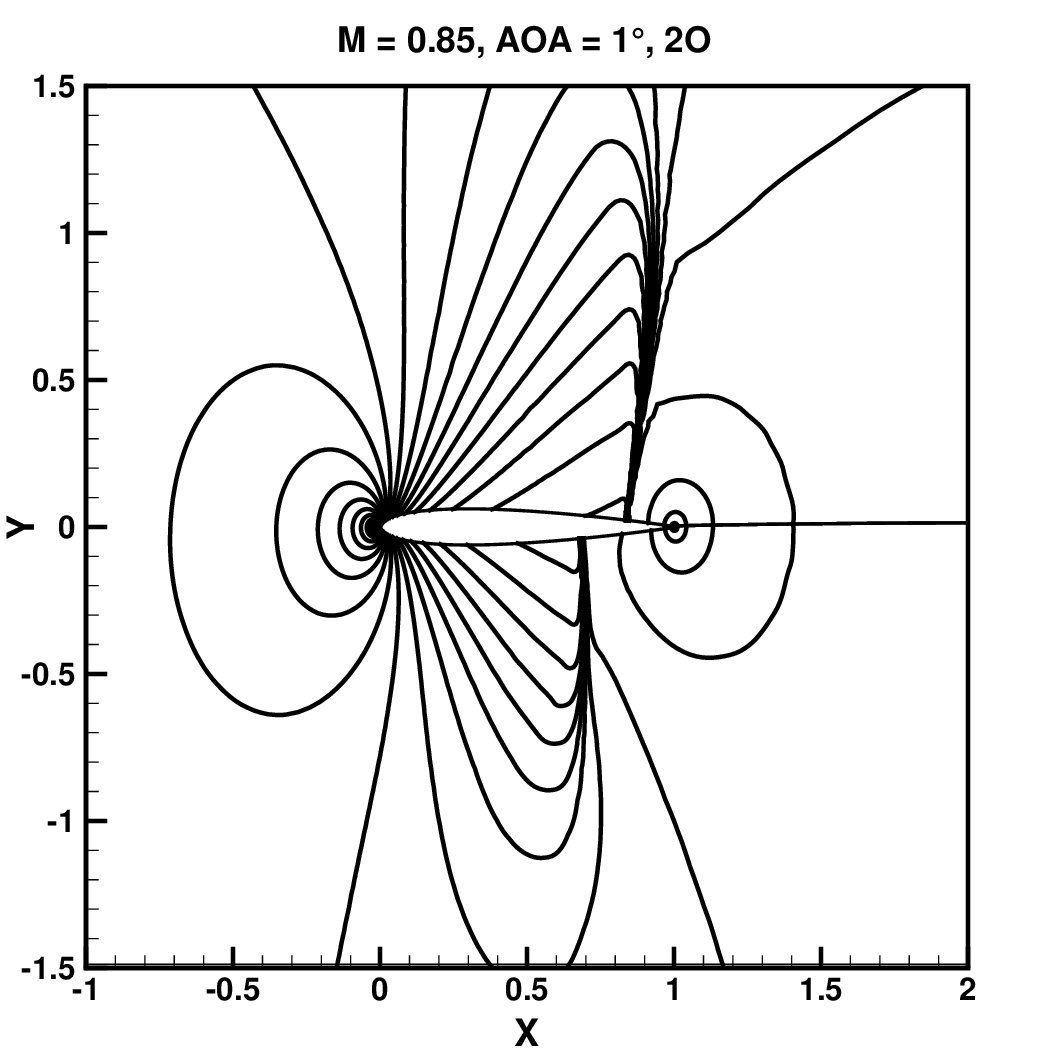} &
\includegraphics[width=0.23\textwidth]{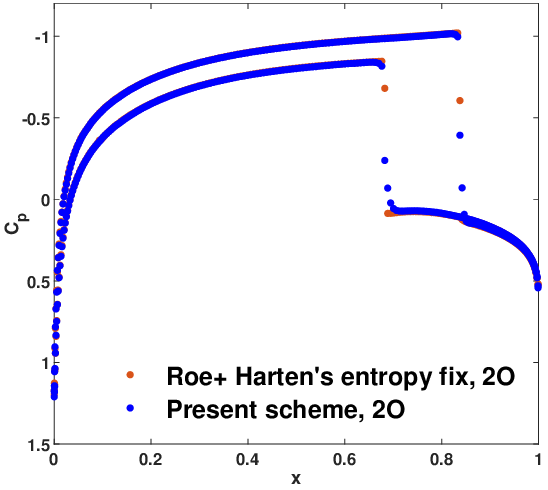} & \includegraphics[width=0.23\textwidth]{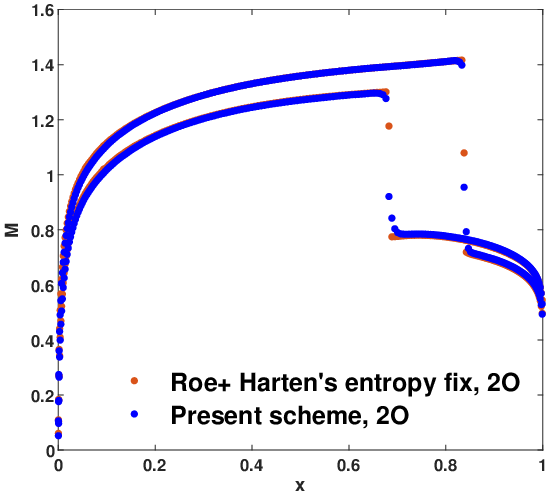} 
\end{tabular}
\caption{NACA0012, $M_{\infty}=$ 0.85, A.O.A$= 1^\circ$, $438 \times 107$ grid: a) I order, b) II order,  pressure contours (0.4:0.05:2.0), c)~$C_{p}$~vs~$x$, d) $M$ vs $x$ along the top and bottom surfaces of the airfoil.}
\label{fig:2d_euler_14}
\end{figure}
\begin{figure}[h!] 
\begin{tabular}{cccc}
\centering
\includegraphics[width=0.22\textwidth]{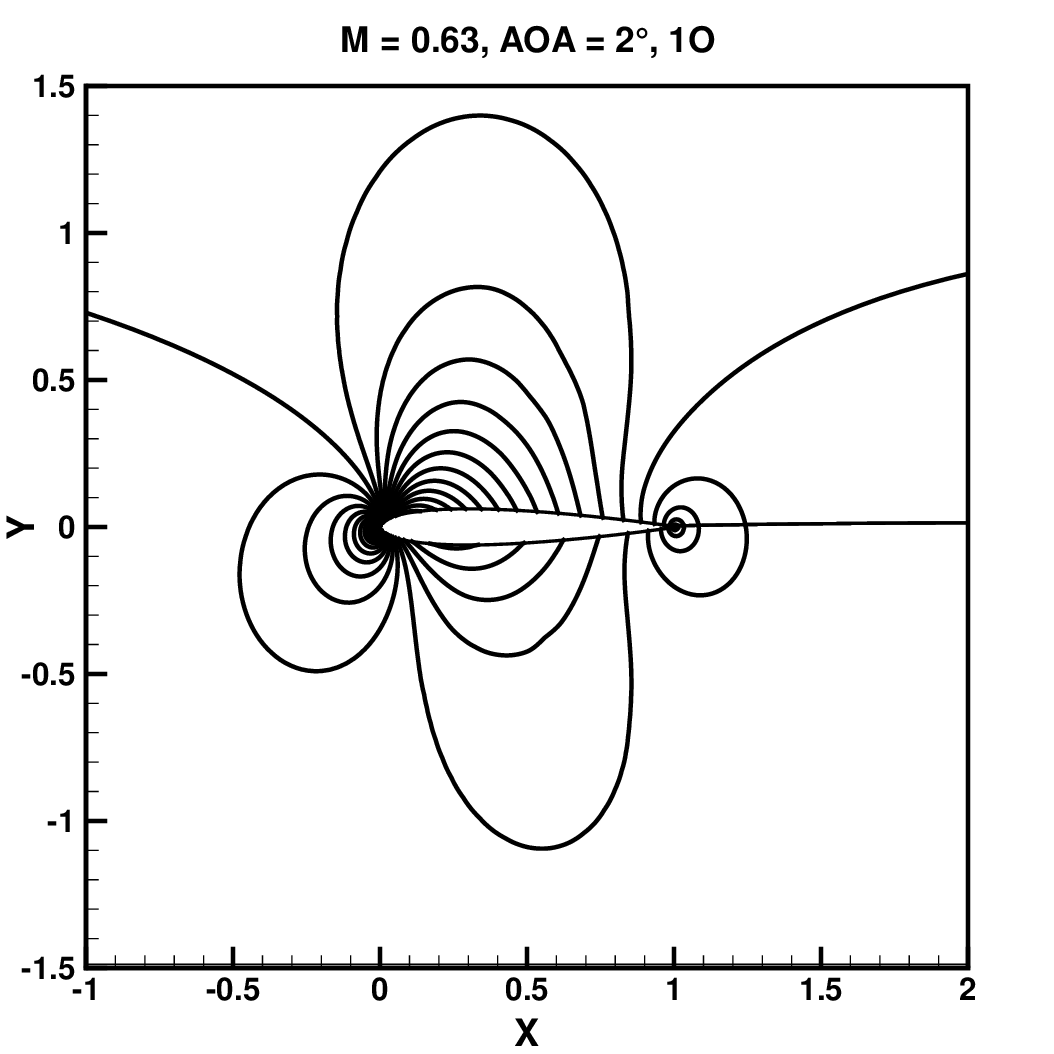} & \includegraphics[width=0.22\textwidth]{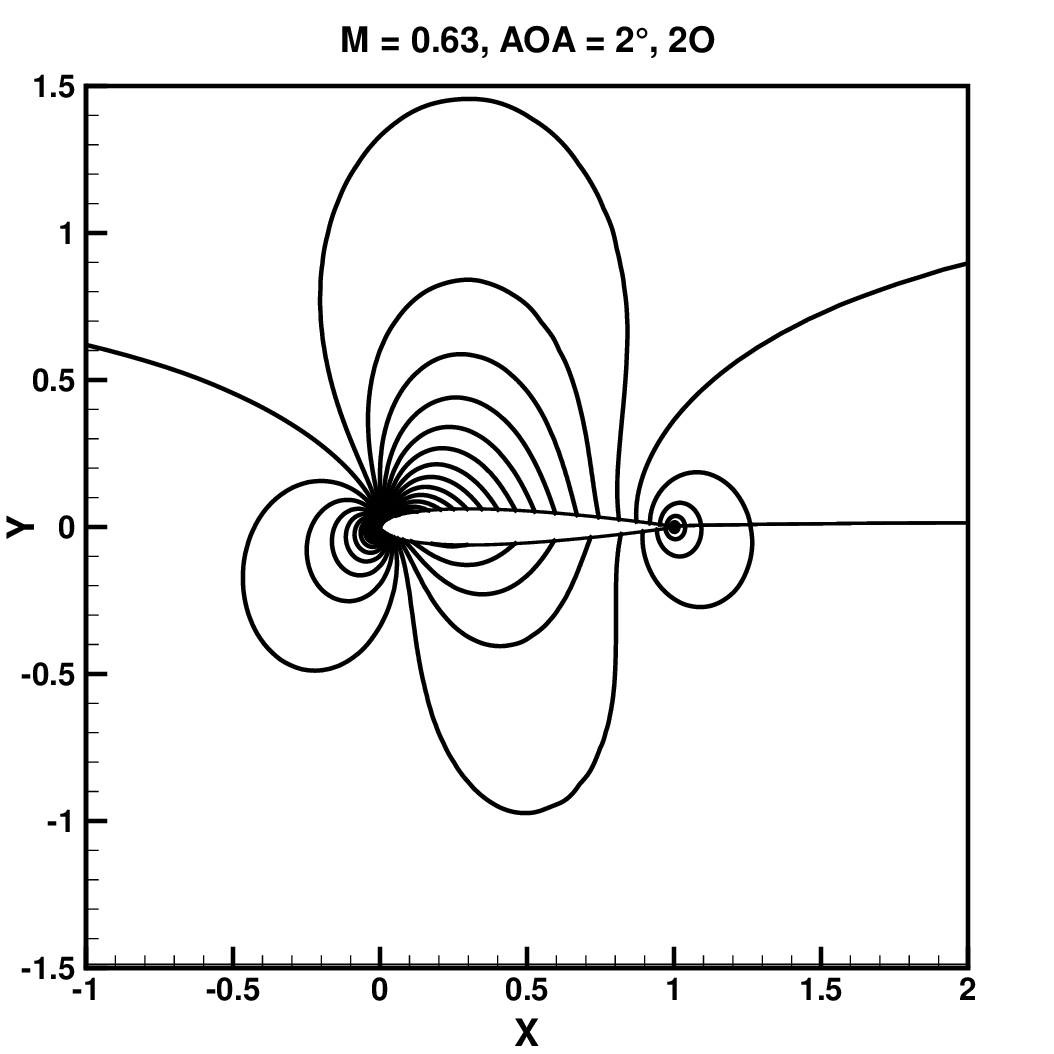} &
\includegraphics[width=0.23\textwidth]{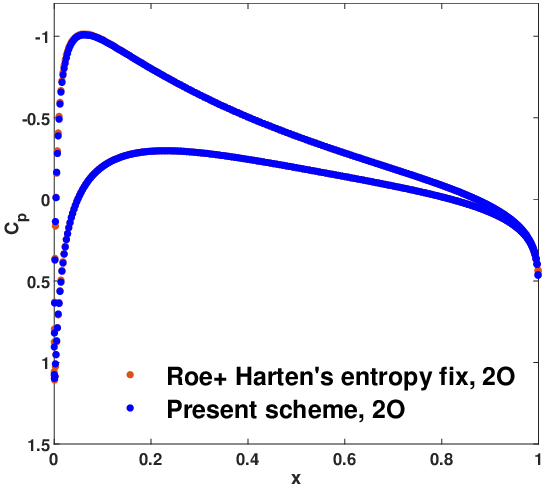} & \includegraphics[width=0.23\textwidth]{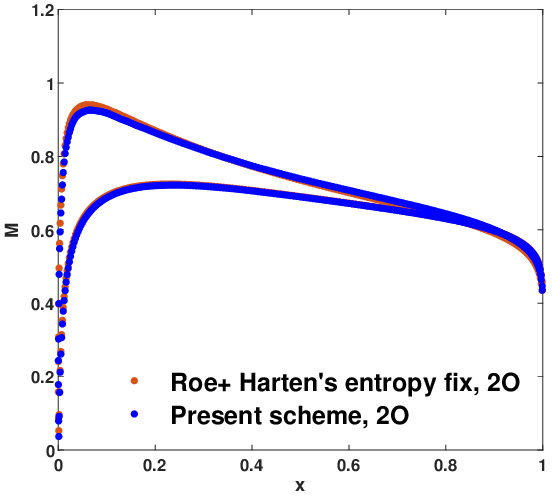} 
\end{tabular}
\caption{NACA0012, $M_{\infty}=$ 0.63, A.O.A$= 2^\circ$, $438 \times 107$ grid: a) I order, b) II order,  pressure contours (0.7:0.02:1.4), c)~$C_{p}$~vs~$x$, d) $M$ vs $x$ along the top and bottom surfaces of the airfoil.}
\label{fig:2d_euler_15}
\end{figure}

 Numerical results for these NACA0012 airfoil test cases are shown in Figures \eqref{fig:2d_euler_11} to \eqref{fig:2d_euler_15}. The pressure contours for first and second order accuracy are shown. Further, our second order accurate results for the variation of pressure coefficient $C_{p}$(=$\frac{p- p_{\infty}}{0.5\rho_{\infty} |\textbf{u}|^{2}_{\infty}}$) and Mach no. $M$(= $\frac{|\textbf{u}|}{a}$) along the top and bottom surfaces of the airfoil are plotted along with results for Roe's scheme with Harten's entropy fix on the same grid.

\subsection{2D Viscous tests}
Some viscous test cases are solved to demonstrate the ability of our numerical scheme to solve viscous equations and resolve viscous flow features such as boundary layers. For these problems, the gradient terms in the viscous fluxes are computed using auxiliary volume method.

\subsubsection{Plane Couette flow}
\begin{figure}[h!] 
\centering
\begin{tabular}{cc}
\includegraphics[width=0.3\textwidth]{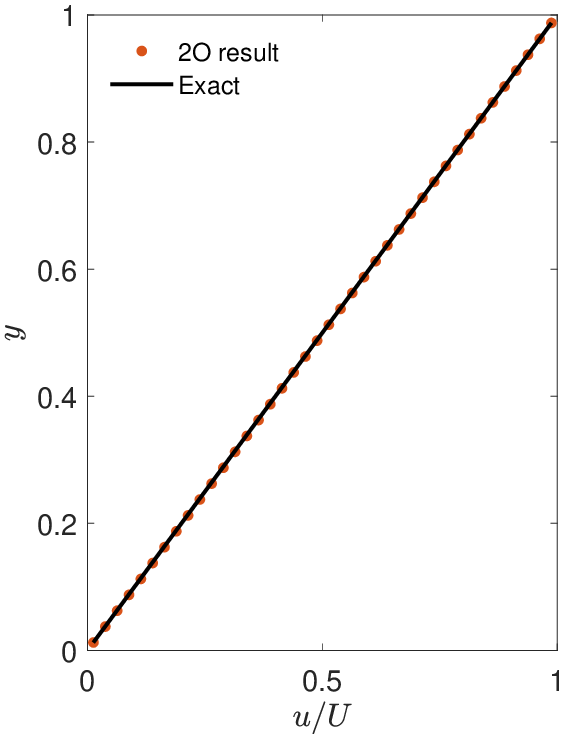} & \includegraphics[width=0.3\textwidth]{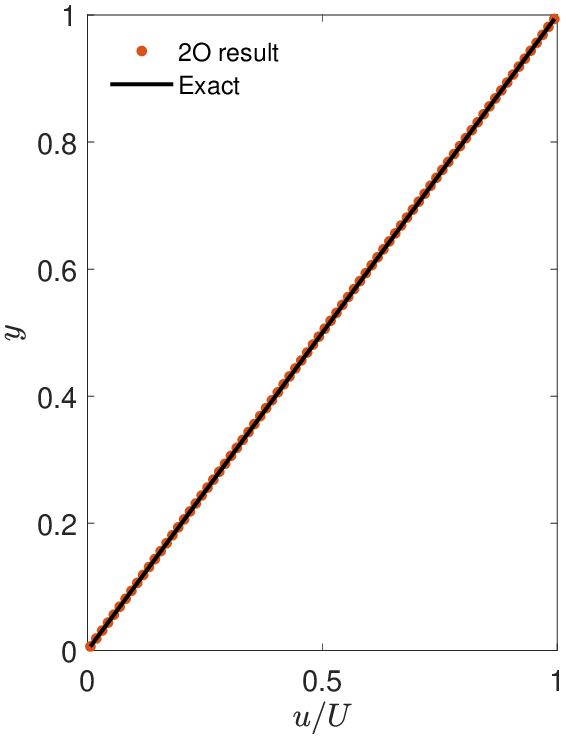}
\end{tabular}
\caption{Plane Couette flow, $y$ vs $\frac{u}{U}$ for $i= \frac{Nx}{2}$  on, Left) $40 \times 40$ grid, Right) $80 \times 80$ grid}
\label{fig:2d_ns_tc_couette}
\end{figure}

Plane Couette flow is an incompressible viscous flow between two infinite parallel plates, sustained by the relative velocity between the plates and without any imposed pressure gradient. The lower horizontal plate at $y$= 0 is stationary, while the upper horizontal plate at $y$= $h$ moves with speed U in the $x$-direction. Under the assumption that the flow is steady and fully developed, its exact solution exists and is given by $u= u(y)=U\frac{y}{h}$. To solve this flow numerically, we take a computational domain with dimension $[0,1]\times[0,1]$. A uniform Cartesian grid is used to discretize the domain. For this low speed flow, the values assumed (at the upper plate) are Mach no M= 0.05 and Reynolds no Re= 10. Following conditions are applied at the top and bottom boundaries.
\begin{eqnarray}
\text{At $y$= 0: } \ u= 0, \ v= 0\nonumber\\
\text{At $y$= 1: } \ u = U, \ v= 0
\label{eq:2d_ns_tc_couette}
\end{eqnarray}
Periodic conditions are applied at the left and right boundaries. The fluid is assumed to be initially at rest. Steady state results are obtained for $40 \times 40$ and $80 \times 80$ grids and are shown in Figure \ref{fig:2d_ns_tc_couette} as plots of $y$ vs. $\frac{u}{U}$ for $i= \frac{Nx}{2}$, along with the exact solution.

\subsubsection{Sod's shock tube problem}
This test is the two-dimensional version of the Sod's shock tube problem. The computational domain taken is $[0,1]\times[0,0.3]$. The initial conditions for this unsteady problem are given below.
\begin{eqnarray}
\text{Left state, x$\leq$ 0.5: }&&\rho_{L}= 1, \ (u_{1})_{L}= 0, \ (u_{2})_{L}= 0, \ p_{L}= 1 \nonumber\\
\text{Right state, x$>$ 0.5: }&&\rho_{R}= 0.125, \ (u_{1})_{R}= 0, \ (u_{2})_{R}= 0, \ p_{R}= 0.1
\label{eq:2d_vis_1}
\end{eqnarray}

\begin{figure}[h!] 
\centering
\begin{tabular}{cc}
\includegraphics[width=0.45\textwidth]{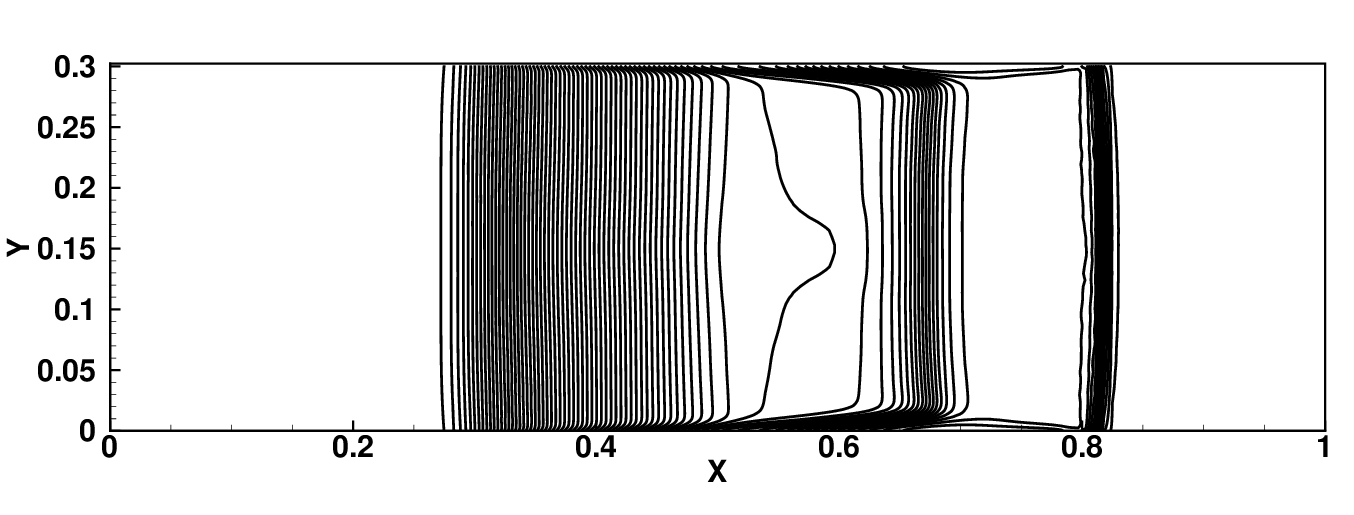} & \includegraphics[width=0.45\textwidth]{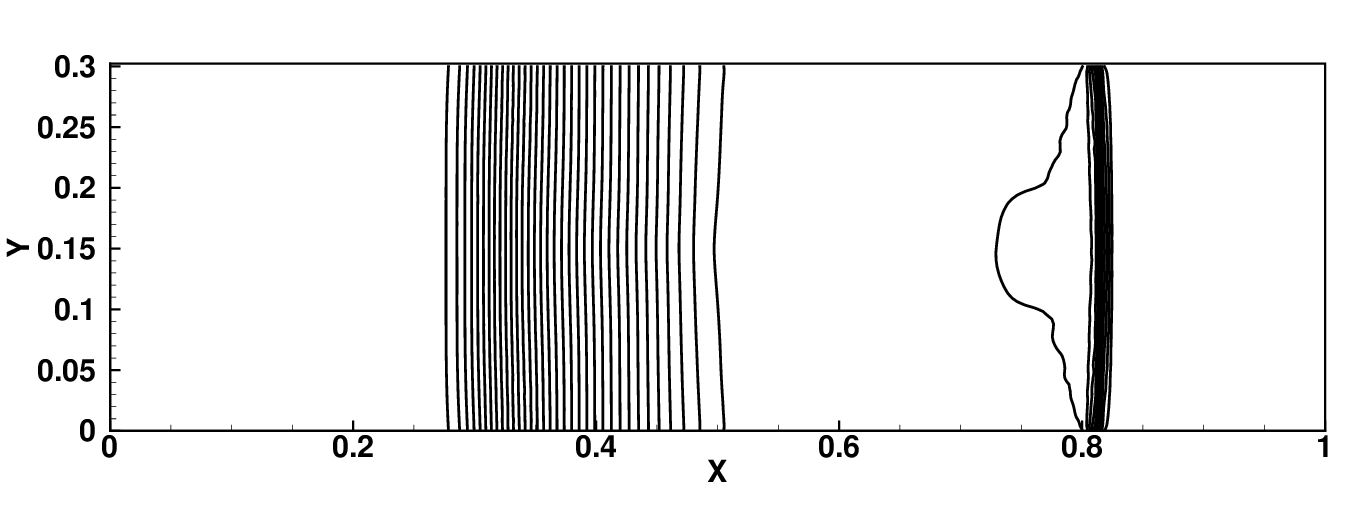}\\
\includegraphics[width=0.45\textwidth]{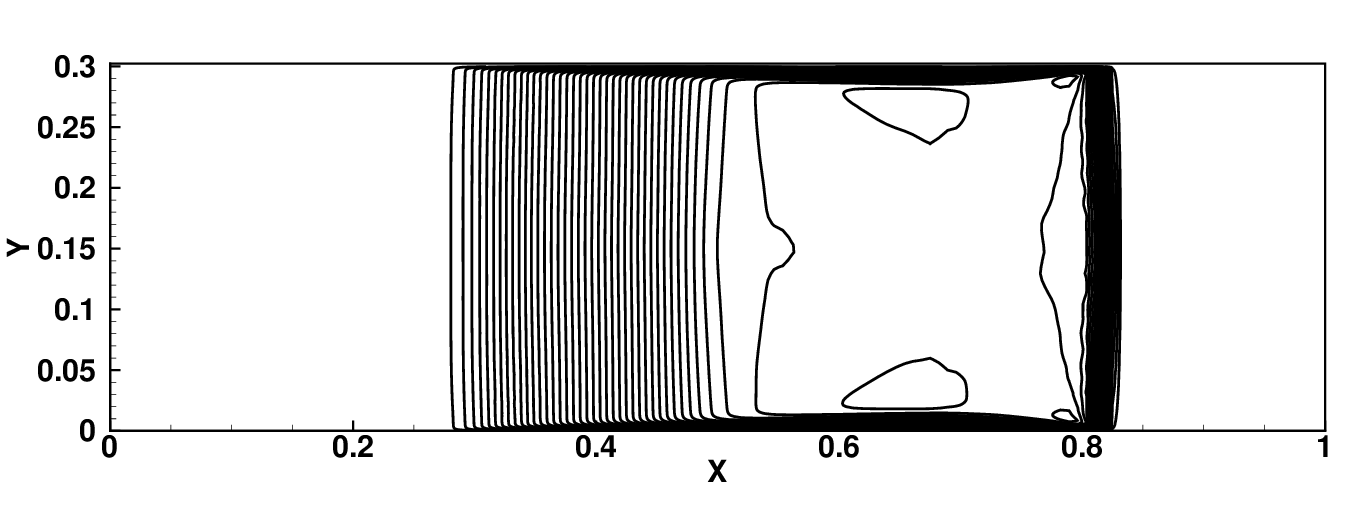} & \includegraphics[width=0.45\textwidth]{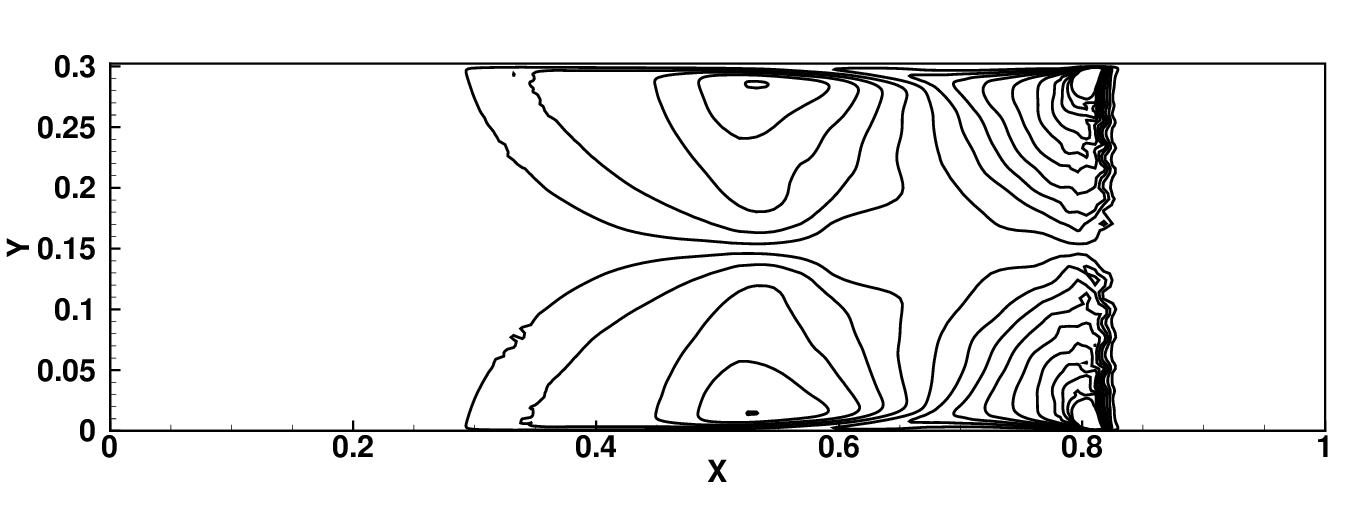} 
\end{tabular}
\caption{Sod's shock tube problem (140x 140): Top) Density and pressure contours, Bottom) $u_{1}$ and $u_{2}$ velocity contours.}
\label{fig:2d_vis_1}
\end{figure}

\begin{figure}%
\centering
\includegraphics[width=0.35\textwidth]{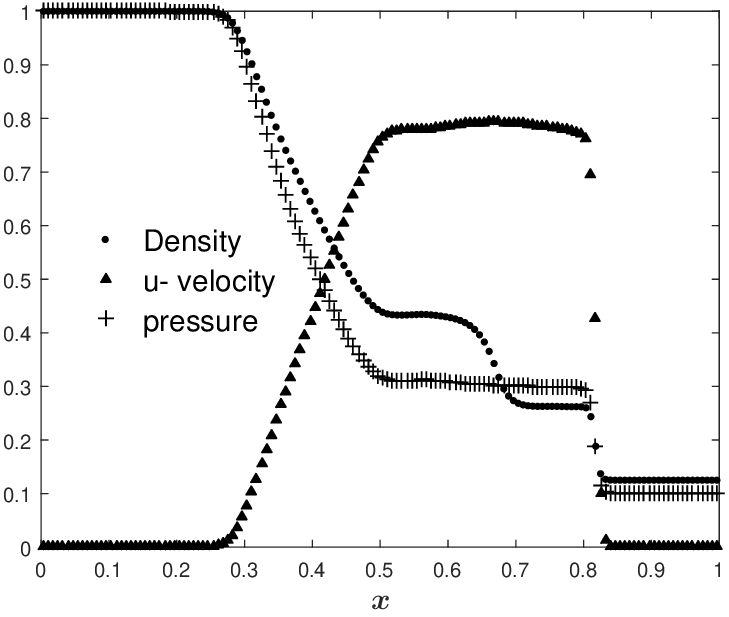}%
\caption{Sod's shock tube problem (140x 140), flow variables along the center-line}%
\label{fig:2d_vis_2}%
\end{figure}
The prescribed initial Reynolds no., Re= 25000, and Prandtl no., Pr= 0.72, are used. No-slip conditions (viscous wall) are applied at the top and bottom boundaries, and Neumann conditions are applied at the left and right boundaries. The domain is discretized into $140\times140$ cells, with constant cell size along the x-direction. Along the y-direction, the cells are geometrically stretched from the walls to the center with a regular increment of $4 \%$. The initial discontinuity leads to flow from left to right, with a shock wave and a contact-discontinuity traveling to the right, while rarefaction waves propagate to the left. The flow behind the shock results in the formation of boundary layers at the top and bottom, which bring in non-uniformity of flow along the transverse direction. Figure \ref{fig:2d_vis_1} shows the contours of pressure, density, $u_{1}$ and $u_{2}$ for the second order accurate results at time t= 0.2136. Flow properties at the center-line have been plotted against the length of the shock tube in Figure \ref{fig:2d_vis_2}. Flow features like shock, contact-discontinuity, and expansion are captured well. The $u_{2}$ contours show that the boundary layers at the walls behind the shock are also resolved by our scheme.  

\subsubsection{Shock-boundary layer interaction}
\begin{figure}
\centering
\begin{tabular}{cc}
\includegraphics[width=0.45\textwidth]{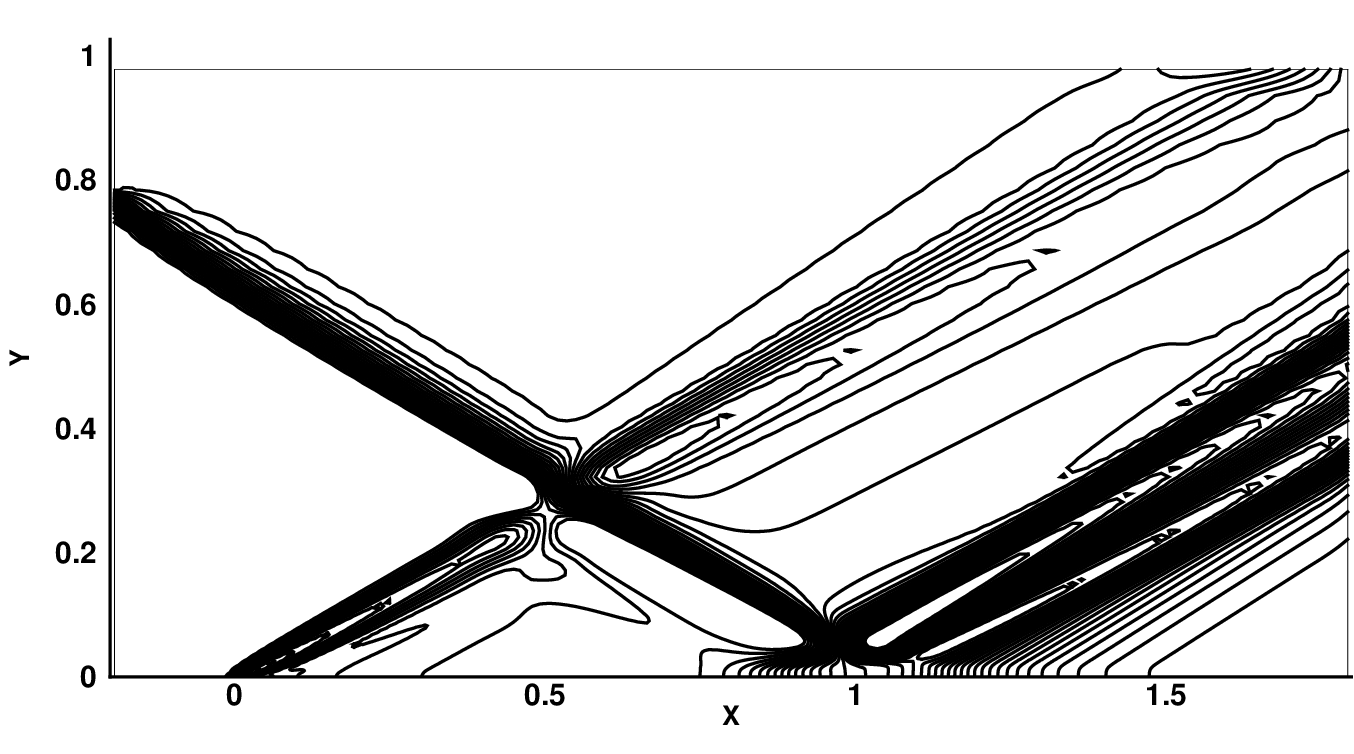} & \includegraphics[width=0.45\textwidth]{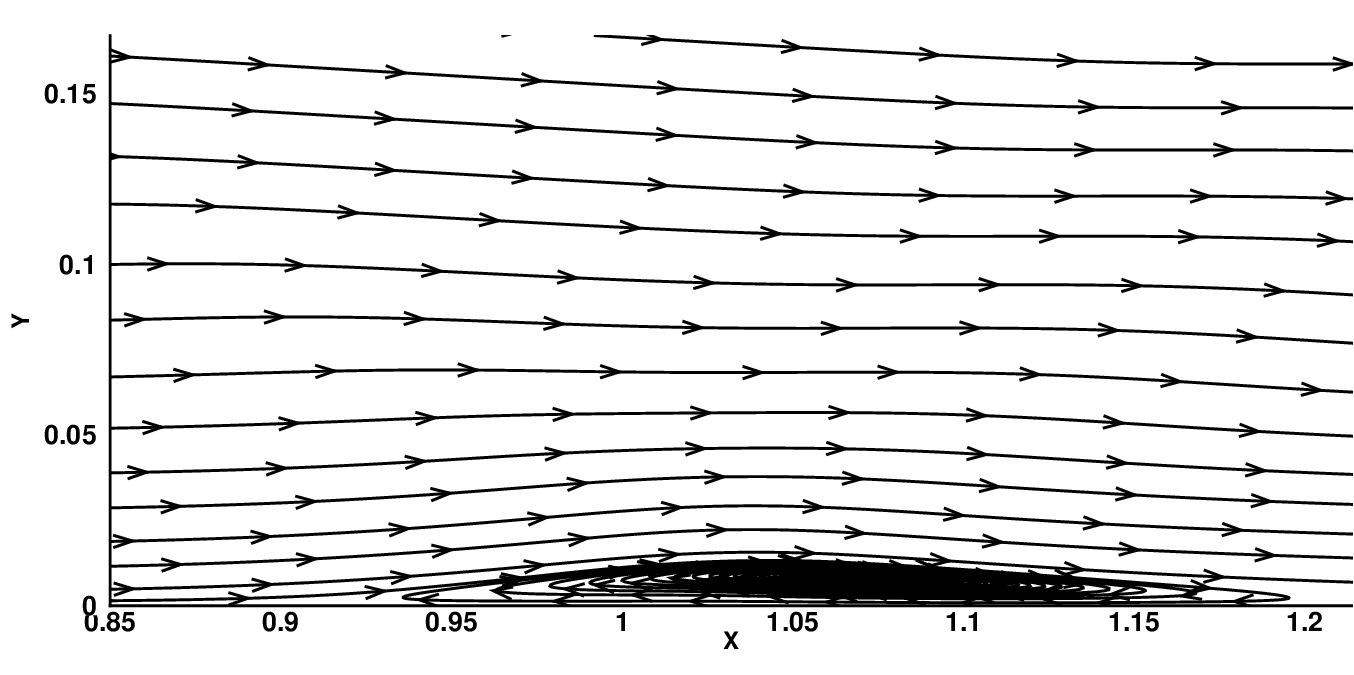}
\end{tabular}
\caption{Test case: Shock wave- boundary layer interaction (140x 120), Left) 2O Pressure contours, Right) Streamlines showing the recirculation zone}
\label{fig:2d_vis_3}
\end{figure}

\begin{figure}
\centering
\begin{tabular}{cc}
\includegraphics[width=0.35\textwidth]{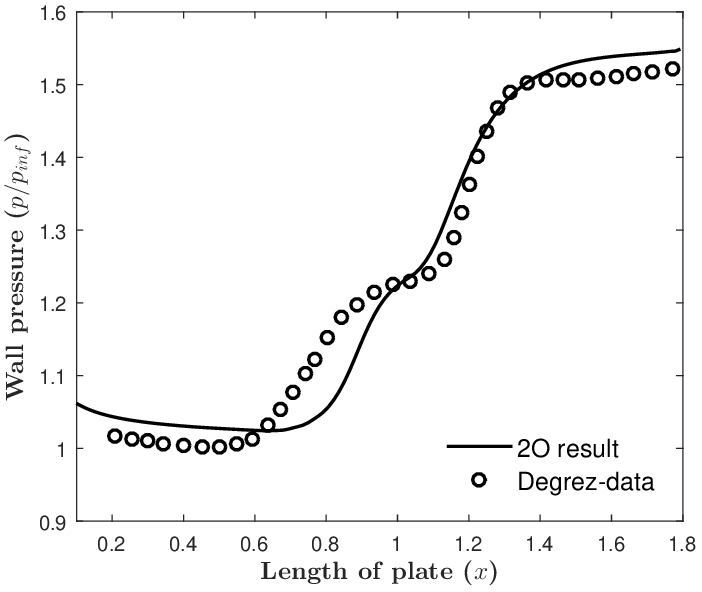} & \includegraphics[width=0.35\textwidth]{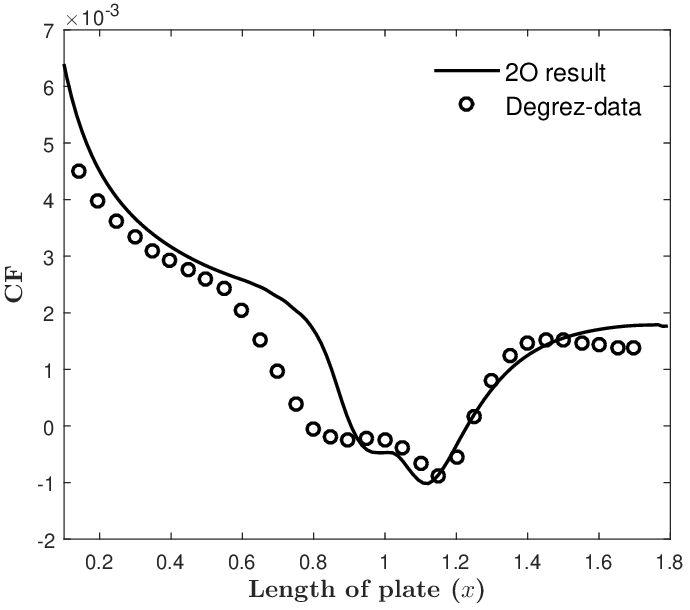}
\end{tabular}
\caption{Test case: Shock wave-boundary layer interaction (140x 120), Left) Wall pressure, Right) Skin friction coefficient along the length of the plate}
\label{fig:2d_vis_4}
\end{figure}
This viscous problem comprises an oblique shock, formed by a supersonic freestream flow with Mach no. $M_{\infty}$= 2.15 and shock angle of $30.8^\circ$, striking a flat plate at the bottom on which a laminar boundary layer is evolving \cite{degrez1987interaction}. It causes the flow to locally separate and then reattach to the surface. The reflected waves consist of compression waves converging into a shock, then expansion waves, followed again by compression waves. The freestream Reynolds no., $Re= 10^{5}$, and Prandtl no., Pr= 0.72, are prescribed. The computational domain taken is $[-0.2,1.8]\times[0,1]$. Supersonic inflow conditions are applied at the left boundary for y$\leq$ 0.765. Whereas post-shock conditions are applied for y$>$0.765 at the left boundary, as well as for the entire top boundary. At the bottom, we use flow symmetry conditions for x $\leq 0.2$, and viscous wall conditions for x $>$0.2. Supersonic outflow conditions are applied at the right boundary. We have discretized the domain into $140\times120$ cells, with constant grid spacing along the x direction and a geometrically stretched grid with a regular $4.5 \%$ increment in grid spacing along the y direction. Additionally, for this test, the limiter functions in the inviscid fluxes are set to 1. Figure \ref{fig:2d_vis_3} shows the pressure contours and streamlines of our second order accurate steady state results. Our results show that the reflected compression and expansion waves as well as the recirculated flow in the separated flow region are properly captured. The wall pressure $\frac{p_{w}}{p_{\infty}}$ and skin friction coefficient $c_{f}$ along the length of the plate are plotted in Figure \ref{fig:2d_vis_4}. Our results match reasonably well with the data from Degrez {\em et al.} \cite{degrez1987interaction}.  

\subsubsection{Supersonic flow over a bump}
\begin{figure}%
\centering
\includegraphics[width=0.8\textwidth]{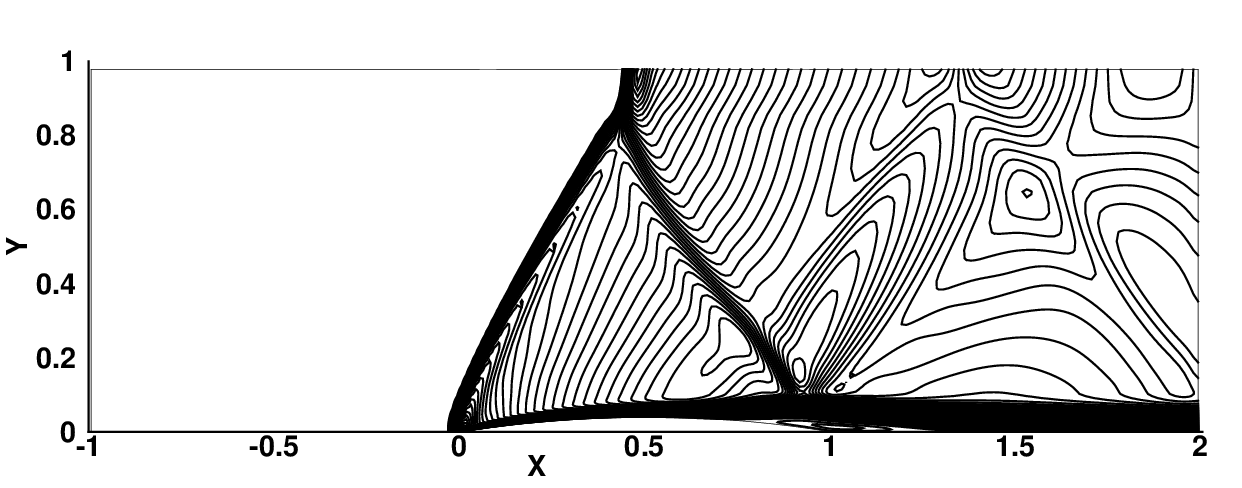}%
\caption{Test case: Supersonic flow over a bump (240x 80), 2nd order, Mach contours}%
\label{fig:2d_vis_5}%
\end{figure}
\begin{figure}%
\centering
\includegraphics[width=0.35\textwidth]{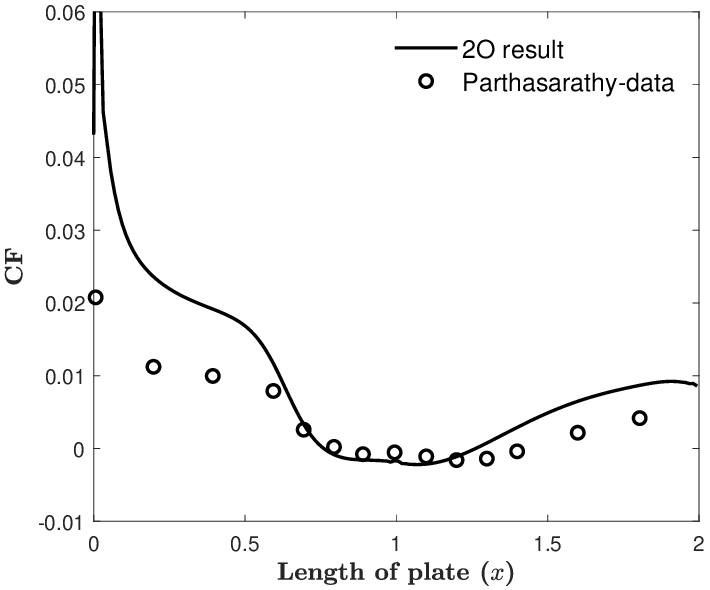}%
\caption{Test case: Supersonic flow over a bump (240x 80), Skin friction coefficient along wall}%
\label{fig:2d_vis_6}%
\end{figure}
In this test case, we consider a Mach 1.4 flow over a circular bump in a channel \cite{parthasarathy1995directional}. The computational domain taken is $[-1,2]\times[0,1]$ with a $4 \%$ circular arc at the bottom from x= 0 to x= 1. The freestream Reynolds no., Re= 8000, and Prandtl no., Pr= 0.72, are prescribed. Supersonic inflow conditions are applied at the left boundary and supersonic outflow conditions are applied at the right boundary. At the bottom, flow symmetry conditions are imposed for x $\leq$0, whereas no-slip (viscous wall) conditions are applied for x$>$0. Flow tangency conditions are applied at the top wall. The domain is discretized into $240\times80$ cells, with constant grid spacing along the x direction and a geometrically stretched grid with a regular $4.5 \%$ increment in grid spacing along the y direction. The interface inviscid fluxes have limiters set to 1 for this test as well. The Mach contours for the second order accurate steady state result (Figure \ref{fig:2d_vis_5}) show an oblique shock forming at the leading edge of the bump. This shock reflects from the top wall and then interacts with the separated flow at the end of the bump and reflects from it. In Figure \ref{fig:2d_vis_6}, the skin friction coefficient is plotted along the length of the bottom wall and compared with the data from Parthasarathy \& Kallenderis \cite{parthasarathy1995directional}. 

\section{Conclusions}
We have developed a new kinetic model for the Euler equations, consisting of two flexible velocities in 1D that satisfy specific positivity preservation conditions. Our asymmetrical model (in terms of normal velocity at the interface) captures a steady shock exactly without requiring a shock sensor, while our more diffusive symmetrical model generates entropic solutions. Using our version of kinetic relative entropy and an additional criterion, we identify smoothly varying flow regions and switch from the asymmetrical model to the symmetrical model in these areas. We have also derived a constraint on the time step which ensures both positivity preservation and numerical stability for the resulting numerical scheme. We have extended our basic scheme to second order accuracy using a flux-limited approach and a higher order Runge-Kutta method. In 2D, we have formulated a novel three-velocity kinetic model, with the velocities aligned to the cell-interface such that the resulting macroscopic normal flux takes a locally one-dimensional form and the grid-aligned steady shocks are captured exactly. Additionally, we have obtained inviscid normal boundary fluxes in flux difference split form at certain boundaries for our kinetic model. Finally, we have used our inviscid scheme along with the viscous fluxes to numerically solve the viscous equations. We have solved benchmark 1D and 2D compressible flow problems to demonstrate the robustness and accuracy of our numerical scheme.  

\section*{CRediT author statement}
\textbf{Shashi Shekhar Roy}: Conceptualization, Methodology, Investigation, Software, Validation, Formal analysis, Writing- Original draft.\\

\textbf{S. V. Raghurama Rao}: Conceptualization, Investigation, Supervision, Resources, Writing- Review \& Editing.

\section*{Declaration of competing interest}
The authors declare that they have no known financial interests or personal relationships with any other people or organizations that could influence the work presented here.

\section*{Data availability}
No data was used for research described in this article.

\section*{Acknowledgments}
This research was carried out as part of Shashi Shekhar Roy's PhD thesis. It did not receive any specific grant from funding agencies in the public, commercial, or not-for-profit sectors.  

\appendix
\section{Positivity condition}
\label{appendix:a0}
We consider the requirement of positivity of the term $\left(\lambda_{p}\right)_{j+\frac{1}{2}}\textbf{U}_{j+1} - \textbf{G}_{j+1}$. That is,
\begin{equation}
\left(\lambda_{p}\right)_{j+\frac{1}{2}}\textbf{U}_{j+1} - \textbf{G}_{j+1}=  \begin{bmatrix} G_{1} & G_{2} & G_{3} \end{bmatrix}^{T} \in \textbf{W}
\label{eq:a0_1}
\end{equation}
Since $\lambda_{p}\geq$0, the condition \eqref{eq:a0_1} can be restated as
\begin{equation}
\textbf{U}_{j+1} - \textbf{G}_{j+1}/\left(\lambda_{p}\right)_{j+\frac{1}{2}}=  \begin{bmatrix} U_{1} & U_{2} & U_{3} \end{bmatrix}^{T} \in \textbf{W}
\label{eq:a0_2}
\end{equation}
Now, the requirement of non-negative density and pressure in \eqref{eq:a0_2} implies that,
\begin{equation}
U_{1}\geq 0 \text{ and } 2 U_{1}U_{3}- U^{2}_{2}\geq 0 \ \left(\text{equivalently, }G_{1}\geq 0 \text{ and } 2 G_{1}G_{3}- G^{2}_{2}\geq 0\right)
\label{eq:a0_3}
\end{equation}
The condition $G_{1}\geq 0$ gives us
\begin{equation}
\rho_{j+1}\left(\lambda_{p}\right)_{j+\frac{1}{2}} -(\rho u)_{j+1}\geq 0 \Rightarrow \left(\lambda_{p}\right)_{j+\frac{1}{2}} \geq u_{j+1}
\label{eq:a0_4}
\end{equation}
Similarly, the condition $2 G_{1}G_{3}- G^{2}_{2}\geq 0$ leads to
\begin{equation}
\begin{split}
&2\left[\rho_{j+1}\left(\lambda_{p}\right)_{j+\frac{1}{2}} -(\rho u)_{j+1}\right]\left[(\rho E)_{j+1}\left(\lambda_{p}\right)_{j+\frac{1}{2}} -\left\{\left(\rho E+ p\right)u\right\}_{j+1}\right] - \left[(\rho u)_{j+1}\left(\lambda_{p}\right)_{j+\frac{1}{2}} - (\rho u^{2}+ p)_{j+1}\right]^{2} \geq 0 \\
&\Rightarrow p_{j+1} \left[ \frac{2}{\gamma -1}\rho_{j+1}\left\{\left(\lambda_{p}\right)_{j+\frac{1}{2}}- u_{j+1}\right\}^{2} -p_{j+1}\right]\geq 0\text{ (on simplifying)}\\
&\Rightarrow \left(\lambda_{p}\right)_{j+\frac{1}{2}} \geq u_{j+1}+ \sqrt{\frac{\gamma -1}{2 \gamma}} a_{j+1}
\end{split}
\label{eq:a0_5}
\end{equation}
We note that the condition \eqref{eq:a0_5} automatically satisfies \eqref{eq:a0_4}. Thus, the positivity condition \eqref{eq:a0_1} leads to a limitation on $\lambda_{p}$ as specified in \eqref{eq:a0_5}.

\section{Kinetic relative entropy $d^{2}$}
\label{appendix:a3} 
We have defined relative entropy\cite{ROY2023106016,SHASHI_HYP} as the kinetic entropy distance $d^{2}$, given by
\begin{equation}
d^{2}= \left\langle \Delta \left(\frac{\partial H(f^{eq})}{\partial f^{eq}}\right)\Delta f^{eq}\right\rangle; \ \Delta=()_{R}- ()_{L}
\label{eq:a3_1}
\end{equation}
Here H is the kinetic entropy function, and $\left\langle \right\rangle$ refers to taking moment. For the classical case with continuous velocity, H refers to \eqref{eq:gkt_7a} without the moments, {\em i.e.}, $H= f^{eq} \ln f^{eq}$. Substituting it in \eqref{eq:a3_1}, we get:
\begin{equation}
d^{2}= \left\langle \Delta ln f^{eq} \Delta f^{eq} \right\rangle= \left\langle ln\frac{f^{eq}_{R}}{f^{eq}_{L}}(f^{eq}_{R}- f^{eq}_{L})\right\rangle
\label{eq:a3_2}
\end{equation}
which is the Kullback-Leibler divergence \cite{kullback1951information}. However, our kinetic model is closer to the discrete case. Hence, we use Bouchut's kinetic entropy function formulated for the discrete velocity Boltzmann equation \cite{bouchut1999construction}. To begin, we rewrite $\textbf{f}^{eq}$ for our kinetic model as follows.
\begin{eqnarray}
\textbf{f}_{i}^{eq}&&=\begin{bmatrix} f^{eq}_{1i} \\ f^{eq}_{2i} \end{bmatrix}= \begin{bmatrix} \frac{-\lambda_{m}}{\lambda_{p}- \lambda_{m}}U_{i} + \frac{1}{\lambda_{p}- \lambda_{m}}G_{i}\\ \frac{\lambda_{p}}{\lambda_{p}- \lambda_{m}}U_{i} - \frac{1}{\lambda_{p}- \lambda_{m}}G_{i}\end{bmatrix} = \begin{bmatrix} \frac{-\lambda_{m}}{\lambda_{p}- \lambda_{m}} \\ \frac{\lambda_{p}}{\lambda_{p}- \lambda_{m}} \end{bmatrix}U_{i} + \begin{bmatrix} \frac{1}{\lambda_{p}- \lambda_{m}} \\ - \frac{1}{\lambda_{p}- \lambda_{m}} \end{bmatrix}G_{i}
\label{eq:a3_3}
\end{eqnarray}
or,
\begin{eqnarray}
\textbf{f}^{eq}&&= \begin{bmatrix} f^{eq}_{11} \\ f^{eq}_{21} \\ f^{eq}_{12} \\ f^{eq}_{22} \\ f^{eq}_{13} \\ f^{eq}_{23} \\ \end{bmatrix}= \begin{bmatrix} \frac{-\lambda_{m}}{\lambda_{p}- \lambda_{m}}&0&0\\ \frac{\lambda_{p}}{\lambda_{p}- \lambda_{m}}&0&0\\ 0&\frac{-\lambda_{m}}{\lambda_{p}- \lambda_{m}} &0 \\ 0& \frac{\lambda_{p}}{\lambda_{p}- \lambda_{m}} &0 \\ 0&0&\frac{-\lambda_{m}}{\lambda_{p}- \lambda_{m}}\\ 0&0&\frac{\lambda_{p}}{\lambda_{p}- \lambda_{m}} \end{bmatrix}\textbf{U}+ \begin{bmatrix} \frac{1}{\lambda_{p}- \lambda_{m}} &0 &0\\ -\frac{1}{\lambda_{p}- \lambda_{m}} &0 &0\\ 0& \frac{1}{\lambda_{p}- \lambda_{m}} &0 \\ 0& - \frac{1}{\lambda_{p}- \lambda_{m}} &0 \\ 0& 0& \frac{1}{\lambda_{p}- \lambda_{m}}\\ 0& 0& - \frac{1}{\lambda_{p}- \lambda_{m}} \end{bmatrix}\textbf{G}\nonumber\\
&&= \boldsymbol{\alpha}_{0}\textbf{U}+ \boldsymbol{\alpha}_{1}\textbf{G}
\label{eq:a3_4}
\end{eqnarray}
The moment relations for $\textbf{f}^{eq}$ can then be rewritten as
\begin{subequations}
\label{eq:a3_5}
\begin{equation}
\left\langle \textbf{f}^{eq} \right\rangle= \textbf{P}\textbf{f}^{eq}=\textbf{U} \Rightarrow \textbf{P}\boldsymbol{\alpha}_{0}= \textbf{I}, \textbf{P}\boldsymbol{\alpha}_{1}= 0
\label{eq:a3_5a}
\end{equation}
\begin{equation}
\left\langle \boldsymbol{\Lambda}\textbf{f}^{eq} \right\rangle= \textbf{P}\boldsymbol{\Lambda}\textbf{f}^{eq}=\textbf{G} \Rightarrow \textbf{P}\boldsymbol{\Lambda}\boldsymbol{\alpha}_{0}= 0, \textbf{P}\boldsymbol{\Lambda}\boldsymbol{\alpha}_{1}= \textbf{I}
\label{eq:a3_5b}
\end{equation}
\end{subequations}
Here, $\textbf{P}= \begin{bmatrix}\textbf{P}_{i} & & \\ & \textbf{P}_{i} & \\ & & \textbf{P}_{i}\end{bmatrix}_{3\times6}$, $\boldsymbol{\Lambda}= \begin{bmatrix}\Lambda & & \\ & \Lambda & \\ & & \Lambda \end{bmatrix}_{6\times6}$ and $\textbf{I}$ is the identity matrix. Thus the moment relations are satisfied as $\boldsymbol{\alpha}_{0}$ and $\boldsymbol{\alpha}_{1}$ satisfy the compatibility conditions in \eqref{eq:a3_5}. We then define the following Bouchut-type kinetic entropy function $H$.
\begin{equation}
H= \boldsymbol{\alpha}_{0}\eta+ \boldsymbol{\alpha}_{1} \psi
\label{eq:a3_6}
\end{equation}
where ($\eta, \psi$) are the macroscopic entropy-entropy flux pairs. For Euler equations, $\eta= -\rho s$  and $\psi= -\rho u s$, where $s$= $c_{v} \ln \frac{p}{\rho^{\gamma}} + \textrm{constant}$. It can then easily be shown that $\left\langle H\right\rangle= \eta$ and $ \left\langle \Lambda H\right\rangle= \psi$. On substituting for $H$, the relative entropy becomes
\begin{equation}
d^{2}= \left\langle \Delta \frac{\partial H(\textbf{f}^{eq})}{\partial \textbf{f}^{eq}}. \Delta \textbf{f}^{eq}\right\rangle= \left\langle \frac{\partial \Delta H(\textbf{f}^{eq})}{\partial \textbf{f}^{eq}}. \Delta \textbf{f}^{eq}\right\rangle = \left\langle \frac{\partial \Delta \left(\boldsymbol{\alpha}_{0}\eta+ \boldsymbol{\alpha}_{1} \psi\right)}{\partial \textbf{f}^{eq}}. \Delta \textbf{f}^{eq}\right\rangle
\label{eq:a3_7}
\end{equation}
Next, we use the relation: $\Delta \left(x y\right)= \Delta x\overline{y}+ \overline{x}\Delta y$, where $\overline{x}$ refers to the average of a quantity $x$. Thus,
\begin{eqnarray}
	\Delta H &&= \Delta\left(\boldsymbol{\alpha}_{0}\eta\right)+ \Delta\left(\boldsymbol{\alpha}_{1} \psi\right) \nonumber\\
	         &&= \Delta\boldsymbol{\alpha}_{0} \overline{\eta}+ \boldsymbol{\overline{\alpha}}_{0}\Delta\eta+ \Delta\boldsymbol{\alpha}_{1} \overline{\psi}+ \boldsymbol{\overline{\alpha}}_{1}\Delta\psi
	\label{eq:a3_8}
	\end{eqnarray}
	We note that $\boldsymbol{\overline{\alpha}}_{0}$ and $\boldsymbol{\overline{\alpha}}_{1}$ satisfy the compatibility conditions \ref{eq:a3_5a} as well. Now,
	\begin{equation}
	\frac{\partial \Delta H(\textbf{f}^{eq})}{\partial \textbf{f}^{eq}}= \Delta\boldsymbol{\alpha}_{0}\frac{\partial \overline{\eta}}{\partial \textbf{f}^{eq}}+ \boldsymbol{\overline{\alpha}}_{0}\frac{\partial \Delta\eta}{\partial \textbf{f}^{eq}}+ \Delta\boldsymbol{\alpha}_{1}\frac{\partial \overline{\psi}}{\partial \textbf{f}^{eq}}+ \boldsymbol{\overline{\alpha}}_{1}\frac{\partial \Delta\psi}{\partial \textbf{f}^{eq}}
	\label{eq:a3_9}
	\end{equation}
	\begin{equation}
	\frac{\partial \Delta H(\textbf{f}^{eq})}{\partial \textbf{f}^{eq}}. \Delta \textbf{f}^{eq}= \Delta\boldsymbol{\alpha}_{0}\frac{\partial \overline{\eta}}{\partial \textbf{f}^{eq}}. \Delta \textbf{f}^{eq}+ \boldsymbol{\overline{\alpha}}_{0}\frac{\partial \Delta\eta}{\partial \textbf{f}^{eq}}. \Delta \textbf{f}^{eq}+ \Delta\boldsymbol{\alpha}_{1}\frac{\partial \overline{\psi}}{\partial \textbf{f}^{eq}}. \Delta \textbf{f}^{eq}+ \boldsymbol{\overline{\alpha}}_{1}\frac{\partial \Delta\psi}{\partial \textbf{f}^{eq}}. \Delta \textbf{f}^{eq}
	\label{eq:a3_10}
	\end{equation}
	Then,
\begin{eqnarray}
	d^{2}&&= \left\langle \frac{\partial \Delta H(\textbf{f}^{eq})}{\partial \textbf{f}^{eq}}. \Delta \textbf{f}^{eq}\right\rangle= \textbf{P}\frac{\partial \Delta H(\textbf{f}^{eq})}{\partial \textbf{f}^{eq}}. \Delta \textbf{f}^{eq} \nonumber\\
	     &&= \textbf{P}\Delta\boldsymbol{\alpha}_{0}\frac{\partial \overline{\eta}}{\partial \textbf{f}^{eq}}. \Delta \textbf{f}^{eq}+ \textbf{P}\boldsymbol{\overline{\alpha}}_{0}\frac{\partial \Delta\eta}{\partial \textbf{f}^{eq}}. \Delta \textbf{f}^{eq}+ \textbf{P}\Delta\boldsymbol{\alpha}_{1}\frac{\partial \overline{\psi}}{\partial \textbf{f}^{eq}}. \Delta \textbf{f}^{eq}+ \textbf{P}\boldsymbol{\overline{\alpha}}_{1}\frac{\partial \Delta\psi}{\partial \textbf{f}^{eq}}. \Delta \textbf{f}^{eq}\nonumber\\
			&&= \Delta \left(\textbf{P}\boldsymbol{\alpha}_{0}\right)\frac{\partial \overline{\eta}}{\partial \textbf{f}^{eq}}. \Delta \textbf{f}^{eq}+ \textbf{P}\boldsymbol{\overline{\alpha}}_{0}\frac{\partial \Delta\eta}{\partial \textbf{f}^{eq}}. \Delta \textbf{f}^{eq}+ \Delta \left(\textbf{P}\boldsymbol{\alpha}_{1}\right)\frac{\partial \overline{\psi}}{\partial \textbf{f}^{eq}}. \Delta \textbf{f}^{eq}+ \textbf{P}\boldsymbol{\overline{\alpha}}_{1}\frac{\partial \Delta\psi}{\partial \textbf{f}^{eq}}. \Delta \textbf{f}^{eq}\nonumber\\
			&&=\frac{\partial \Delta\eta}{\partial \textbf{f}^{eq}}. \Delta \textbf{f}^{eq} \ \left(\text{since }\textbf{P}\boldsymbol{\alpha}_{0}= \textbf{P}\boldsymbol{\overline{\alpha}}_{0}= \textbf{I}, \Delta \left(\textbf{P}\boldsymbol{\alpha}_{0}\right)= 0, \textbf{P}\boldsymbol{\alpha}_{1}= \textbf{P}\boldsymbol{\overline{\alpha}}_{1}= \Delta \left(\textbf{P}\boldsymbol{\alpha}_{1}\right)= 0 \right) \nonumber\\
			&&= \Delta \frac{\partial \eta}{\partial \textbf{f}^{eq}}. \Delta \textbf{f}^{eq}
			\label{eq:a3_11}
	\end{eqnarray}
	
\begin{eqnarray}
d^{2}&&=\Delta\begin{bmatrix}\frac{\partial \eta}{\partial U_{1}}\frac{\partial U_{1}}{\partial f^{eq}_{11}}+ \frac{\partial \eta}{\partial U_{2}}\frac{\partial U_{2}}{\partial f^{eq}_{11}}+\frac{\partial \eta}{\partial U_{3}}\frac{\partial U_{3}}{\partial f^{eq}_{11}}\\
\frac{\partial \eta}{\partial U_{1}}\frac{\partial U_{1}}{\partial f^{eq}_{21}}+ \frac{\partial \eta}{\partial U_{2}}\frac{\partial U_{2}}{\partial f^{eq}_{21}}+\frac{\partial \eta}{\partial U_{3}}\frac{\partial U_{3}}{\partial f^{eq}_{21}}\\
\frac{\partial \eta}{\partial U_{1}}\frac{\partial U_{1}}{\partial f^{eq}_{12}}+ \frac{\partial \eta}{\partial U_{2}}\frac{\partial U_{2}}{\partial f^{eq}_{12}}+\frac{\partial \eta}{\partial U_{3}}\frac{\partial U_{3}}{\partial f^{eq}_{12}}\\
\frac{\partial \eta}{\partial U_{1}}\frac{\partial U_{1}}{\partial f^{eq}_{22}}+ \frac{\partial \eta}{\partial U_{2}}\frac{\partial U_{2}}{\partial f^{eq}_{22}}+\frac{\partial \eta}{\partial U_{3}}\frac{\partial U_{3}}{\partial f^{eq}_{22}}\\
\frac{\partial \eta}{\partial U_{1}}\frac{\partial U_{1}}{\partial f^{eq}_{13}}+ \frac{\partial \eta}{\partial U_{2}}\frac{\partial U_{2}}{\partial f^{eq}_{13}}+\frac{\partial \eta}{\partial U_{3}}\frac{\partial U_{3}}{\partial f^{eq}_{13}}\\
\frac{\partial \eta}{\partial U_{1}}\frac{\partial U_{1}}{\partial f^{eq}_{23}}+ \frac{\partial \eta}{\partial U_{2}}\frac{\partial U_{2}}{\partial f^{eq}_{23}}+\frac{\partial \eta}{\partial U_{3}}\frac{\partial U_{3}}{\partial f^{eq}_{23}}\end{bmatrix}. \Delta\begin{bmatrix} f^{eq}_{11} \\ f^{eq}_{21} \\ f^{eq}_{12} \\ f^{eq}_{22} \\ f^{eq}_{13} \\ f^{eq}_{23} \end{bmatrix} =\Delta\begin{bmatrix}\frac{\partial \eta}{\partial U_{1}} \\\frac{\partial \eta}{\partial U_{1}} \\
\frac{\partial \eta}{\partial U_{2}} \\\frac{\partial \eta}{\partial U_{2}} \\ \frac{\partial \eta}{\partial U_{3}} \\\frac{\partial \eta}{\partial U_{3}} \end{bmatrix}. \Delta\begin{bmatrix} f^{eq}_{11} \\ f^{eq}_{21} \\ f^{eq}_{12} \\ f^{eq}_{22} \\ f^{eq}_{13} \\ f^{eq}_{23} \end{bmatrix} \nonumber\\
&&=\Delta \frac{\partial \eta}{\partial U_{1}} \Delta(f^{eq}_{11}+f^{eq}_{21} )+\Delta\frac{\partial \eta}{\partial U_{2}} \Delta(f^{eq}_{12}+f^{eq}_{22})+\Delta\frac{\partial \eta}{\partial U_{3}} \Delta(f^{eq}_{13}+f^{eq}_{23})\nonumber\\
&&=  \Delta \frac{\partial \eta}{\partial U_{1}} \Delta U_{1}+\Delta \frac{\partial \eta}{\partial U_{2}} \Delta U_{2}+\Delta \frac{\partial \eta}{\partial U_{3}} \Delta U_{3}\nonumber\\
&&= \left(\Delta \frac{\partial \eta}{\partial \textbf{U}}\right) \cdot \Delta \textbf{U}\nonumber\\
&&=\Delta \left\{R\left(\frac{\gamma -s/c_{v}}{\gamma -1} - \frac{\rho u^{2}}{2p}\right)\right\} \Delta \left(\rho\right)+ \Delta \left(R\frac{\rho u}{p}\right) \Delta \left(\rho u\right)+ \Delta \left(-R\frac{\rho}{p}\right) \Delta \left(\rho E\right)
\label{eq:a3_12}
\end{eqnarray}
Here, $R$ is the gas constant.

\section{Linear Stability Analysis}
\label{appendix:a1}
A von Neumann linear stability analysis of the advective part of the 1D Boltzmann equations is done for first order accuracy. We consider the scalar linear advection equation as the macroscopic governing equation for simplicity. The advective part of the 1D Boltzmann equation can then be written as
	
	\begin{equation}
	\frac{\partial}{\partial t} \begin{bmatrix}f_{1} \\ f_{2}\end{bmatrix}+ \frac{\partial}{\partial x} \left\{ \begin{bmatrix} \lambda_{p} &0 \\ 0& \lambda_{m}\end{bmatrix} \begin{bmatrix}f_{1} \\ f_{2}\end{bmatrix} \right\}= 0
	\label{eq:a1_1}
	\end{equation}
	
	We discretize \eqref{eq:a1_1} for first order accuracy, while noting that for modeling the macroscopic linear advection equation , $\lambda_{p}$ and $\lambda_{m}$ are taken as constants.
	\begin{eqnarray}
	\frac{1}{\Delta t} \left\{\begin{bmatrix} (f_{1})^{n+1}_{j} \\ (f_{2})^{n+1}_{j} \end{bmatrix} - \begin{bmatrix} (f_{1})^{n}_{j} \\ (f_{2})^{n}_{j} \end{bmatrix} \right\} + \frac{1}{\Delta x} &&\left\{ \begin{bmatrix} \frac{1}{2}\left(\lambda_{p}(f_{1})^{n}_{j}+ \lambda_{p}(f_{1})^{n}_{j+1}\right)- \frac{\lambda_{p}}{2}\left((f_{1})^{n}_{j+1}- (f_{1})^{n}_{j}\right) \\ \frac{1}{2}\left(\lambda_{m} (f_{2})^{n}_{j}+\lambda_{m} (f_{2})^{n}_{j+1}\right)+ \frac{\lambda_{m}}{2}\left((f_{2})^{n}_{j+1}- (f_{2})^{n}_{j}\right) \end{bmatrix}- \right. \nonumber\\
	&& \left. \begin{bmatrix} \frac{1}{2}\left(\lambda_{p}(f_{1})^{n}_{j-1}+ \lambda_{p}(f_{1})^{n}_{j}\right)- \frac{\lambda_{p}}{2}\left((f_{1})^{n}_{j}- (f_{1})^{n}_{j-1}\right) \\ \frac{1}{2}\left(\lambda_{m} (f_{2})^{n}_{j-1}+\lambda_{m} (f_{2})^{n}_{j}\right)+ \frac{\lambda_{m}}{2}\left((f_{2})^{n}_{j}- (f_{2})^{n}_{j-1}\right) \end{bmatrix}\right\} = 0
	\label{eq:a1_2}
	\end{eqnarray}
	or,
	\begin{equation}
	\frac{1}{\Delta t} \left\{\begin{bmatrix} (f_{1})^{n+1}_{j} \\ (f_{2})^{n+1}_{j} \end{bmatrix} - \begin{bmatrix} (f_{1})^{n}_{j} \\ (f_{2})^{n}_{j} \end{bmatrix} \right\} + \frac{1}{\Delta x}\left\{ \begin{bmatrix} \lambda_{p}(f_{1})^{n}_{j} \\ \lambda_{m} (f_{2})^{n}_{j+1} \end{bmatrix}- \begin{bmatrix} \lambda_{p}(f_{1})^{n}_{j-1} \\ \lambda_{m} (f_{2})^{n}_{j} \end{bmatrix}\right\} = 0
	\label{eq:a1_3}
	\end{equation}
Thus, for the linear case, the flux difference formulation is equivalent to an upwind method with Courant-type splitting. Next, we introduce Fourier expansions for $f_{1}$ and $f_{2}$ as follows.  
	\begin{eqnarray}
	(f_{1})^{n}_{j}&=& (\widetilde{f}_{1})^{n}e^{Ij\theta}\nonumber\\
	(f_{2})^{n}_{j}&=& (\widetilde{f}_{2})^{n}e^{Ij\theta}
	\label{eq:a1_4}
	\end{eqnarray}
	where I=$\sqrt{-1}$. Substituting \eqref{eq:a1_4} in \eqref{eq:a1_3} and simplifying, we get
	\begin{equation}
	\begin{bmatrix} (\widetilde{f}_{1})^{n+1} \\ (\widetilde{f}_{2})^{n+1} \end{bmatrix} = \begin{bmatrix} \left\{1 + \frac{\lambda_{p} \Delta t}{\Delta x}(e^{-I\theta}-1)\right\} & 0\\ 0& \left\{1 - \frac{\lambda_{m} \Delta t}{\Delta x}(e^{I\theta}-1)\right\} \end{bmatrix} \begin{bmatrix} (\widetilde{f}_{1})^{n} \\ (\widetilde{f}_{2})^{n} \end{bmatrix}
	\label{eq:a1_5}
	\end{equation}
	or
	\begin{equation}
	\begin{bmatrix} \widetilde{f}_{1} \\ \widetilde{f}_{2} \end{bmatrix}^{n+1}= \textbf{A} \begin{bmatrix} \widetilde{f}_{1} \\ \widetilde{f}_{2} \end{bmatrix}^{n}
	\label{eq:a1_6}
	\end{equation}
	Now, for the above scheme to be stable, {\em i.e.}, for the amplification factor to be less than or equal to 1, the absolute value of the eigenvalues of $\textbf{A}$ should be less than or equal to 1. Therefore, 
	\begin{equation}
	|1 + \frac{\lambda_{p} \Delta t}{\Delta x}(e^{-I\theta}-1)| \leq 1,\text{ and } |1 - \frac{\lambda_{m} \Delta t}{\Delta x}(e^{I\theta}-1)| \leq 1
	\label{eq:a1_7}
	\end{equation}
	From \eqref{eq:a1_7} we get,
	\begin{equation}
	\frac{\lambda_{p} \Delta t}{\Delta x} \leq 1,\text{ and } -\frac{\lambda_{m} \Delta t}{\Delta x} \leq 1
	\label{eq:a1_8}
	\end{equation}
	or,
	\begin{equation}
	\frac{max (\lambda_{p}, -\lambda_{m} ) \Delta t}{\Delta x} \leq 1
	\label{eq:a1_9}
	\end{equation}
Thus, for Linear Advection Equation, Model 1 with $\lambda_{p}$ and $\lambda_{m}$ is linearly stable if Model 2 with scalar numerical diffusion $\lambda= max (\lambda_{p}, -\lambda_{m})$ is linearly stable.

\section{Kinetic Boundary Conditions for 2D Euler Equations}
\label{appendix:a2}
In this section, we obtain simplified expressions for inviscid normal fluxes at different types of boundaries for our 2D kinetic model. We start this effort by describing our kinetic model in a continuous molecular velocity framework. We model only in the velocity $v$ space, and introduce the following truncated equilibrium distributions by integrating w.r.t. the internal energy variable $I$.  
\begin{equation}
\breve{f}^{eq} = \int^{\infty}_{0} \ f^{eq} dI, \ \ \hat{f}^{eq}_{i} = \Psi_{i} \breve{f}^{eq}
\label{eq:1d_euler_1}
\end{equation} 
Then, the moment relations in 2-D become
\begin{subequations}
\label{eq:1d_euler_2}
\begin{equation}
U_{i} = \int_{-\infty}^{\infty} dv_{1} \int_{-\infty}^{\infty} dv_{2} \ \Psi_{i} \breve{f}^{eq} = \int_{-\infty}^{\infty} dv_{1} \int_{-\infty}^{\infty} dv_{2} \ \hat{f}^{eq}_{i} = \langle \hat{f}^{eq}_{i} \rangle
\end{equation}
\begin{equation}
G_{1i} = \int_{-\infty}^{\infty} v_{1}dv_{1} \int_{-\infty}^{\infty} dv_{2} \Psi_{i} \breve{f}^{eq} = \int_{-\infty}^{\infty} v_{1}dv_{1} \int_{-\infty}^{\infty} dv_{2} \hat{f}^{eq}_{i} = \langle v_{1} \hat{f}^{eq}_{i} \rangle
\end{equation}
\begin{equation}
G_{2i} = \int_{-\infty}^{\infty} dv_{1} \int_{-\infty}^{\infty} v_{2}dv_{2} \Psi_{i} \breve{f}^{eq} = \int_{-\infty}^{\infty} dv_{1} \int_{-\infty}^{\infty} v_{2}dv_{2} \hat{f}^{eq}_{i} = \langle v_{2} \hat{f}^{eq}_{i} \rangle
\end{equation}
\end{subequations}
For our kinetic model, the 2D equilibrium distribution function is given by
\begin{equation}
\hat{f}^{eq}_{i}= f^{eq}_{1i}\delta\left(v_{\perp}-\lambda_{p,\perp}\right)\delta\left(v_{\parallel}\right)+ f^{eq}_{2i}\delta\left(v_{\perp}-\lambda_{m,\perp}\right)\delta\left(v_{\parallel}-\lambda_{\parallel}\right)+ f^{eq}_{3i}\delta\left(v_{\perp}-\lambda_{m,\perp}\right)\delta\left(v_{\parallel}+\lambda_{\parallel}\right)
\label{eq:a2_3}
\end{equation}
The individual weights ($f^{eq}_{1i}$, $f^{eq}_{2i}$ and $f^{eq}_{3i}$) each satisfy a Boltzmann equation while being advected by their corresponding velocity. Next, we obtain normal fluxes at certain boundaries for our kinetic model, assuming that the unit normal vector at the boundary surfaces points outward.

\textbf{\textit{Normal flux at wall:}} We utilize kinetic theory to obtain normal fluxes at a wall for the Euler equations. We begin with the use of the equilibrium distribution $f^{eq}$ and not just the truncated distributions $\hat{f}^{eq}_{i}$. We define the normal fluxes $G_{\perp i}$ at a boundary $b$ between the interior state $int$ and exterior state $ext$ in a flux difference split form in the following way.
\begin{equation}
(G_{\perp i})_{b}= \frac{1}{2}\left[(G_{\perp i})_{int}+ (G_{\perp i})_{ext}\right]- \frac{1}{2}\left[(\Delta G^{+}_{\perp i})_{b}- (\Delta G^{-}_{\perp i})_{b}\right]
\label{eq:a2_1}
\end{equation}
where
\begin{subequations}
\label{eq:a2_2}
\begin{equation}
(G_{\perp i})_{int \ (or \ ext)}= \int^{\infty}_{-\infty}v_{\perp}dv_{\perp}\int^{\infty}_{-\infty}dv_{\parallel}\underbrace{\int^{\infty}_{0}dI \ \Psi_{i}f^{eq}_{int \ (or \ ext)}(v_{\perp},v_{\parallel})}_{\hat{f}^{eq}_{i}}
\end{equation}
\begin{equation}
(\Delta G^{+}_{\perp i})_{b}= \int^{\infty}_{0}v_{\perp}dv_{\perp}\int^{\infty}_{-\infty}dv_{\parallel}\int^{\infty}_{0}dI \ \Psi_{i}\left\{f^{eq}_{ext}(v_{\perp},v_{\parallel})- f^{eq}_{int}(v_{\perp},v_{\parallel})\right\}
\end{equation}
\begin{equation}
(\Delta G^{-}_{\perp i})_{b}= \int^{0}_{-\infty}v_{\perp}dv_{\perp}\int^{\infty}_{-\infty}dv_{\parallel}\int^{\infty}_{0}dI \ \Psi_{i}\left\{f^{eq}_{ext}(v_{\perp},v_{\parallel})- f^{eq}_{int}(v_{\perp},v_{\parallel})\right\}
\end{equation}
\end{subequations}
For flow tangency at the wall, we utilize the specular reflection model of kinetic theory of gases to define $f^{eq}_{ext}$ as,
\begin{equation}
f^{eq}_{ext}= f^{eq}_{int}(-v_{\perp},v_{\parallel})
\label{eq:a2_4}
\end{equation}
Then, we have
\begin{eqnarray}
(G_{\perp 1})_{ext} &&= \int^{\infty}_{-\infty}v_{\perp}dv_{\perp}\int^{\infty}_{-\infty}dv_{\parallel}\int^{\infty}_{0}dI \ f^{eq}_{int}(-v_{\perp},v_{\parallel}) \nonumber\\
&&= -\int^{\infty}_{-\infty}(-v_{\perp})d(-v_{\perp})\int^{\infty}_{-\infty}dv_{\parallel}\int^{\infty}_{0}dI \ f^{eq}_{int}(-v_{\perp},v_{\parallel}) \nonumber\\
&&= -(G_{\perp 1})_{int}
\label{eq:a2_5}
\end{eqnarray}
\begin{eqnarray}
(G_{\perp 2})_{ext} &&= \int^{\infty}_{-\infty}v_{\perp}dv_{\perp}\int^{\infty}_{-\infty}dv_{\parallel}\int^{\infty}_{0}dI \ v_{1} f^{eq}_{int}(-v_{\perp},v_{\parallel}) \nonumber\\
&&= \int^{\infty}_{-\infty}v_{\perp}dv_{\perp}\int^{\infty}_{-\infty}dv_{\parallel}\int^{\infty}_{0}dI \ \left(v_{\perp}n_{1}- v_{\parallel}n_{2}\right) f^{eq}_{int}(-v_{\perp},v_{\parallel}) \nonumber\\
&&=\int^{\infty}_{-\infty}v_{\perp}dv_{\perp}\int^{\infty}_{-\infty}dv_{\parallel}\int^{\infty}_{0}dI \ \left(-v_{\perp}n_{1}- v_{\parallel}n_{2}\right) f^{eq}_{int}(-v_{\perp},v_{\parallel})+ \nonumber\\
&& 2n_{1}\int^{\infty}_{-\infty}dv_{\perp}\int^{\infty}_{-\infty}dv_{\parallel}\int^{\infty}_{0}dI \ v^{2}_{\perp} f^{eq}_{int}(-v_{\perp},v_{\parallel}) \nonumber\\
&&= -(G_{\perp 2})_{int}+ 2n_{1}\int^{\infty}_{-\infty}dv_{\perp}\int^{\infty}_{-\infty}dv_{\parallel}\int^{\infty}_{0}dI \ v^{2}_{\perp} f^{eq}_{int}(-v_{\perp},v_{\parallel})
\label{eq:a2_6}
\end{eqnarray}
\begin{equation}
\text{Similarly, }(G_{\perp 3})_{ext}= -(G_{\perp 3})_{int}+ 2n_{2}\int^{\infty}_{-\infty}dv_{\perp}\int^{\infty}_{-\infty}dv_{\parallel}\int^{\infty}_{0}dI \ v^{2}_{\perp} f^{eq}_{int}(-v_{\perp},v_{\parallel})
\label{eq:a2_7}
\end{equation}
\begin{eqnarray}
(G_{\perp 4})_{ext} &&= \int^{\infty}_{-\infty}v_{\perp}dv_{\perp}\int^{\infty}_{-\infty}dv_{\parallel}\int^{\infty}_{0}dI \ \left[I_{0}+ \frac{v^{2}_{\perp}+ v^{2}_{\parallel}}{2}\right]f^{eq}_{int}(-v_{\perp},v_{\parallel}) \nonumber\\
&&= -\int^{\infty}_{-\infty}(-v_{\perp})d(-v_{\perp})\int^{\infty}_{-\infty}dv_{\parallel}\int^{\infty}_{0}dI \ \left[I_{0}+ \frac{(-v_{\perp})^{2}+ v^{2}_{\parallel}}{2}\right]f^{eq}_{int}(-v_{\perp},v_{\parallel}) \nonumber\\
&&= -(G_{\perp 4})_{int}
\label{eq:a2_8}
\end{eqnarray}
The flux differences are evaluated next.
\begin{eqnarray}
(\Delta G^{+}_{\perp 1})_{b}&&= \int^{\infty}_{0}v_{\perp}dv_{\perp}\int^{\infty}_{-\infty}dv_{\parallel}\int^{\infty}_{0}dI \ \left\{f^{eq}_{int}(-v_{\perp},v_{\parallel})- f^{eq}_{int}(v_{\perp},v_{\parallel})\right\} \nonumber\\
&&= -\lambda_{m,\perp}(f^{eq}_{21}+ f^{eq}_{31})- \lambda_{p,\perp}f^{eq}_{11}
\label{eq:a2_9}
\end{eqnarray}
\begin{eqnarray}
(\Delta G^{+}_{\perp 2})_{b}&&= \int^{\infty}_{0}v_{\perp}dv_{\perp}\int^{\infty}_{-\infty}dv_{\parallel}\int^{\infty}_{0}dI \ \left\{v_{1}f^{eq}_{int}(-v_{\perp},v_{\parallel})- v_{1}f^{eq}_{int}(v_{\perp},v_{\parallel})\right\} \nonumber\\
&&=\int^{\infty}_{0}v_{\perp}dv_{\perp}\int^{\infty}_{-\infty}dv_{\parallel}\int^{\infty}_{0}dI \ \left\{\left(-v_{\perp}n_{1}- v_{\parallel}n_{2}+ 2v_{\perp}n_{1}\right)f^{eq}_{int}(-v_{\perp},v_{\parallel})- v_{1}f^{eq}_{int}(v_{\perp},v_{\parallel})\right\} \nonumber\\
&&= -\lambda_{m,\perp}(f^{eq}_{22}+ f^{eq}_{32})- \lambda_{p,\perp}f^{eq}_{12}+ 2n_{1}\int^{\infty}_{0}dv_{\perp}\int^{\infty}_{-\infty}dv_{\parallel}\int^{\infty}_{0}dI \ v^{2}_{\perp} f^{eq}_{int}(-v_{\perp},v_{\parallel})
\label{eq:a2_10}
\end{eqnarray}
\begin{equation}
(\Delta G^{+}_{\perp 3})_{b}= -\lambda_{m,\perp}(f^{eq}_{23}+ f^{eq}_{33})- \lambda_{p,\perp}f^{eq}_{13}+ 2n_{2}\int^{\infty}_{0}dv_{\perp}\int^{\infty}_{-\infty}dv_{\parallel}\int^{\infty}_{0}dI \ v^{2}_{\perp} f^{eq}_{int}(-v_{\perp},v_{\parallel})
\label{eq:a2_11}
\end{equation}
\begin{eqnarray}
(\Delta G^{+}_{\perp 4})_{b}&&= \int^{\infty}_{0}v_{\perp}dv_{\perp}\int^{\infty}_{-\infty}dv_{\parallel}\int^{\infty}_{0}dI \ \left[I_{0}+ \frac{v^{2}_{\perp}+ v^{2}_{\parallel}}{2}\right]\left\{f^{eq}_{int}(-v_{\perp},v_{\parallel})- f^{eq}_{int}(v_{\perp},v_{\parallel})\right\} \nonumber\\
&&= -\lambda_{m,\perp}(f^{eq}_{24}+ f^{eq}_{34})- \lambda_{p,\perp}f^{eq}_{14}
\label{eq:a2_12}
\end{eqnarray}
Similarly,
\begin{equation}
(\Delta G^{-}_{\perp 1})_{b}= - \lambda_{p,\perp}f^{eq}_{11}- \lambda_{m,\perp}(f^{eq}_{21}+ f^{eq}_{31})
\label{eq:a2_13}
\end{equation}
\begin{equation}
(\Delta G^{-}_{\perp 2})_{b}= - \lambda_{p,\perp}f^{eq}_{12}- \lambda_{m,\perp}(f^{eq}_{22}+ f^{eq}_{32})+ 2n_{1}\int^{0}_{-\infty}dv_{\perp}\int^{\infty}_{-\infty}dv_{\parallel}\int^{\infty}_{0}dI \ v^{2}_{\perp} f^{eq}_{int}(-v_{\perp},v_{\parallel})
\label{eq:a2_14}
\end{equation}
\begin{equation}
(\Delta G^{-}_{\perp 3})_{b}= - \lambda_{p,\perp}f^{eq}_{13}- \lambda_{m,\perp}(f^{eq}_{23}+ f^{eq}_{33})+ 2n_{2}\int^{0}_{-\infty}dv_{\perp}\int^{\infty}_{-\infty}dv_{\parallel}\int^{\infty}_{0}dI \ v^{2}_{\perp} f^{eq}_{int}(-v_{\perp},v_{\parallel})
\label{eq:a2_15}
\end{equation}
\begin{equation}
(\Delta G^{-}_{\perp 4})_{b}= - \lambda_{p,\perp}f^{eq}_{14}- \lambda_{m,\perp}(f^{eq}_{24}+ f^{eq}_{34})
\label{eq:a2_16}
\end{equation}
Substituting the obtained expressions for $(G_{\perp i})_{ext}$ and $(\Delta G^{\pm}_{\perp i})_{b}$ into Equation \eqref{eq:a2_1} and simplifying, we get
\begin{equation}
(G_{\perp 1})_{b}= 0
\label{eq:a2_17}
\end{equation}
\begin{eqnarray}
(G_{\perp 2})_{b} &&= 2n_{1}\int^{0}_{-\infty}dv_{\perp}\int^{\infty}_{-\infty}dv_{\parallel}\int^{\infty}_{0}dI \ v^{2}_{\perp} f^{eq}_{int}(-v_{\perp},v_{\parallel}) \nonumber\\
&&= -2n^{2}_{1}\int^{0}_{-\infty}v_{\perp}dv_{\perp}\int^{\infty}_{-\infty}dv_{\parallel}\underbrace{\int^{\infty}_{0}dI \ \left(-v_{\perp}n_{1}- v_{\parallel}n_{2}\right) f^{eq}_{int}(-v_{\perp},v_{\parallel})}_{\hat{f}^{eq}_{2}(-v_{\perp},v_{\parallel})}-\nonumber\\
&& 2n_{1}n_{2}\int^{0}_{-\infty}v_{\perp}dv_{\perp}\int^{\infty}_{-\infty}dv_{\parallel}\underbrace{\int^{\infty}_{0}dI \ \left(-v_{\perp}n_{2}+ v_{\parallel}n_{1}\right) f^{eq}_{int}(-v_{\perp},v_{\parallel})}_{\hat{f}^{eq}_{3}(-v_{\perp},v_{\parallel})} \nonumber\\
&&= -2n^{2}_{1}\left(-\lambda_{p,\perp}f^{eq}_{12}\right) -2n_{1}n_{2}\left(-\lambda_{p,\perp}f^{eq}_{13}\right) \nonumber\\
&&= 2n_{1}\left(\frac{\lambda_{p,\perp}}{\lambda_{p,\perp}- \lambda_{m,\perp}}\right)\left(-\lambda_{m,\perp} \rho u_{\perp}+ \rho u^{2}_{\perp}+ p\right) \nonumber\\
&&= 2n_{1}\left(\frac{\lambda_{p,\perp}}{\lambda_{p,\perp}- \lambda_{m,\perp}}\right)p\text{  ($u_{\perp}$= 0 at wall)}\nonumber\\
&&= n_{1} p\text{  (simplification by assuming $\lambda_{p,\perp}$= -$\lambda_{m,\perp}$= $\lambda_{\perp}$)}
\label{eq:a2_18}
\end{eqnarray}
Similarly,
\begin{equation}
(G_{\perp 3})_{b} = 2n_{2}\left(\frac{\lambda_{p,\perp}}{\lambda_{p,\perp}- \lambda_{m,\perp}}\right)p= n_{2} p\text{  (simplification by assuming $\lambda_{p,\perp}$= -$\lambda_{m,\perp}$= $\lambda_{\perp}$)}
\label{eq:a2_19}
\end{equation}
\begin{equation}
(G_{\perp 4})_{b}= 0
\label{eq:a2_20}
\end{equation}

\textbf{\textit{Normal flux at farfield boundary:}} Farfield boundary is a boundary far from the body. We are computing the normal flux at that boundary using the same scheme as that at the interior interfaces. That is, we are using Equation \eqref{eq:2d_euler_9b}, where the left state is the interior state (L= $int$) and the right state has freestream conditions (R= $freestream$).

\textbf{\textit{Normal flux at supersonic outflow and inflow boundary:}} Our expression for interface normal flux, given by Equation \eqref{eq:2d_euler_9b} gets simplified at a supersonic outflow and inflow boundary if we make two simplifying assumptions. The first assumption is that not just the flow, but the normal flow at the boundary is supersonic as well. Our second assumption is that the supersonic flow does not change between the interior and exterior states at the boundary $b$, {\em i.e.}, $int$= $ext$ at the boundary $b$. Now, let us consider the outflow ($u_{\perp}>$0) case. Since the normal flow is supersonic,
\begin{equation}
u_{\perp}- a >0
\label{eq:a2_21}
\end{equation}
Therefore,
\begin{equation}
\left(u_{\perp}- \sqrt{\frac{\gamma -1}{2\gamma}}a\right)_{int}> \left(u_{\perp}- a\right)_{int}> 0
\label{eq:a2_22}
\end{equation}
Across the boundary, since the flow is assumed uniform,
\begin{equation}
(\lambda_{RH})_{b}= 0
\label{eq:a2_23}
\end{equation}
Therefore,
\begin{equation}
(\lambda_{m,\perp})_{b}= min\left((\lambda_{RH})_{b},(u_{\perp}-\sqrt{\frac{\gamma -1}{2\gamma}}a)_{int}\right)= min\left(0,(u_{\perp}-\sqrt{\frac{\gamma -1}{2\gamma}}a)_{int}\right)=0
\label{eq:a2_24}
\end{equation}
whereas $(\lambda_{p,\perp})_{b}>$ 0. Thus, we get,
\begin{equation}
(\textbf{G}_{\perp})_{b}= (\textbf{G}_{\perp})_{int}
\label{eq:a2_25}
\end{equation}
That is, at a boundary with supersonic outflow, the flow conditions are extrapolated from the interior of the domain. Similarly, we can show that at a supersonic inflow boundary,
\begin{equation}
(\textbf{G}_{\perp})_{b}= (\textbf{G}_{\perp})_{ext}
\label{eq:a2_26}
\end{equation}
Thus, at a supersonic inflow boundary, the supersonic conditions are externally imposed.


\begin{thebibliography}{10}
\expandafter\ifx\csname url\endcsname\relax
  \def\url#1{\texttt{#1}}\fi
\expandafter\ifx\csname urlprefix\endcsname\relax\def\urlprefix{URL }\fi
\expandafter\ifx\csname href\endcsname\relax
  \def\href#1#2{#2} \def\path#1{#1}\fi

\bibitem{chu1965kinetic}
C.~Chu, Kinetic-theoretic description of the formation of a shock wave, Physics
  of Fluids 8~(1) (1965) 12--22.

\bibitem{sanders1974possible}
R.~Sanders, K.~H. Prendergast, The possible relation of the 3-kiloparsec arm to
  explosions in the galactic nucleus, The Astrophysical Journal 188 (1974)
  489--500.

\bibitem{pullin1980direct}
D.~Pullin, Direct simulation methods for compressible inviscid ideal-gas flow,
  Journal of Computational Physics 34~(2) (1980) 231--244.

\bibitem{reitz1981one}
R.~D. Reitz, One-dimensional compressible gas dynamics calculations using the
  {Boltzmann} equation, Journal of Computational Physics 42~(1) (1981)
  108--123.

\bibitem{deshpande1986second}
S.~M. Deshpande, A second-order accurate kinetic-theory-based method for
  inviscid compressible flows, Tech. Rep. NASA TP 2613 (1986).

\bibitem{deshpande1986kinetic}
S.~M. Deshpande, Kinetic theory based new upwind methods for inviscid
  compressible flows, in: AIAA $24^{th}$ Aerospace Sciences Meeting, no.
  AIAA-86-0275, 1986.

\bibitem{mandal1994kinetic}
J.~C. Mandal, S.~M. Deshpande, Kinetic {Flux} {Vector} {Splitting} for {Euler}
  equations, Computers \& Fluids 23~(2) (1994) 447--478.

\bibitem{kaniel1988kinetic}
S.~Kaniel, A kinetic model for the compressible flow equations, Indiana
  University Mathematics Journal 37~(3) (1988) 537--563.

\bibitem{perthame1990boltzmann}
B.~Perthame, Boltzmann type schemes for gas dynamics and the entropy property,
  SIAM Journal on Numerical Analysis 27~(6) (1990) 1405--1421.

\bibitem{prendergast1993numerical}
K.~H. Prendergast, K.~Xu, Numerical hydrodynamics from gas-kinetic theory,
  Journal of Computational Physics 109~(1) (1993) 53--66.

\bibitem{raghurama1995peculiar}
S.~V. Raghurama~Rao, S.~M. Deshpande, Peculiar velocity based upwind method for
  inviscid compressible flows, Computational Fluid Dynamics Journal 3 (1995)
  415--432.

\bibitem{natalini1998discrete}
R.~Natalini, A discrete kinetic approximation of entropy solutions to
  multidimensional scalar conservation laws, Journal of differential equations
  148~(2) (1998) 292--317.

\bibitem{aregba2000discrete}
D.~Aregba-Driollet, R.~Natalini, Discrete kinetic schemes for multidimensional
  systems of conservation laws, SIAM Journal on Numerical Analysis 37~(6)
  (2000) 1973--2004.

\bibitem{shrinath2023kinetic}
K.~Shrinath, N.~Maruthi, S.~V. Raghurama~Rao, V.~Vasudev~Rao, A kinetic flux
  difference splitting method for compressible flows, Computers \& Fluids 250
  (2023) 105702.

\bibitem{ROY2023106016}
S.~S. Roy, S.~V. Raghurama~Rao, A kinetic scheme with variable velocities and
  relative entropy, Computers \& Fluids 265 (2023) 106016.

\bibitem{estivalezes1996high}
J.~Estivalezes, P.~Villedieu, High-order positivity-preserving kinetic schemes
  for the compressible euler equations, SIAM journal on numerical analysis
  33~(5) (1996) 2050--2067.

\bibitem{GRESSIER1999199}
J.~Gressier, P.~Villedieu, J.-M. Moschetta, Positivity of flux vector splitting
  schemes, Journal of Computational Physics 155~(1) (1999) 199--220.

\bibitem{EINFELDT1991273}
B.~Einfeldt, C.~Munz, P.~Roe, B.~Sjögreen, On godunov-type methods near low
  densities, Journal of Computational Physics 92~(2) (1991) 273--295.

\bibitem{PARENT2013194}
B.~Parent, Positivity-preserving flux difference splitting schemes, Journal of
  Computational Physics 243 (2013) 194--209.

\bibitem{quirk1997contribution}
J.~J. Quirk, A contribution to the great {Riemann} solver debate, International
  Journal of Numerical Methods in Fluids 18 (1994) 555--574.

\bibitem{bhatnagar1954model}
P.~L. Bhatnagar, E.~P. Gross, M.~Krook, A model for collision processes in
  gases. i. small amplitude processes in charged and neutral one-component
  systems, Physical Review 94~(3) (1954) 511--525.

\bibitem{PARENT2011238}
B.~Parent, Positivity-preserving flux-limited method for compressible fluid
  flow, Computers \& Fluids 44~(1) (2011) 238--247.

\bibitem{EINFELDT1988}
B.~Einfeldt, On godunov-type methods for gas dynamics, SIAM Journal on
  Numerical Analysis 25~(2) (1988) 294--318.

\bibitem{SHASHI_HYP}
S.~S. Roy, S.~V. Raghurama~Rao, An entropic kinetic scheme with compactly
  supported velocities, in: Hyperbolic Problems: Theory, Numerics,
  Applications. Volume II, Springer Nature Switzerland, Cham, 2024, pp.
  433--443.

\bibitem{bouchut1999construction}
F.~Bouchut, Construction of {BGK} models with a family of kinetic entropies for
  a given system of conservation laws, Journal of Statistical Physics 95~(1)
  (1999) 113--170.

\bibitem{gottlieb2001strong}
S.~Gottlieb, C.-W. Shu, E.~Tadmor, Strong stability-preserving high-order time
  discretization methods, SIAM review 43~(1) (2001) 89--112.

\bibitem{kumar}
R.~Kumar, A.~K. Dass, A new flux-limiting approach–based kinetic scheme for
  the euler equations of gas dynamics, International Journal for Numerical
  Methods in Fluids 90~(1) (2019) 22--56.

\bibitem{1987PhDT_Martinelli}
L.~{Martinelli}, {Calculations of viscous flows with a multigrid method}, Ph.D.
  thesis, Princeton University, New Jersey (Jan. 1987).

\bibitem{toro2013riemann}
E.~F. Toro, Riemann solvers and numerical methods for fluid dynamics: a
  practical introduction, Springer Science \& Business Media, 2013.

\bibitem{yee1982high}
H.~Yee, R.~Warming, A.~Harten, A high-resolution numerical technique for
  inviscid gas-dynamic problems with weak solutions, in: Eighth International
  Conference on Numerical Methods in Fluid Dynamics: Proceedings of the
  Conference, Rheinisch-Westf{\"a}lische Technische Hochschule Aachen, Germany,
  June 28--July 2, 1982, Springer, 1982, pp. 546--552.

\bibitem{levy1993use}
D.~W. Levy, K.~G. Powell, B.~van Leer, Use of a rotated {Riemann} solver for
  the two-dimensional {Euler} equations, Journal of Computational Physics
  106~(2) (1993) 201--214.

\bibitem{woodward1984numerical}
P.~Woodward, P.~Colella, The numerical simulation of two-dimensional fluid flow
  with strong shocks, Journal of computational physics 54~(1) (1984) 115--173.

\bibitem{PARENT2012173}
B.~Parent, Positivity-preserving high-resolution schemes for systems of
  conservation laws, Journal of Computational Physics 231~(1) (2012) 173--189.

\bibitem{jones1985reference}
D.~J. Jones, Test Cases for Inviscid Flow Field Methods, 1985, Ch. Reference
  test cases and contributors, {AGARD} Advisory Report No. AGARD-AR-211.

\bibitem{dervieux1989numerical}
A.~Dervieux, B.~van Leer, J.~Periaux, A.~Rizzi, Numerical simulation of
  compressible {Euler} flows: A {GAMM} workshop, Notes on Numerical Fluid
  Mechanics, Vieweg Verlag, 1989, (Proceedings of the GAMM Workshop on
  Numerical simulation of compressible Euler flows, held at {INRIA},
  {Rocquencourt}, June 10--13, 1986).

\bibitem{degrez1987interaction}
G.~Degrez, C.~Boccadoro, J.~F. Wendt, The interaction of an oblique shock wave
  with a laminar boundary layer revisited: {An} experimental and numerical
  study, Journal of Fluid Mechanics 177 (1987) 247--263.

\bibitem{parthasarathy1995directional}
V.~Parthasarathy, Y.~Kallinderis, Directional viscous multigrid using adaptive
  prismatic meshes, AIAA Journal 33~(1) (1995) 69--78.

\bibitem{kullback1951information}
S.~Kullback, R.~Leibler, On information and sufficiency, Annals of Mathematical
  Statistics 22 (1951) 79--86.

\end{thebibliography}

\end{document}